\documentclass[a4paper,fleqn,usenatbib,useAMS]{mnras}

\usepackage{graphicx}	% Including figure files
\usepackage{amsmath}	% Advanced maths commands
\usepackage{amssymb}	% Extra maths symbols
\usepackage{multicol}        % Multi-column entries in tables
\usepackage{bm}		% Bold maths symbols, including upright Greek
\usepackage{pdflscape}	% Landscape pagess
\usepackage[encapsulated]{CJK}
\usepackage{ucs}
\usepackage[utf8x]{inputenc}
\usepackage{multirow}
\usepackage{booktabs,caption}
\usepackage[flushleft]{threeparttable}
\usepackage[dvipsnames]{xcolor}
\usepackage{natbib}
\usepackage{times}
\usepackage{calc}
\usepackage{rotating}
\usepackage{lscape}
\bibpunct{(}{)}{;}{a}{}{,} % to follow the A&A style
\usepackage{longtable}
\usepackage[referable]{threeparttablex}
\usepackage{threeparttable}
\usepackage{enumitem} 
\usepackage{hyperref}
\usepackage{pifont}
\usepackage{tikz}
\usepackage{xcolor}
\usepackage{soul}

\usepackage[T1]{fontenc}
\usepackage{ae,aecompl}

\usepackage{newtxtext,newtxmath}
\title[Hot-WD candidates from the IGAPS-GALEX cross-match]{Hot white dwarf candidates from the IGAPS-GALEX cross-match}
\author[G{\'o}mez-Mu{\~n}oz, et al.] 
{G\'omez-Mu$\mathrm{\Tilde{n}}$oz, M.~A.$^1$\thanks{E-mail:mgomez\_astro@outlook.com}, Sabin, L.$^1$, Raddi, R.$^2$, and Wells, R.~D.$^3$\\
$^1$Instituto de Astronom\'\i a, Universidad Nacional Aut{\'o}noma de M{\'e}xico, Apdo.\ Postal 877, Ensenada 22860, B.C., Mexico \\
$^2$Departament de F\'\i sica, Universitat Polit{\`e}cnica de Catalunya, c/Esteve Terrades 5, E-08860 Castelldefels, Spain \\
$^3$Center for Space and Habitability, University of Bern, Gesellschaftsstrasse 6, 3012, Bern, Switzerland \\
}

\pubyear{2022}

\begin{document}
\label{firstpage}
\pagerange{\pageref{firstpage}--\pageref{lastpage}}
\maketitle

\begin{abstract}

%-Context
White dwarf (WD) stars are often associated with the central stars of planetary nebulae (CSPNe) on their way to the cooling track.  A large number of WD star candidates have been identified thanks to optical large-scale surveys such as Gaia DR2 and EDR3. However, hot-WD/CSPNe stars are quite elusive in optical bands due to their high temperatures and low optical luminosities.
%- Aims
The Galaxy Evolution Explorer (\textit{GALEX}) matched with the INT Galactic Plane Survey (IGAPS) allowed us to identify hot-WD candidates by combining the \textit{GALEX} far-UV (\textit{FUV}) and near-UV (\textit{NUV}) with optical photometric bands (\textit{g, r, i} and H$\alpha$). After accounting for source confusion and filtering bad photometric data, a total of 236\,485 sources were found in the \textit{GALEX} and IGAPS footprint (GaPHAS). 
%- Results
A preliminary selection of hot stellar sources was made using the \textit{GALEX} colour cut on \textit{FUV}$-$\textit{NUV}$>-$0.53, yielding 74 hot-WD candidates. We analysed their spectral energy distribution (SED) by developing a fitting program for single- and two-body SED using an MCMC algorithm; 41 are probably binary systems (a binary fraction of $\sim$55\% was estimated).
Additionally, we classified the WD star candidates using different infrared (IR) colours available for our sample obtaining similar results as in the SED analysis for the single and binary systems.
%- Conclusions
This supports the strength of the fitting method and the advantages of the combination of \textit{GALEX} UV with optical photometry. Ground-based time-series photometry and spectra are required in order to confirm the nature of the WD star candidates.

\end{abstract}


%%%%%%%%%%%%%%%%%%%%%%%%%%%%%%%%%%%%%%%%%%%%%%%%%%

%%%%%%%%%%%%%%%%% BODY OF PAPER %%%%%%%%%%%%%%%%%%

% The MNRAS class isn't designed to include a table of contents, but for this document one is useful.
% I therefore have to do some kludging to make it work without masses of blank space.
%\begingroup
%\let\clearpage\relax
%\tableofcontents
%\endgroup
%\newpage

\section{Introduction}

White dwarfs (WDs) represent the late stage of stellar evolution for low- and intermediate-mass stars ($\sim$0.8-8 M$\sun$ stars) and are often associated with Planetary Nebulae (PNe) as being their central stars (CSPNe) on their way to the cooling track (see \cite{Koester1990, Weidmann2020, Jones2020}).

PNe are ionised shells of gas and dust that will eventually merge with the interstellar medium (ISM) after $\sim$10${^4}$ years \citep{Kwitter2022}. Hence, the most evolved CSPNe are less likely to still be surrounded by a bright and well visible nebula. It is, therefore (extremely) difficult to detect these old PNe via traditional optical imaging analysis through the mapping of H$\alpha$ emission for example \citep{Parker2005,Parker2006,Sabin2014}.
Thus, in order to trace the population of PNe in their most advanced stage of evolution and by extension to improve the PN census, the detection of their central stars is an alternative. Indeed, the identification of the ''youngest'' and therefore hottest WDs i.e. at the tip of the cooling track (T$_\mathrm{eff}$)$\gtrapprox$ 50 kK), would likely point towards an associated old and low surface brightness PN.

Various galactic optical large-scale surveys providing a deep scanning have been carried out, allowing the retrieval of a large number of WDs candidates. We can cite in particular the use of Gaia DR2 data \citep{Gaia2018} by \citet[See citations therein for a full summary of WD researches]{Gentile2019} and Gaia EDR3 \citep{EDR32021} by \citet{Gentile2021} as well as those from the Sloan Digital Sky Survey \citep{York2000} by \citet{Kepler2016,Kepler2019} for example.

However, \citet{Gentile2019} pointed out that their catalogue of white dwarfs lacked completeness in the areas close to the Galactic plane and in crowded regions due to their selection criteria, although there were some improvements using Gaia EDR3. It is also important to notice the inherent faintness of WDs/CSPNe which makes their detection difficult.

Therefore, in an attempt for completeness, we propose to use the latest deep optical surveys performed in Galactic Plane namely the \textit{INT and VST Galactic Plane Surveys} \citep[IGAPS;][]{Monguio2020} and \citep[VPHAS+;][]{Drew2014}) associated to the \textit{Galaxy Evolution EXplorer} \citep[\textit{GALEX};][]{Martin2005} survey. This will offer a unique opportunity to investigate hot stellar sources in the spectral range from UV to optical. 

We propose to assemble a comprehensive matched catalogue, named GaPHAS, of unique sources that are in the footprint of the \textit{GALEX} and the IGAPS/VPHAS+ catalogues. For our purpose, the sources of interest are hot objects (with $T_\mathrm{eff}$ $\geq$50 000 K) as they are more likely to be the nuclei of old/mature PNe. The outcome of this work would therefore set the stage for a deep imaging investigation aiming at unveiling the possibly still present surrounding PNe.

The article is organised as follows. In Section \ref{sec:sky_surveys} we present the surveys used to detect the WDs as well as the cross-match process and its result. In Section \ref{sec:analysis_selection} we describe the hot-WD selection process and its outcome. The stellar analysis of these hot-WDs via their spectral energy distribution is discussed in Section \ref{sec:SED_analysis}. Finally our discussion and conclusion are presented in Section \ref{discussion} and Section \ref{conclusion} respectively.

\section{The sky surveys}\label{sec:sky_surveys}

\subsection{The IGAPS and VPHAS+ surveys}\label{subsec:igaps_vphas}

The INT Galactic Plane Surveys\footnote{\url{http://www.star.ucl.ac.uk/IGAPS/}} is the latest release of two surveys of the Northern Galactic plane conducted with the Wide Field Camera (WFC) mounted on the Isaac Newton Telescope (INT). IGAPS combines the recalibrated data from the INT Photometric H$\alpha$ survey \citep[IPHAS;][]{Drew2005, Barentsen2014} and the UV-Excess survey \citep[UVEX;][]{Groot2009} with a new astrometric solution that has been tied to the {\em Gaia} DR2 reference frame. Both surveys cover the range |b| < 5${\degr}$ and 30${\degr}$ < l < 215${\degr}$, which was scanned with the broadband filters Sloan i, r, g, and U$_{RGO}$, and narrow-band H$\alpha$.

The VST Photometric H$\alpha$ Survey\footnote{\url{https://www.vphasplus.org/}}  has scanned the Southern Galactic Plane in the same latitude the range as IPHAS and UVEX and the Bulge between the range |b| < 10${\degr}$. This survey was performed with the Omega-CAM imager \citep[][]{Kuijken2002} on the VLT Survey Telescope \citep[VST;][]{Capaccioli2012} and uses (nearly) the same set of filters as its northern counterparts (Sloan broadband u, g, r and i; and narrow-band H$\alpha$).

Finally, it is worth noting that IGAPS and VPHAS+ are $\sim$1 arcsec angular resolution CCD surveys going down to $\sim$21 mag and provide photometric data for $\sim$300 million stars each.

\subsection{The Galaxy Evolution Explorer}\label{subsec:GALEX}

The \textit{Galaxy Evolution Explorer} \citep[\textit{GALEX};][]{Martin2005} imaged the sky in
far-UV (\textit{\textit{FUV}}, 1344--1786{\AA}, $\lambda_\text{eff}=$1538.6{\AA})
and near-UV (\textit{NUV}, 1771--2831{\AA}, $\lambda_\text{eff}=$2315.7{\AA}),
simultaneously, with a field-of-view of 1{\fdg}2.
The image resolution in \textit{\textit{FUV}} and \textit{NUV} is 4{\farcs}2 and 5{\farcs}3 \citep{Morrissey2007}, respectively,
sampled with virtual pixels of 1{\farcs}5~pixel$^{-1}$ as reconstructed from
photon counting recordings \citep{Bianchi2009}.
\textit{GALEX} contains a few main surveys with different coverage and depths,of which we use the All-sky
Imaging Survey (AIS) and the Medium [depth] Imaging Survey (MIS), the widest sky
surveys in \textit{GALEX}, reach a typical depth of 19.9/20.8 (\textit{\textit{FUV}}/\textit{NUV}) and 22.6/22.7 (\textit{\textit{FUV}}/\textit{NUV}),
respectively, in the AB magnitude system \citep[see][for a description of data products]{Morrissey2007,Bianchi2009apss}.

We used the \textit{GALEX} sixth and seventh releases (GR6+7)
AIS and MIS surveys, including a total of $\sim$83$\times$10$^{6}$
unique sources in the \textit{GALEX} catalogue of UV sources \citep[GUVcat;][]{Bianchi2017}.
The typical depth of GUVcat is 19.9 and 20.8 in \textit{\textit{FUV}} and \textit{NUV}, respectively.

\subsection{Matching IGAPS/VPHAS+ with \textit{GALEX}}\label{subsec:matchgin_GALEX_igaps}

\begin{figure*}
    \centering
    % \vspace{0.2cm}
    \includegraphics[width=0.7\textwidth]{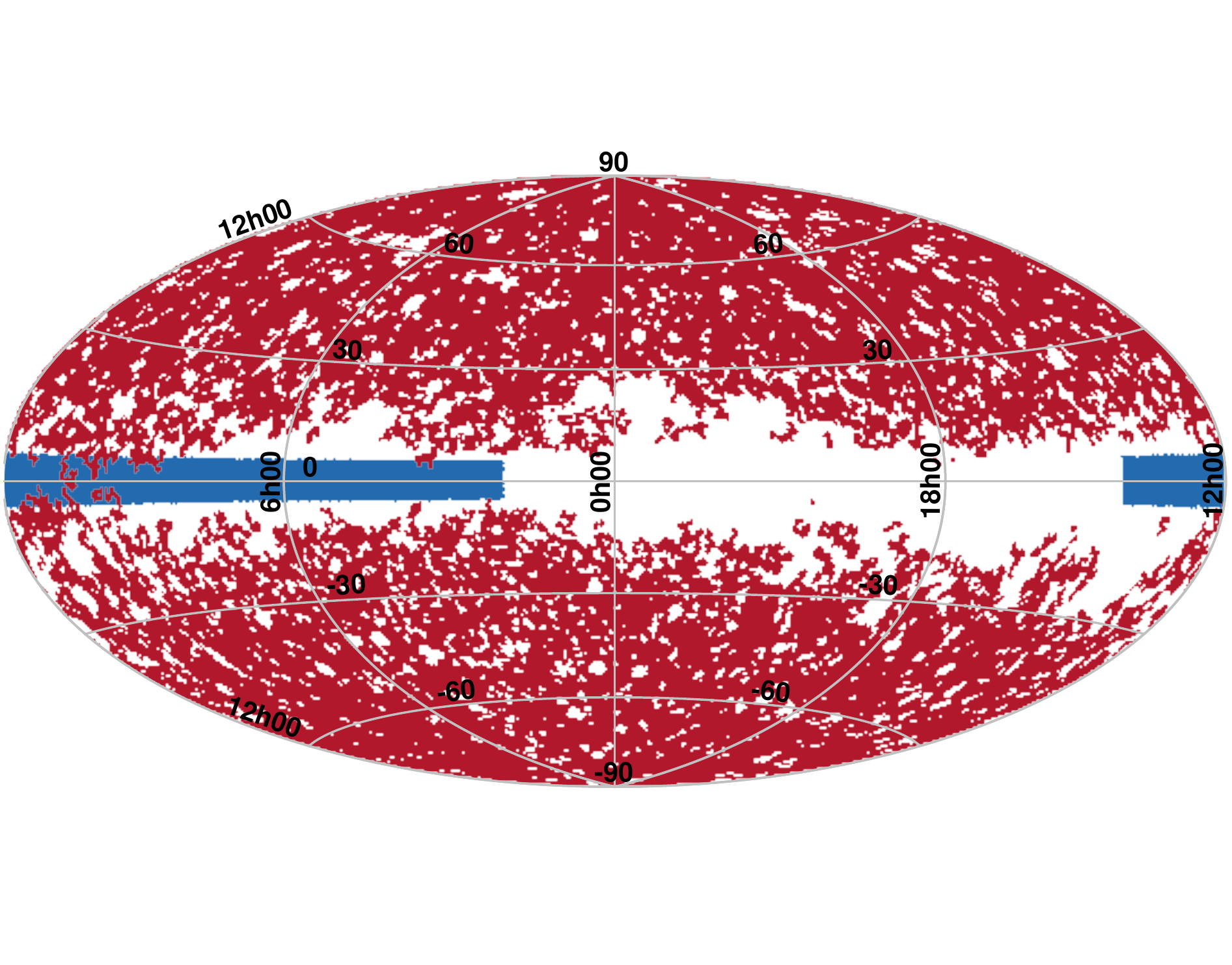}
    \vspace{-1.2cm}
    \caption{Red: Sky coverage of the \textit{GALEX} catalogue GUVcat in Galactic coordinates. Blue: Sky coverage
    of IGAPS catalogue in Galactic coordinates. Only a small portion of the sky is in the footprint
    of both catalogues.}
    \label{fig:aitoff_footprint}
\end{figure*}

IGAPs and GUVcat catalogues were matched using the SIMBAD
TAP VizieR service\footnote{\url{http://tapvizier.u-strasbg.fr/adql/}}, which provides
access to all VizieR tables using the Astronomical Data Query Language \citep[ADQL;][]{adql},
by using the JOIN method around a matching radius of 5{\arcsec}.
A total of 250\,180 cross matches were found (hereafter, the GaPHAS catalogue).
Note that although the \textit{GALEX} sky coverage is fairly complete, there are very few observed fields
located at low Galactic latitudes (between $|b|<5$ in which IGAPS and VPHAS+ are defined)
due to \textit{GALEX} brightness safety limits \citep[see][]{Bianchi2017}, which explains the small
number of sources between IGAPS and GUVcat, and the null sources between VPHAS+
and GUVcat (see Figure~\ref{fig:aitoff_footprint}).
Therefore, the VPHAS+ catalogue will not be part of this investigation.

The IGAPS catalogue also contains a flag, \texttt{Class}, to separate between point-like (=$-$1), extended (=1), and noise (=0) sources; for which GaPHAS catalogue contains 146\,663, 39\,658, and 13\,695 sources, respectively. It can also indicate probable point-like (=$-2$; 1456 in GaPHAS) and probably extended (=$-3$; 0 in GaPHAS) sources. For the purpose of this work we only considered all sources without \texttt{Class}=0 set, leaving a total of 236\,485 sources in the GaPHAS catalogue.

Multiple sources from the IGAPS catalogue can match the same source in \textit{GALEX} due to differences in resolution. In such cases, the TAP query returns multiple rows with the same \textit{GALEX} source and different IGAPS matches. We created a tag, \texttt{distancerank}, to classify these objects according to the distance between IGAPS and \textit{GALEX} coordinates \citep[following the same method as described in Table~1 of][]{bianchi2020}. If for a given \textit{GALEX} source there is only one IGAPS match we set \texttt{distancerank}=0, otherwise we set \texttt{distancerank}=1 for the closest match whereas the other matches were noted (\texttt{distancerank}>1) and ordered by distance. In the same way, we created a tab \texttt{inversedistancerank}, to classify those matches with multiple rows with the same IGAPS name and different GALEX matches \citep[algouth these are rare cases; see Table~3 and 4 of][]{bianchi2020}.

According to the work of \citet{Bianchi2011}, and more recently of \citet{bianchi2020}, the number of spurious matches increase considerably toward the Galactic disc. To investigate the probability of spurious matches we follow the same method as \citet{Bianchi2011,bianchi2020}. We offset the GUVcat catalogue by 30-arcsec in (RA and DEC) and cross-matched it against the IGAPS catalogue (only \texttt{Class}=$-$1 were taken). We found just 4189 matches that correspond to a 3\% of false positives.

\section{WD candidate selection} \label{sec:analysis_selection}
\label{subsec:wd_selection}

\begin{figure*}
    \centering
    \includegraphics[width=0.45\textwidth]{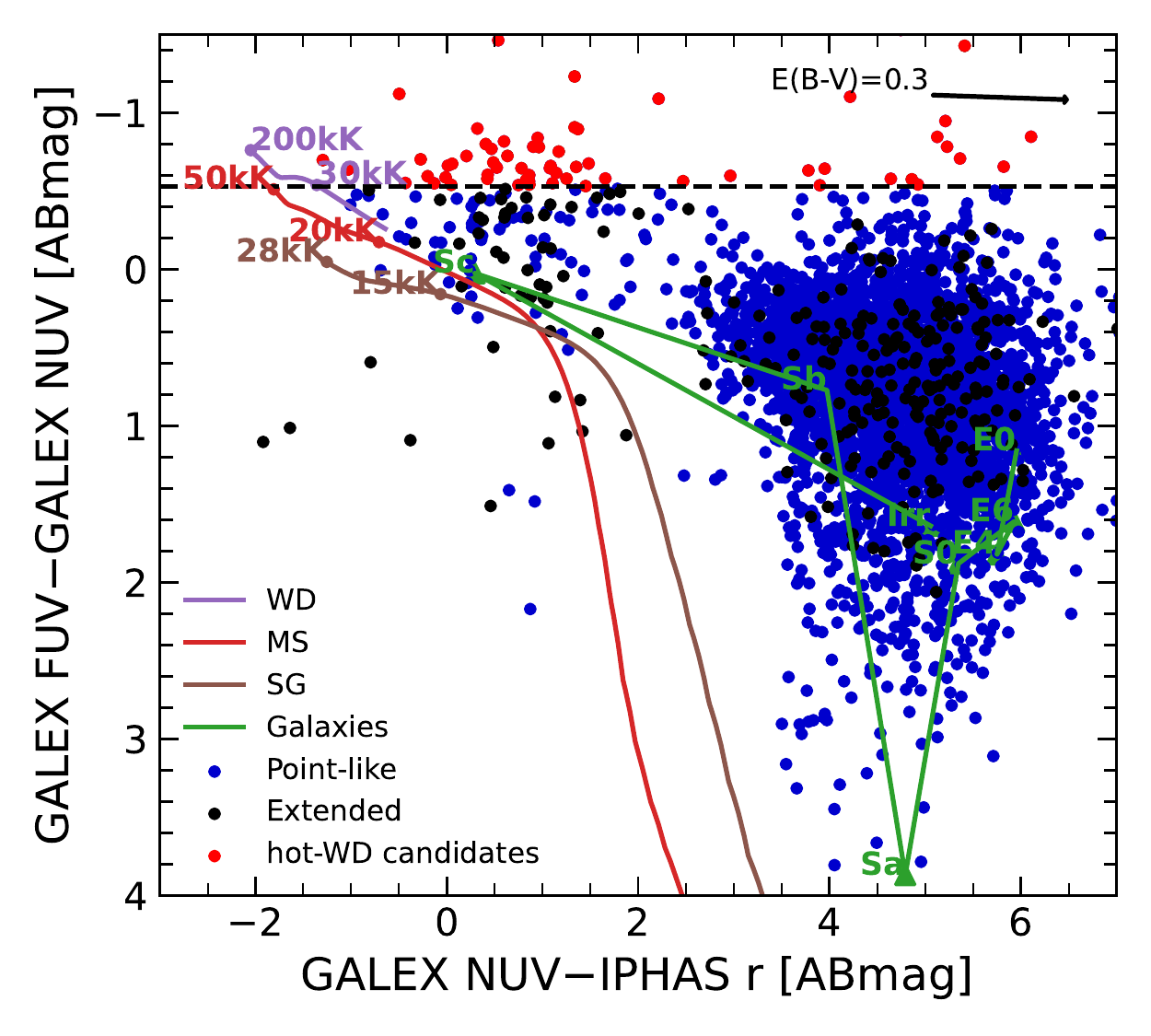}
    \includegraphics[width=0.45\textwidth]{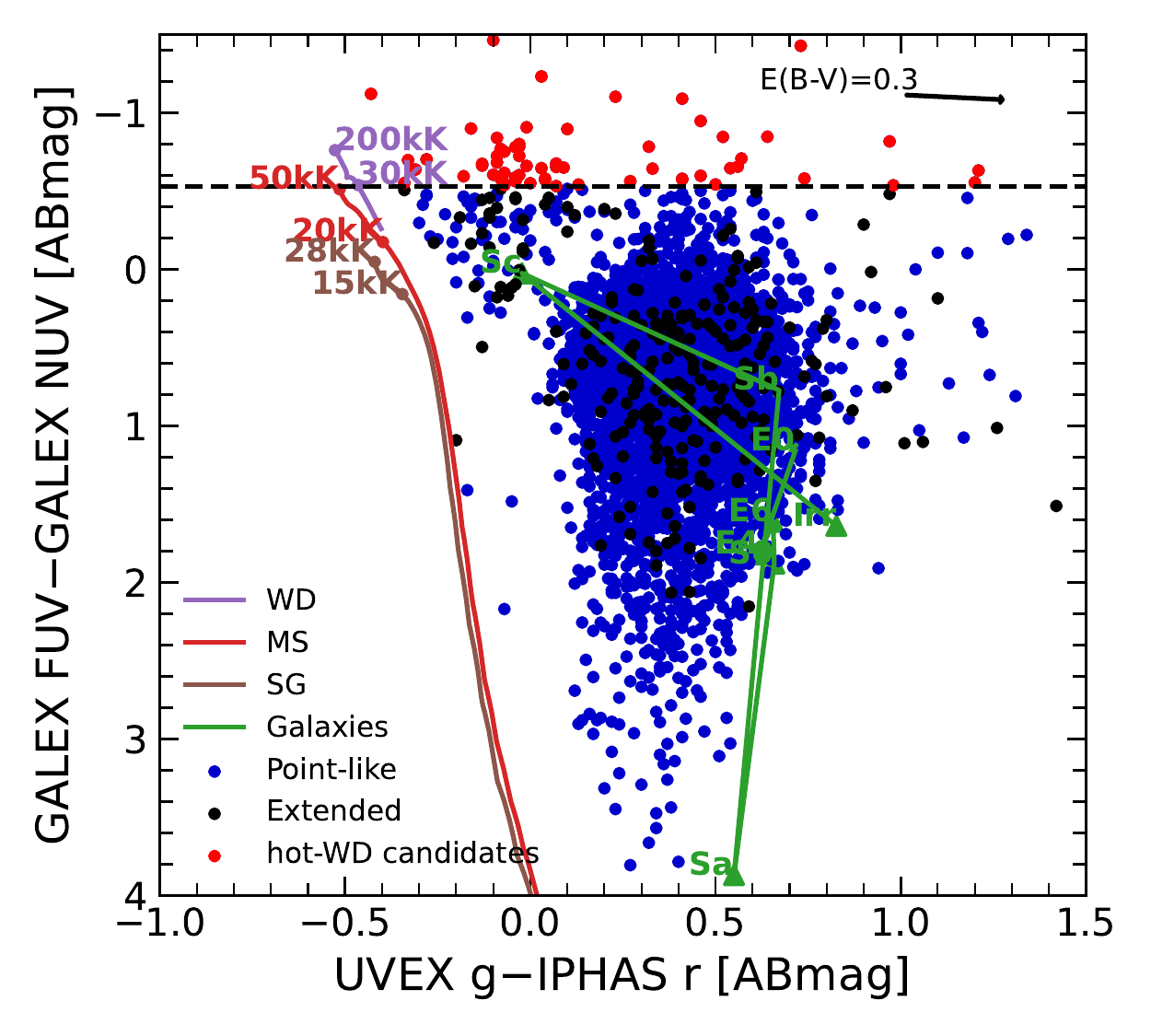}
    \includegraphics[width=0.45\textwidth]{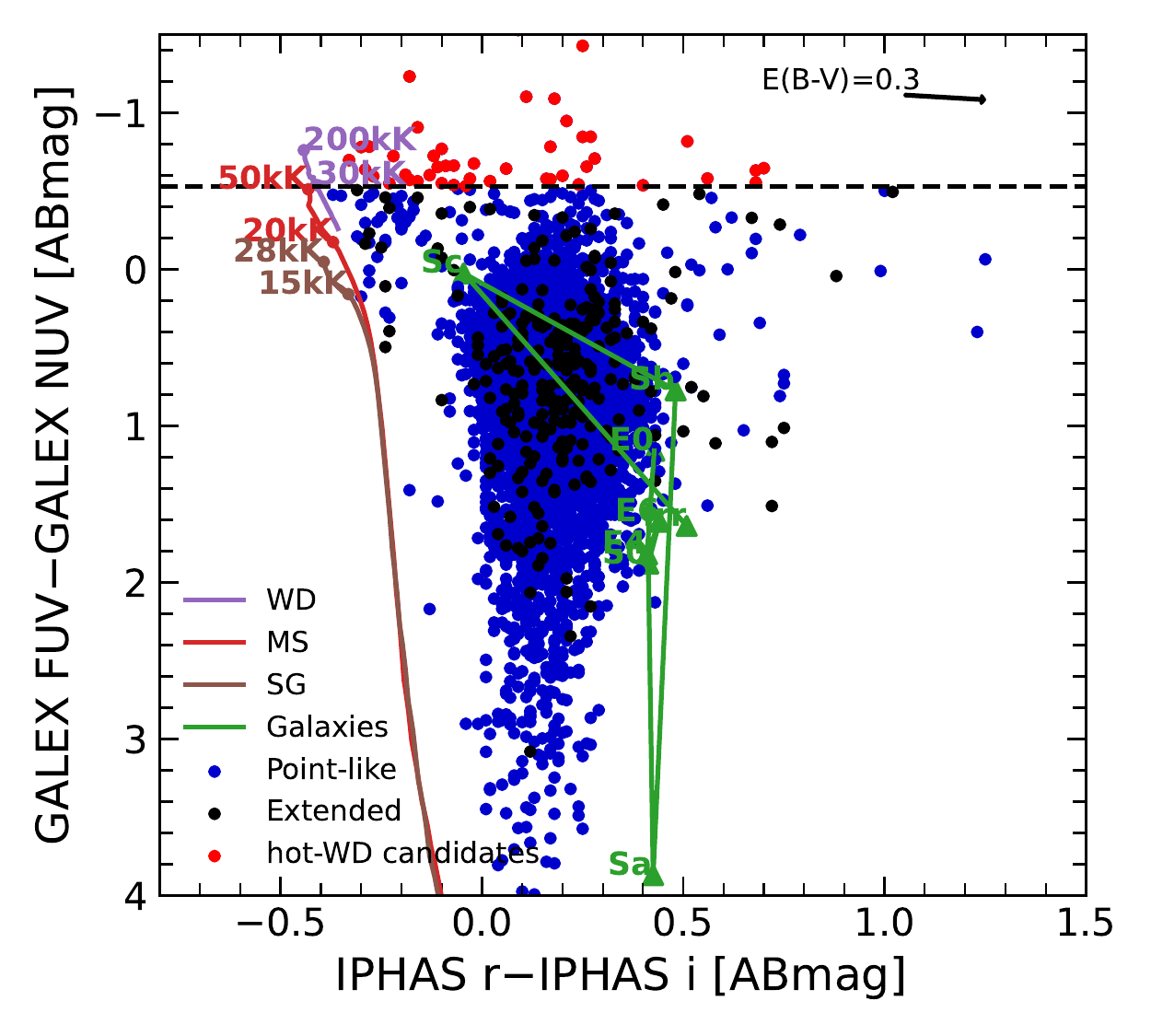}
    \includegraphics[width=0.45\textwidth]{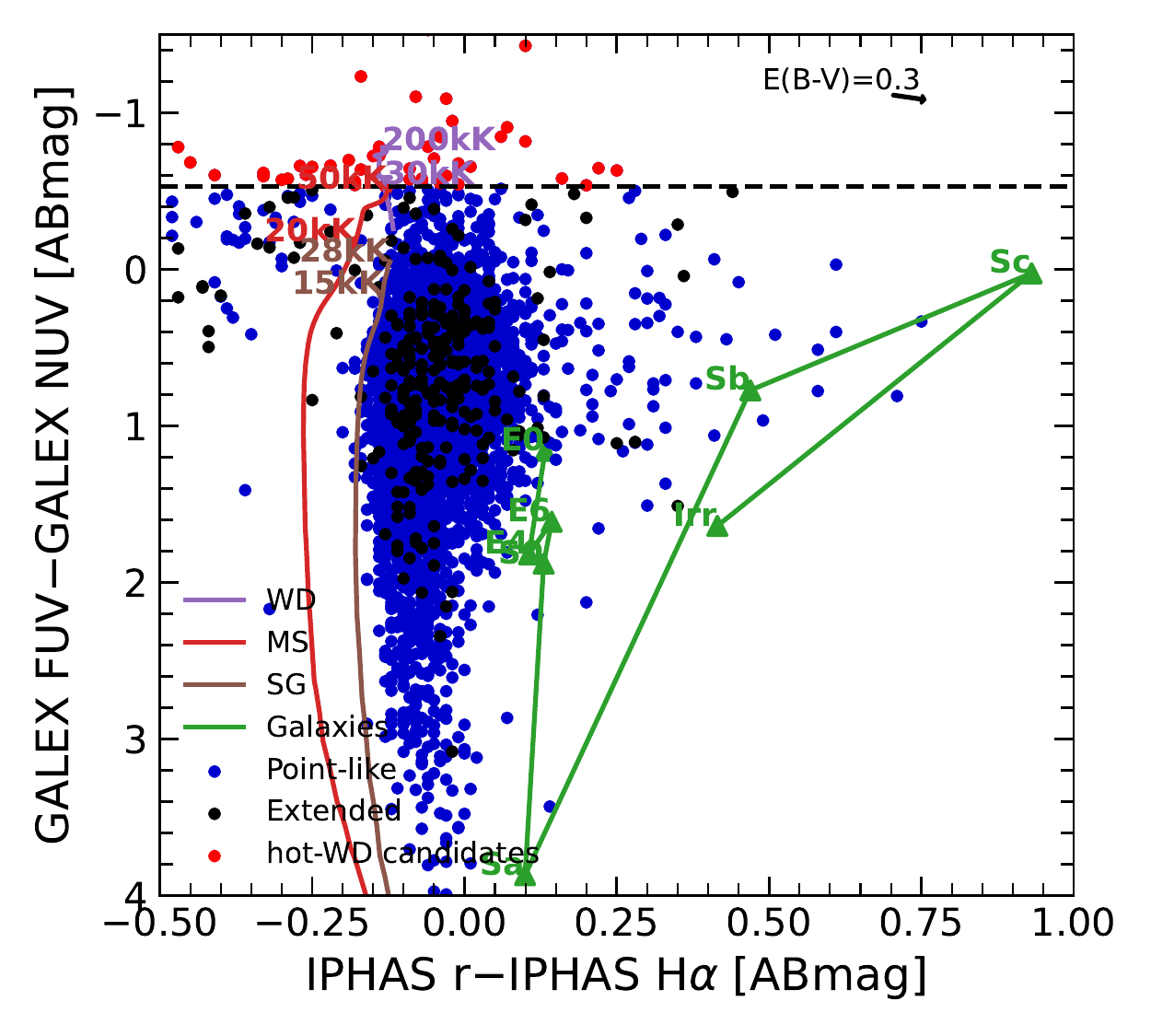}
    \caption{Color-color diagrams for IGAPS sources matched with \textit{GALEX} UV sources.
    IGAPS point-like and extended sources are blue and black density points, respectively.
    The hot WD cooling sequence (purple line), the main-sequence (MS; red line), and the supergiant (SG; brown line)
    stellar sequences are for log(g) = 7.0, 5.0 and 3.0, respectively. Galaxy templates
    are shown in green for different morphologies labelled in the figure (E0, E4, E6, Sa, Sb, Sc, and Irr).
    Reddening vector for $E(B-V)=0.3$~mag is also shown in the upper-right side on each panel. The black dashed line indicates the cut of the hot WD candidates (red dots) using the \textit{FUV}$-$\textit{NUV}$-<$0.53 colour cut.
    }
    \label{fig:cc_plots_1}
\end{figure*}

We selected WD candidates via colour-colour cuts that are defined by the intrinsic colours of hot-WD \citep[e.g.,][]{Bianchi2011,bianchi2020}. In order to do so, we analysed a different set of UV-optical colour combinations. Figure~\ref{fig:cc_plots_1} shows different colour-colour diagrams of the point-like (blue dots) and extended (black dots) sources, as defined in the tag \texttt{Class} from IGAPS, of the GaPHAS catalogue. 
We computed  theoretical stellar colours for main-sequence (MS; red line) and supergiants (SG; brown line) stars using the stellar atmosphere models of \citet{Castelli2003} for solar metallicity with $\log g=5.0$ and $\log g=3.0$, respectively, to guide the eye in the interpretation of the GaPHAS colours. We also calculated WD model colours (purple line) using the H-He non-LTE atmosphere models computed by \citet{Bianchi2009apss,Bianchi2011}, with the \textsc{Tlusty} code \citep{Hubeny1995}, with solar metallicity and $\log g=7.0$.
A reddening vector with $E(B-V)=0.3$~mag is also shown in Figure~\ref{fig:cc_plots_1} for a typical Milky-Way type dust, with $R_{V}=3.1$, using the \citet[][hereafter CCM89]{Cardelli1989} extinction law.

In order to select reliable photometry of hot stellar sources from the GaPHAS catalogue, we selected the sources with the \texttt{distancerank}=0 and \texttt{inversedistancerank}=0 and with error cuts in optical and UV bands of 0.2~mag and 0.3~mag, respectively. We also removed the sources with the \texttt{saturated}=1 (from IGAPS catalogue) which indicates a saturated source in one or more than one optical bands. Similarly, we only selected \textit{FUV} and \textit{NUV} with magnitudes fainter than 13.73 and 13.85~mag, respectively \citep[see][for more information related to non-linearity limits]{bianchi2018}.
According to the work of \citet{Bianchi2011} the \textit{FUV}$-$\textit{NUV}$<-$0.13 colour cut corresponds to stellar $T_{\rm eff}$ hotter than 18\,000~K (exact value might vary with gravity).
Based on the Figure~7 of \citet{Bianchi2011}, we set \textit{FUV}$-$\textit{NUV}$<-$0.53 colour cut in order to select stellar sources with $T_{\rm eff}$ hotter than 30\,000~K (black dashed line in Figure~\ref{fig:cc_plots_1}) which covers the cases in which the hot-WD is young enough to still be likely to display a circumstellar ionised shell.
The \textit{FUV}$-$\textit{NUV} colour selection also limits the contaminant sources such as galaxies and MS stars \citep[e.g.,][and references therein; see Figure~\ref{fig:cc_plots_1}]{bianchi2020}.
A total of 74 sources were selected from the GaPHAS catalogue as probable hot-WD stellar sources (see Table~\ref{tab:example_wd}; red dots in Figure~\ref{fig:cc_plots_1}) according to the \textit{FUV}$-$\textit{NUV}$<-$0.53 criteria. We also added a tag \texttt{t50} indicating the sources with $T_{\rm eff}$ hotter than 50\,000~K according to the \textit{FUV}$-$\textit{NUV}$<-$0.60 colour; a total of 52 sources are part of this group.
Note that we only selected hot-WD candidates based on the \textit{FUV}$-$\textit{NUV} colour as it is usually less affected by the interstellar reddening (see $E(B-V)$ in Figure~\ref{fig:cc_plots_1}) and that other contaminants such as binary systems are not discarded \citep[e.g.,][for a review]{Bianchi2011,Raddi2017,bianchi2020}.

\subsection{Matching WD candidates with Gaia EDR3}
\label{subsec:matching_gaiaedr3}

Distances from Gaia EDR3 \citep{GaiaColab2020} were included for the WD star candidates shown in Table~\ref{tab:example_wd} when available.
Because of the difference in epochs between IGAPS (which is in the Gaia DR2 reference frame) and Gaia EDR3 ( up to 12~yr) we need to account for stellar proper motions \citep[see section~2.2 and section~4 of][, respectively]{Raddi2017,Monguio2020}. First, we matched the list of WD candidates shown in Table~\ref{tab:example_wd} to Gaia EDR3 \citep{GaiaColab2020} using a matching radius of 1{\arcmin}. The Gaia coordinates were converted to IGAPS observational epoch using the {\sc topcat} \emph{Gaia} command \texttt{epochProp}. Then, we re-calculated the sky distance between the Gaia propagated and IGAPS set of coordinates. Finally, the closest Gaia EDR3 match to the IGAPS coordinates was selected.
The Gaia EDR3 \texttt{Source} column was used to match the WD stars with the Bayesian statistical distances from \citet{BailerJones2021}.
Reported distances are only those with errors $\leq$30\% resulting in 40 WD stars with reliable distances (see Table~\ref{tab:example_wd}).
Of these, only 30 were previously classified as WD candidates by \citet{Gentile2021} (\texttt{detGF} in Table~\ref{tab:example_wd}) by means of Gaia colour-colour diagrams and SDSS spectra.

In addition, we also tagged all WD stars that are reported with H$\alpha$ flux excess by matching the WD stars with the \citet[][Population-based identification of H {\ensuremath{\alpha}}-excess sources in the Gaia DR2 and IPHAS catalogues]{Fratta2021} by matching the Gaia EDR3 object ids.
The Ha flux excess flag is also included in Table~\ref{tab:example_wd} when available; 17 WD star candidates are reported to have H$\alpha$ excess.

\subsection{Matching WD candidates with 2MASS and UKIDDS}
\label{subsec:match_wd_2massukidds}

We matched the WD candidates with the Two-Micron All-Sky Survey Point Source Catalogue \citep[2MASS PSC;][]{Skrustskie2006} and with the UKIRT Infrared Deep Sky Survey \citep[UKIDSS DR11;][]{Lawrence2007,Almaini2007} using a matching radius of 1{\arcsec}, resulting in a 23 and 27 WD candidates with IR measurements, respectively. Note that we only account for matches with the information on the three J, H, and K bands, simultaneously.

\begin{table*}
\caption{WD star candidates found between GALEX and IGAPS catalogues. \label{tab:example_wd}}
\resizebox{0.85\textwidth}{!}{
\begin{tabular}{ccccccccccccc}
\hline
IGAPS Name & RAJ2000 & DEJ2000 & FUV & NUV & g & r & i & H$\alpha$ & $r_{\rm geo}$ & detGF & t50 & H$\alpha^{\rm ex}$ \\ 
\cline{4-9}
      &        &         & \multicolumn{6}{c}{(mag)}         & (pc)          &        &     &        \\ 
\hline 
J012902.52+644429.1	&	01:29:02.52	&	+64:44:29.1	&	17.86	&	18.46	&	18.48	&	18.66	&	18.93	&	18.99	&	246.94	&	1	&	0	&	1 \\ 
J014150.56+650546.9	&	01:41:50.56	&	+65:05:46.9	&	18.64	&	19.31	&	17.90	&	17.83	&	17.85	&	17.84	&	2175.63	&	0	&	1	&	0 \\ 
J015418.01+654741.6	&	01:54:18.01	&	+65:47:41.6	&	19.81	&	20.59	&	19.59	&	19.63	&	19.93	&	20.10	&	320.34	&	1	&	1	&	1 \\ 
J023215.29+632115.9	&	02:32:15.29	&	+63:21:15.9	&	19.96	&	20.50	&	16.08	&	15.58	&	15.34	&	15.59	&	5715.61	&	0	&	0	&	0 \\ 
J023223.07+632040.9	&	02:32:23.07	&	+63:20:40.9	&	20.27	&	20.84	&	16.39	&	15.98	&	15.81	&	16.07	&	3837.86	&	0	&	0	&	1 \\ 
J023500.57+632737.0	&	02:35:00.57	&	+63:27:37.0	&	20.09	&	20.84	&	19.60	&	19.67	&	19.99	&	19.87	&	371.75	&	1	&	1	&	0 \\ 
J023527.80+632738.4	&	02:35:27.80	&	+63:27:38.4	&	20.67	&	21.52	&	16.91	&	16.39	&	16.14	&	16.43	&	7173.46	&	0	&	1	&	0 \\ 
J024130.89+632923.7	&	02:41:30.89	&	+63:29:23.7	&	18.85	&	19.58	&	19.28	&	19.37	&	19.59	&	19.52	&	967.14	&	1	&	1	&	0 \\ 
J024713.51+644122.5	&	02:47:13.51	&	+64:41:22.5	&	19.60	&	20.38	&	19.45	&	19.48	&	19.76	&	19.62	&	731.35	&	1	&	1	&	0 \\ 
J024849.31+640930.4	&	02:48:49.31	&	+64:09:30.4	&	18.87	&	19.43	&	18.53	&	18.57	&	18.73	&	18.75	&	390.29	&	1	&	0	&	0 \\ 
J024933.98+643643.0	&	02:49:33.98	&	+64:36:43.0	&	18.44	&	20.65	&	\,.\,.\,.	&	21.71	&	20.25	&	\,.\,.\,.	&	\,.\,.\,.	&	0	&	1	&	0 \\ 
J025236.54+544506.9	&	02:52:36.54	&	+54:45:06.9	&	20.45	&	21.35	&	20.08	&	19.98	&	19.83	&	20.51	&	308.61	&	1	&	1	&	0 \\ 
J025319.52+544414.1	&	02:53:19.52	&	+54:44:14.1	&	18.99	&	19.57	&	18.65	&	17.91	&	17.35	&	17.75	&	378.97	&	0	&	0	&	1 \\ 
J025931.05+550806.3	&	02:59:31.05	&	+55:08:06.3	&	19.53	&	20.10	&	19.31	&	19.27	&	19.45	&	19.57	&	492.87	&	1	&	0	&	0 \\ 
J031048.65+530022.7	&	03:10:48.65	&	+53:00:22.7	&	20.08	&	20.82	&	19.26	&	19.15	&	19.03	&	19.17	&	739.68	&	1	&	1	&	0 \\ 
J031936.06+531425.1	&	03:19:36.06	&	+53:14:25.1	&	19.90	&	20.55	&	19.27	&	19.20	&	19.31	&	19.45	&	474.87	&	1	&	1	&	1 \\ 
J032807.05+525737.0	&	03:28:07.05	&	+52:57:37.0	&	18.24	&	18.85	&	18.32	&	18.42	&	18.61	&	18.68	&	289.92	&	1	&	1	&	1 \\ 
J033300.67+604529.5	&	03:33:00.67	&	+60:45:29.5	&	19.56	&	20.29	&	20.71	&	21.16	&	\,.\,.\,.	&	\,.\,.\,.	&	3592.07	&	0	&	1	&	0 \\ 
J033300.81+524748.0	&	03:33:00.81	&	+52:47:48.0	&	18.38	&	19.10	&	18.44	&	18.47	&	18.59	&	18.61	&	340.38	&	1	&	1	&	1 \\ 
J034015.23+615204.1	&	03:40:15.23	&	+61:52:04.1	&	19.04	&	20.47	&	15.79	&	15.06	&	14.81	&	14.96	&	894.44	&	0	&	1	&	0 \\ 
J034631.63+590920.2	&	03:46:31.63	&	+59:09:20.2	&	18.98	&	19.75	&	19.20	&	19.28	&	19.38	&	19.42	&	628.46	&	1	&	1	&	1 \\ 
J035010.66+600113.9	&	03:50:10.66	&	+60:01:13.9	&	19.46	&	20.00	&	19.98	&	19.95	&	20.23	&	20.19	&	500.34	&	1	&	0	&	0 \\ 
J035110.38+590055.6	&	03:51:10.38	&	+59:00:55.6	&	19.31	&	19.99	&	19.80	&	19.93	&	20.16	&	20.19	&	636.55	&	1	&	1	&	0 \\ 
J035737.37+570303.8	&	03:57:37.37	&	+57:03:03.8	&	19.64	&	21.78	&	15.83	&	15.11	&	14.75	&	14.99	&	8979.38	&	0	&	1	&	0 \\ 
J035934.41+571348.5	&	03:59:34.41	&	+57:13:48.5	&	19.82	&	20.40	&	16.17	&	15.76	&	15.60	&	15.83	&	2872.91	&	0	&	0	&	1 \\ 
J040144.53+461434.1	&	04:01:44.53	&	+46:14:34.1	&	20.44	&	21.05	&	19.84	&	19.91	&	20.07	&	20.24	&	424.76	&	1	&	1	&	0 \\ 
J040522.62+555339.4	&	04:05:22.62	&	+55:53:39.4	&	20.68	&	21.63	&	16.88	&	16.42	&	16.21	&	16.44	&	6377.98	&	0	&	1	&	0 \\ 
J040613.18+561233.9	&	04:06:13.18	&	+56:12:33.9	&	20.98	&	21.63	&	20.88	&	20.85	&	20.87	&	20.99	&	2913.53	&	0	&	1	&	0 \\ 
J041039.52+444840.9	&	04:10:39.52	&	+44:48:40.9	&	20.26	&	21.10	&	20.10	&	20.14	&	\,.\,.\,.	&	20.52	&	1192.50	&	1	&	1	&	0 \\ 
J041531.64+450209.2	&	04:15:31.64	&	+45:02:09.2	&	20.31	&	20.89	&	19.68	&	19.64	&	19.67	&	19.93	&	232.96	&	1	&	0	&	1 \\ 
J042327.71+445223.8	&	04:23:27.71	&	+44:52:23.8	&	15.56	&	16.25	&	17.22	&	17.55	&	17.88	&	17.74	&	163.48	&	1	&	1	&	1 \\ 
J042553.88+441907.5	&	04:25:53.88	&	+44:19:07.5	&	20.67	&	21.21	&	20.48	&	20.35	&	20.54	&	20.59	&	3436.54	&	0	&	0	&	0 \\ 
J042641.65+444404.7	&	04:26:41.65	&	+44:44:04.7	&	18.01	&	18.55	&	15.63	&	14.65	&	14.25	&	14.45	&	336.63	&	0	&	0	&	0 \\ 
J042646.87+444612.3	&	04:26:46.87	&	+44:46:12.3	&	19.59	&	20.30	&	20.29	&	20.57	&	\,.\,.\,.	&	20.85	&	527.77	&	1	&	1	&	0 \\ 
J042703.35+414349.2	&	04:27:03.35	&	+41:43:49.2	&	18.73	&	19.27	&	18.45	&	18.52	&	18.59	&	18.70	&	441.61	&	1	&	0	&	1 \\ 
J042756.39+413000.6	&	04:27:56.39	&	+41:30:00.6	&	20.02	&	20.59	&	20.09	&	20.17	&	\,.\,.\,.	&	20.45	&	1112.97	&	0	&	0	&	0 \\ 
J042837.29+411452.6	&	04:28:37.29	&	+41:14:52.6	&	20.48	&	21.13	&	20.70	&	20.61	&	21.04	&	21.15	&	1265.56	&	0	&	1	&	0 \\ 
J043005.06+481612.0	&	04:30:05.06	&	+48:16:12.0	&	19.52	&	20.19	&	19.09	&	19.10	&	19.19	&	19.37	&	462.32	&	1	&	1	&	1 \\ 
J043137.79+415249.2	&	04:31:37.79	&	+41:52:49.2	&	20.11	&	21.22	&	17.23	&	17.00	&	16.89	&	17.08	&	5853.45	&	0	&	1	&	0 \\ 
J043143.43+415247.0	&	04:31:43.43	&	+41:52:47.0	&	19.48	&	21.22	&	16.38	&	15.78	&	15.48	&	15.76	&	2896.42	&	0	&	1	&	0 \\ 
J043145.95+445458.7	&	04:31:45.95	&	+44:54:58.7	&	19.51	&	20.36	&	14.89	&	14.25	&	13.98	&	14.19	&	824.39	&	0	&	1	&	1 \\ 
J043326.01+413116.1	&	04:33:26.01	&	+41:31:16.1	&	20.86	&	21.50	&	17.88	&	17.55	&	17.49	&	17.64	&	4620.90	&	0	&	1	&	0 \\ 
J043519.17+543932.7	&	04:35:19.17	&	+54:39:32.7	&	20.11	&	20.74	&	18.17	&	16.96	&	16.28	&	16.71	&	650.68	&	0	&	1	&	0 \\ 
J043827.29+473238.2	&	04:38:27.29	&	+47:32:38.2	&	19.20	&	19.80	&	18.91	&	18.94	&	19.07	&	19.35	&	257.75	&	1	&	1	&	1 \\ 
J044326.63+464520.2	&	04:43:26.63	&	+46:45:20.2	&	17.64	&	18.46	&	18.83	&	17.86	&	17.35	&	17.76	&	2776.41	&	0	&	1	&	0 \\ 
J044447.07+394243.0	&	04:44:47.07	&	+39:42:43.0	&	20.90	&	21.55	&	20.88	&	20.78	&	\,.\,.\,.	&	\,.\,.\,.	&	\,.\,.\,.	&	0	&	1	&	0 \\ 
J044554.05+492824.2	&	04:45:54.05	&	+49:28:24.2	&	20.27	&	21.05	&	16.14	&	15.82	&	15.65	&	15.88	&	2982.53	&	0	&	1	&	1 \\ 
J044635.04+480118.6	&	04:46:35.04	&	+48:01:18.6	&	19.95	&	21.42	&	20.78	&	20.88	&	21.06	&	21.08	&	\,.\,.\,.	&	0	&	1	&	0 \\ 
J044639.01+402003.3	&	04:46:39.01	&	+40:20:03.3	&	19.90	&	20.70	&	20.26	&	20.29	&	20.15	&	20.41	&	3275.08	&	0	&	1	&	0 \\ 
J044822.93+483429.1	&	04:48:22.93	&	+48:34:29.1	&	18.31	&	18.90	&	16.40	&	15.94	&	15.74	&	15.97	&	4499.73	&	0	&	0	&	0 \\ 
J044839.59+483606.0	&	04:48:39.59	&	+48:36:06.0	&	16.35	&	17.44	&	15.64	&	15.23	&	15.05	&	15.26	&	5244.07	&	0	&	1	&	0 \\ 
J044923.41+491440.4	&	04:49:23.41	&	+49:14:40.4	&	20.59	&	21.25	&	20.33	&	20.48	&	20.62	&	20.65	&	3046.80	&	0	&	1	&	0 \\ 
J044925.36+490701.4	&	04:49:25.36	&	+49:07:01.4	&	19.53	&	20.09	&	20.18	&	20.52	&	\,.\,.\,.	&	\,.\,.\,.	&	1695.95	&	0	&	0	&	0 \\ 
J045614.60+385509.5	&	04:56:14.60	&	+38:55:09.5	&	20.58	&	21.42	&	20.38	&	20.47	&	20.59	&	20.80	&	1152.69	&	1	&	1	&	0 \\ 
J045703.87+485851.3	&	04:57:03.87	&	+48:58:51.3	&	20.26	&	20.82	&	20.24	&	20.56	&	20.45	&	20.80	&	937.39	&	1	&	0	&	0 \\ 
J045739.49+505757.4	&	04:57:39.49	&	+50:57:57.4	&	15.57	&	16.21	&	16.93	&	17.24	&	17.53	&	17.41	&	296.33	&	1	&	1	&	1 \\ 
J045809.37+390538.7	&	04:58:09.37	&	+39:05:38.7	&	18.34	&	19.00	&	18.86	&	18.99	&	19.06	&	19.21	&	605.61	&	1	&	1	&	0 \\ 
J045841.57+364234.2	&	04:58:41.57	&	+36:42:34.2	&	19.52	&	20.08	&	17.88	&	17.61	&	17.59	&	17.68	&	1850.52	&	0	&	0	&	0 \\ 
J045843.35+390046.5	&	04:58:43.35	&	+39:00:46.5	&	18.16	&	18.72	&	17.61	&	17.61	&	17.71	&	17.65	&	2003.94	&	0	&	0	&	0 \\ 
J045854.82+482831.0	&	04:58:54.82	&	+48:28:31.0	&	18.94	&	19.49	&	19.55	&	19.62	&	19.85	&	19.80	&	1421.53	&	1	&	0	&	0 \\ 
J045930.01+485819.9	&	04:59:30.01	&	+48:58:19.9	&	18.24	&	20.14	&	\,.\,.\,.	&	21.69	&	20.99	&	\,.\,.\,.	&	\,.\,.\,.	&	0	&	1	&	0 \\ 
J045946.22+505231.8	&	04:59:46.22	&	+50:52:31.8	&	20.06	&	20.65	&	20.62	&	20.66	&	\,.\,.\,.	&	\,.\,.\,.	&	4101.53	&	0	&	0	&	0 \\ 
J050117.70+485213.5	&	05:01:17.70	&	+48:52:13.5	&	16.02	&	16.58	&	21.45	&	20.25	&	19.57	&	20.15	&	2948.58	&	0	&	0	&	0 \\ 
J050209.17+492513.9	&	05:02:09.17	&	+49:25:13.9	&	20.39	&	21.30	&	19.95	&	19.96	&	20.12	&	19.89	&	2083.38	&	1	&	1	&	0 \\ 
J050405.54+383324.9	&	05:04:05.54	&	+38:33:24.9	&	20.19	&	20.89	&	16.10	&	15.53	&	15.25	&	15.58	&	5473.55	&	0	&	1	&	0 \\ 
J050933.75+345104.9	&	05:09:33.75	&	+34:51:04.9	&	20.00	&	20.65	&	15.39	&	14.83	&	14.57	&	14.82	&	1308.39	&	0	&	1	&	0 \\ 
J051008.24+321804.2	&	05:10:08.24	&	+32:18:04.2	&	19.75	&	20.44	&	19.86	&	19.95	&	20.27	&	20.40	&	740.52	&	1	&	1	&	0 \\ 
J051112.03+344722.5	&	05:11:12.03	&	+34:47:22.5	&	19.43	&	20.08	&	19.54	&	19.00	&	18.30	&	18.78	&	736.67	&	0	&	1	&	0 \\ 
J052925.68+402700.9	&	05:29:25.68	&	+40:27:00.9	&	19.88	&	20.41	&	19.03	&	18.96	&	19.00	&	19.09	&	5651.06	&	0	&	0	&	0 \\ 
J053207.61+404445.3	&	05:32:07.61	&	+40:44:45.3	&	19.85	&	21.46	&	20.18	&	20.27	&	20.26	&	20.58	&	3120.23	&	0	&	1	&	0 \\ 
J053659.57+395430.0	&	05:36:59.57	&	+39:54:30.0	&	19.61	&	20.73	&	20.80	&	21.23	&	\,.\,.\,.	&	21.17	&	\,.\,.\,.	&	0	&	1	&	0 \\ 
J053955.63+395004.6	&	05:39:55.63	&	+39:50:04.6	&	20.05	&	20.95	&	20.47	&	20.63	&	20.73	&	21.03	&	847.64	&	0	&	1	&	0 \\ 
J190347.51+170140.6	&	19:03:47.51	&	+17:01:40.6	&	19.45	&	20.97	&	\,.\,.\,.	&	16.23	&	16.14	&	16.29	&	3926.18	&	0	&	1	&	0 \\ 
J190443.07+172843.4	&	19:04:43.07	&	+17:28:43.4	&	20.26	&	21.50	&	20.19	&	20.16	&	20.34	&	20.33	&	3770.48	&	0	&	1	&	0 \\ 
\hline 
\end{tabular} 
} 
\end{table*}

\begin{table*}
\caption{Single WDs stellar parameters derived using the MCMC method. The objects in bold are the hot WDs with $T_{\rm eff}\geq 50\,000$~K.}
\label{tab:wd_simple}
% \resizebox{\textwidth}{!}{
\renewcommand{\arraystretch}{1.5}
\begin{tabular}{lrrrrrr}
\toprule
IGAPS Name &           $T_{\rm WD}$ &      $\alpha$ &       d &       $R_{\rm WD}$ &  $\log(L_{\rm WD})$ & $E(B-V)$ \\ 
       &             (10$^3$)   &               &     (kpc) &      (R$\sun$)  & ($L\sun$) & (mag) \\ 
\midrule 
J014150.56+650546.9	&	26$\pm$1	&	-27.039$\pm$0.018	&	2.18$\pm^{0.54}_{0.37}$	&	0.174$\pm^{0.044}_{0.030}$	&	1.094$\pm^{0.220}_{0.151}$	&	0.450$\pm$0.010	\\ 
J015418.01+654741.6	&	26$\pm$1	&	-28.085$\pm$0.054	&	0.32$\pm^{0.03}_{0.03}$	&	0.009$\pm^{0.001}_{0.001}$	&	-1.473$\pm^{0.134}_{0.132}$	&	0.310$\pm$0.035	\\ 
J023500.57+632737.0	&	22$\pm$1	&	-27.977$\pm$0.056	&	0.37$\pm^{0.05}_{0.04}$	&	0.012$\pm^{0.002}_{0.001}$	&	-1.530$\pm^{0.165}_{0.137}$	&	0.290$\pm$0.040	\\ 
\textbf{J024130.89+632923.7}	&	54$\pm$6	&	-28.478$\pm$0.050	&	0.97$\pm^{0.28}_{0.18}$	&	0.018$\pm^{0.005}_{0.004}$	&	0.420$\pm^{0.322}_{0.263}$	&	0.330$\pm$0.007	\\ 
J024713.51+644122.5	&	29$\pm$2	&	-28.051$\pm$0.051	&	0.73$\pm^{0.18}_{0.11}$	&	0.021$\pm^{0.005}_{0.004}$	&	-0.563$\pm^{0.243}_{0.171}$	&	0.341$\pm$0.040	\\ 
J024849.31+640930.4	&	25$\pm$1	&	-27.536$\pm$0.023	&	0.39$\pm^{0.02}_{0.02}$	&	0.019$\pm^{0.001}_{0.001}$	&	-0.898$\pm^{0.064}_{0.067}$	&	0.314$\pm$0.014	\\ 
J035110.38+590055.6	&	39$\pm$5	&	-28.611$\pm$0.085	&	0.64$\pm^{0.17}_{0.12}$	&	0.011$\pm^{0.003}_{0.002}$	&	-0.620$\pm^{0.338}_{0.299}$	&	0.286$\pm$0.016	\\ 
\textbf{J042327.71+445223.8}	&	60$\pm$2	&	-27.969$\pm$0.018	&	0.16$\pm^{0.00}_{0.00}$	&	0.005$\pm^{0.001}_{0.001}$	&	-0.503$\pm^{0.071}_{0.072}$	&	0.083$\pm$0.004	\\ 
J043005.06+481612.0	&	29$\pm$1	&	-27.788$\pm$0.038	&	0.46$\pm^{0.07}_{0.05}$	&	0.018$\pm^{0.003}_{0.002}$	&	-0.691$\pm^{0.148}_{0.126}$	&	0.412$\pm$0.023	\\ 
J043827.29+473238.2	&	47$\pm$6	&	-28.105$\pm$0.082	&	0.26$\pm^{0.01}_{0.01}$	&	0.007$\pm^{0.001}_{0.001}$	&	-0.652$\pm^{0.247}_{0.246}$	&	0.419$\pm$0.005	\\ 
\textbf{J044635.04+480118.6}	&	106$\pm$42	&	-29.517$\pm$0.181	&	\,.\,.\,.	&	\,.\,.\,.	&	\,.\,.\,.	&	0.338$\pm$0.020	\\ 
J044923.41+491440.4	&	25$\pm$2	&	-28.417$\pm$0.062	&	3.05$\pm^{2.29}_{1.67}$	&	\,.\,.\,.	&	\,.\,.\,.	&	0.290$\pm$0.052	\\ 
J045614.60+385509.5	&	38$\pm$16	&	-28.653$\pm$0.167	&	1.15$\pm^{0.76}_{0.49}$	&	\,.\,.\,.	&	\,.\,.\,.	&	0.420$\pm$0.060	\\ 
J045703.87+485851.3	&	27$\pm$2	&	-28.524$\pm$0.078	&	0.94$\pm^{1.43}_{0.39}$	&	\,.\,.\,.	&	\,.\,.\,.	&	0.250$\pm$0.050	\\ 
J045843.35+390046.5	&	26$\pm$1	&	-27.008$\pm$0.015	&	2.00$\pm^{0.65}_{0.37}$	&	\,.\,.\,.	&	\,.\,.\,.	&	0.376$\pm$0.010	\\ 
J050209.17+492513.9	&	27$\pm$2	&	-28.116$\pm$0.056	&	2.08$\pm^{0.77}_{0.97}$	&	\,.\,.\,.	&	\,.\,.\,.	&	0.400$\pm$0.040	\\ 
J051008.24+321804.2	&	49$\pm$14	&	-28.682$\pm$0.130	&	0.74$\pm^{0.46}_{0.21}$	&	\,.\,.\,.	&	\,.\,.\,.	&	0.345$\pm$0.041	\\ 
J053955.63+395004.6	&	49$\pm$42	&	-28.977$\pm$0.262	&	0.85$\pm^{0.61}_{0.38}$	&	\,.\,.\,.	&	\,.\,.\,.	&	0.340$\pm$0.032	\\ 
\textbf{J190443.07+172843.4}	&	97$\pm$47	&	-28.904$\pm$0.227	&	3.77$\pm^{2.05}_{1.45}$	&	\,.\,.\,.	&	\,.\,.\,.	&	0.510$\pm$0.023	\\ 
\bottomrule 
\end{tabular} 
\end{table*}

\begin{table*}
\caption{Binary WD stellar parameters determined with the MCMC method. The objects in bold are the hot WDs with $T_{\rm eff}\geq 50\,000$~K}
\label{tab:wd_binary}
\resizebox{\textwidth}{!}{
\renewcommand{\arraystretch}{1.5}
\begin{tabular}{lrrrrrrrrr}
\toprule
IGAPS Name &           $T_{\rm WD}$ &   $T_{\rm S}$  &  $\alpha$ &   $\beta$  &   d &       $R_{\rm WD}$ &  $R_{\rm S}$ & $\log(L_{\rm WD})$ & $E(B-V)$ \\ 
       &             \multicolumn{2}{c}{(10$^3$)}   &     &     &     (kpc) &      \multicolumn{2}{c}{(R$\sun$)}  & (L$\sun$) &(mag) \\ 
\midrule 
\textbf{J012902.52+644429.1}	&	100$\pm$3	&	9$\pm$1	&	-29.719$\pm$0.006	&	11.767$\pm$2.265	&	0.25$\pm ^{0.01}_{0.01}$	&	0.001$\pm^{0.001}_{0.001}$	&	0.016$\pm^{0.003}_{0.003}$	&	-0.775$\pm^{0.057}_{0.052}$	&	0.008$\pm$0.004	\\ 
\textbf{J023215.29+632115.9}	&	140$\pm$20	&	7$\pm$1	&	-29.507$\pm$0.012	&	104.473$\pm$21.832	&	5.72$\pm ^{1.05}_{0.80}$	&	0.039$\pm^{0.007}_{0.005}$	&	4.059$\pm^{1.129}_{1.023}$	&	2.720$\pm^{0.293}_{0.275}$	&	0.380$\pm$0.085	\\ 
\textbf{J023223.07+632040.9}	&	190$\pm$18	&	8$\pm$1	&	-29.553$\pm$0.003	&	80.582$\pm$14.037	&	3.84$\pm ^{0.64}_{0.46}$	&	0.025$\pm^{0.004}_{0.003}$	&	2.006$\pm^{0.484}_{0.423}$	&	2.862$\pm^{0.218}_{0.192}$	&	0.460$\pm$0.033	\\ 
J023527.80+632738.4	&	28$\pm$2	&	8$\pm$1	&	-28.000$\pm$0.003	&	16.119$\pm$16.160	&	\,.\,.\,.	&	\,.\,.\,.	&	\,.\,.\,.	&	\,.\,.\,.	&	0.495$\pm$0.015	\\ 
\textbf{J025236.54+544506.9}	&	98$\pm$15	&	8$\pm$1	&	-29.677$\pm$0.036	&	9.756$\pm$2.770	&	0.31$\pm ^{0.04}_{0.03}$	&	0.002$\pm^{0.001}_{0.001}$	&	0.017$\pm^{0.005}_{0.005}$	&	-0.580$\pm^{0.289}_{0.275}$	&	0.328$\pm$0.025	\\ 
J025319.52+544414.1	&	27$\pm$1	&	4$\pm$1	&	-28.175$\pm$0.004	&	24.160$\pm$0.623	&	0.38$\pm ^{0.01}_{0.01}$	&	0.010$\pm^{0.001}_{0.001}$	&	0.236$\pm^{0.011}_{0.011}$	&	-1.321$\pm^{0.037}_{0.039}$	&	0.206$\pm$0.007	\\ 
J025931.05+550806.3	&	47$\pm$4	&	9$\pm$1	&	-28.661$\pm$0.027	&	3.635$\pm$0.546	&	0.49$\pm ^{0.07}_{0.06}$	&	0.008$\pm^{0.001}_{0.001}$	&	0.028$\pm^{0.006}_{0.006}$	&	-0.573$\pm^{0.207}_{0.193}$	&	0.330$\pm$0.006	\\ 
J031048.65+530022.7	&	27$\pm$1	&	4$\pm$1	&	-27.772$\pm$0.030	&	5.141$\pm$1.104	&	\,.\,.\,.	&	\,.\,.\,.	&	\,.\,.\,.	&	\,.\,.\,.	&	0.444$\pm$0.050	\\ 
\textbf{J031936.06+531425.1}	&	101$\pm$9	&	9$\pm$1	&	-29.814$\pm$0.013	&	14.388$\pm$5.058	&	0.47$\pm ^{0.06}_{0.07}$	&	0.002$\pm^{0.001}_{0.001}$	&	0.034$\pm^{0.013}_{0.013}$	&	-0.282$\pm^{0.183}_{0.197}$	&	0.226$\pm$0.027	\\ 
\textbf{J032807.05+525737.0}	&	71$\pm$2	&	9$\pm$1	&	-29.229$\pm$0.005	&	9.240$\pm$1.466	&	0.29$\pm ^{0.01}_{0.01}$	&	0.003$\pm^{0.001}_{0.001}$	&	0.024$\pm^{0.004}_{0.004}$	&	-0.814$\pm^{0.062}_{0.062}$	&	0.109$\pm$0.014	\\ 
\textbf{J033300.81+524748.0}	&	77$\pm$3	&	8$\pm$1	&	-29.209$\pm$0.007	&	10.272$\pm$3.801	&	0.34$\pm ^{0.02}_{0.02}$	&	0.003$\pm^{0.001}_{0.001}$	&	0.032$\pm^{0.012}_{0.012}$	&	-0.508$\pm^{0.077}_{0.075}$	&	0.147$\pm$0.042	\\ 
\textbf{J034015.23+615204.1}	&	154$\pm$18	&	6$\pm$1	&	-28.872$\pm$0.012	&	92.356$\pm$13.522	&	0.89$\pm ^{0.02}_{0.02}$	&	0.011$\pm^{0.001}_{0.001}$	&	1.059$\pm^{0.157}_{0.157}$	&	1.825$\pm^{0.209}_{0.209}$	&	0.459$\pm$0.007	\\ 
\textbf{J034631.63+590920.2}	&	87$\pm$3	&	9$\pm$1	&	-29.485$\pm$0.009	&	8.864$\pm$3.435	&	0.63$\pm ^{0.09}_{0.07}$	&	0.004$\pm^{0.001}_{0.001}$	&	0.039$\pm^{0.016}_{0.016}$	&	-0.015$\pm^{0.138}_{0.116}$	&	0.169$\pm$0.049	\\ 
J035010.66+600113.9	&	32$\pm$2	&	8$\pm$2	&	-28.570$\pm$0.055	&	1.535$\pm$3.744	&	0.50$\pm ^{0.13}_{0.08}$	&	0.009$\pm^{0.002}_{0.002}$	&	0.013$\pm^{0.033}_{0.033}$	&	-1.139$\pm^{0.250}_{0.173}$	&	0.230$\pm$0.008	\\ 
\textbf{J035737.37+570303.8}	&	200$\pm$12	&	6$\pm$1	&	-29.895$\pm$0.002	&	265.645$\pm$22.274	&	\,.\,.\,.	&	\,.\,.\,.	&	\,.\,.\,.	&	\,.\,.\,.	&	0.310$\pm$0.030	\\ 
\textbf{J035934.41+571348.5}	&	147$\pm$17	&	7$\pm$1	&	-29.465$\pm$0.012	&	83.334$\pm$14.470	&	2.87$\pm ^{0.31}_{0.25}$	&	0.020$\pm^{0.002}_{0.002}$	&	1.697$\pm^{0.349}_{0.331}$	&	2.242$\pm^{0.222}_{0.215}$	&	0.386$\pm$0.010	\\ 
\textbf{J040144.53+461434.1}	&	130$\pm$25	&	10$\pm$1	&	-30.578$\pm$0.014	&	19.045$\pm$4.821	&	0.42$\pm ^{0.11}_{0.08}$	&	0.001$\pm^{0.001}_{0.001}$	&	0.019$\pm^{0.007}_{0.006}$	&	-0.596$\pm^{0.400}_{0.368}$	&	0.152$\pm$0.041	\\ 
\textbf{J040522.62+555339.4}	&	151$\pm$24	&	7$\pm$1	&	-30.206$\pm$0.003	&	140.513$\pm$23.329	&	6.38$\pm ^{1.46}_{1.32}$	&	0.022$\pm^{0.005}_{0.004}$	&	3.028$\pm^{0.857}_{0.805}$	&	2.338$\pm^{0.337}_{0.327}$	&	0.313$\pm$0.031	\\ 
J040613.18+561233.9	&	43$\pm$7	&	8$\pm$1	&	-29.840$\pm$0.039	&	7.162$\pm$5.358	&	\,.\,.\,.	&	\,.\,.\,.	&	\,.\,.\,.	&	\,.\,.\,.	&	0.190$\pm$0.100	\\ 
\textbf{J041531.64+450209.2}	&	122$\pm$20	&	9$\pm$1	&	-30.096$\pm$0.017	&	15.296$\pm$4.383	&	0.23$\pm ^{0.03}_{0.02}$	&	0.001$\pm^{0.001}_{0.001}$	&	0.013$\pm^{0.004}_{0.004}$	&	-0.818$\pm^{0.314}_{0.303}$	&	0.236$\pm$0.035	\\ 
J042553.88+441907.5	&	48$\pm$8	&	8$\pm$1	&	-30.120$\pm$0.026	&	11.777$\pm$6.844	&	\,.\,.\,.	&	\,.\,.\,.	&	\,.\,.\,.	&	\,.\,.\,.	&	0.100$\pm$0.060	\\ 
J042641.65+444404.7	&	41$\pm$1	&	4$\pm$1	&	-29.060$\pm$0.012	&	182.122$\pm$51.271	&	0.34$\pm ^{0.01}_{0.01}$	&	0.004$\pm^{0.001}_{0.001}$	&	0.651$\pm^{0.184}_{0.184}$	&	-1.498$\pm^{0.054}_{0.053}$	&	0.021$\pm$0.021	\\ 
J042703.35+414349.2	&	50$\pm$3	&	9$\pm$1	&	-29.161$\pm$0.006	&	9.109$\pm$2.279	&	0.44$\pm ^{0.04}_{0.04}$	&	0.004$\pm^{0.001}_{0.001}$	&	0.039$\pm^{0.010}_{0.010}$	&	-1.001$\pm^{0.118}_{0.113}$	&	0.122$\pm$0.020	\\ 
\textbf{J042837.29+411452.6}	&	94$\pm$14	&	8$\pm$1	&	-30.514$\pm$0.030	&	12.903$\pm$4.285	&	\,.\,.\,.	&	\,.\,.\,.	&	\,.\,.\,.	&	\,.\,.\,.	&	0.100$\pm$0.070	\\ 
\textbf{J043137.79+415249.2}	&	198$\pm$15	&	8$\pm$1	&	-30.239$\pm$0.002	&	82.734$\pm$11.700	&	\,.\,.\,.	&	\,.\,.\,.	&	\,.\,.\,.	&	\,.\,.\,.	&	0.260$\pm$0.020	\\ 
\textbf{J043143.43+415247.0}	&	172$\pm$13	&	6$\pm$1	&	-29.574$\pm$0.011	&	124.632$\pm$11.675	&	2.90$\pm ^{0.36}_{0.34}$	&	0.018$\pm^{0.002}_{0.002}$	&	2.293$\pm^{0.358}_{0.345}$	&	2.427$\pm^{0.172}_{0.168}$	&	0.343$\pm$0.010	\\ 
\textbf{J043145.95+445458.7}	&	152$\pm$19	&	6$\pm$1	&	-29.870$\pm$0.011	&	345.047$\pm$51.821	&	0.82$\pm ^{0.02}_{0.02}$	&	0.004$\pm^{0.001}_{0.001}$	&	1.344$\pm^{0.204}_{0.205}$	&	0.863$\pm^{0.218}_{0.218}$	&	0.248$\pm$0.086	\\ 
\textbf{J043326.01+413116.1}	&	195$\pm$20	&	7$\pm$1	&	-30.768$\pm$0.003	&	114.280$\pm$18.479	&	\,.\,.\,.	&	\,.\,.\,.	&	\,.\,.\,.	&	\,.\,.\,.	&	0.210$\pm$0.040	\\ 
\textbf{J043519.17+543932.7}	&	121$\pm$11	&	4$\pm$1	&	-30.431$\pm$0.003	&	377.559$\pm$99.229	&	0.65$\pm ^{0.05}_{0.05}$	&	0.002$\pm^{0.001}_{0.001}$	&	0.662$\pm^{0.182}_{0.182}$	&	-0.221$\pm^{0.171}_{0.172}$	&	0.123$\pm$0.040	\\ 
\textbf{J044554.05+492824.2}	&	143$\pm$23	&	8$\pm$1	&	-29.372$\pm$0.004	&	73.031$\pm$13.000	&	2.98$\pm ^{0.31}_{0.23}$	&	0.023$\pm^{0.002}_{0.002}$	&	1.694$\pm^{0.349}_{0.329}$	&	2.309$\pm^{0.296}_{0.290}$	&	0.474$\pm$0.029	\\ 
J044639.01+402003.3	&	30$\pm$2	&	4$\pm$2	&	-28.675$\pm$0.055	&	10.425$\pm$3.378	&	\,.\,.\,.	&	\,.\,.\,.	&	\,.\,.\,.	&	\,.\,.\,.	&	0.240$\pm$0.114	\\ 
J044822.93+483429.1	&	22$\pm$1	&	6$\pm$1	&	-26.894$\pm$0.013	&	6.679$\pm$0.849	&	4.50$\pm ^{1.34}_{0.82}$	&	0.417$\pm^{0.125}_{0.077}$	&	2.786$\pm^{0.905}_{0.622}$	&	1.564$\pm^{0.262}_{0.163}$	&	0.350$\pm$0.019	\\ 
J044839.59+483606.0	&	49$\pm$2	&	5$\pm$1	&	-26.555$\pm$0.020	&	8.391$\pm$0.874	&	5.24$\pm ^{1.11}_{0.76}$	&	0.682$\pm^{0.145}_{0.100}$	&	5.722$\pm^{1.353}_{1.030}$	&	3.368$\pm^{0.200}_{0.149}$	&	0.523$\pm$0.009	\\ 
\textbf{J045739.49+505757.4}	&	62$\pm$1	&	9$\pm$1	&	-28.205$\pm$0.006	&	4.171$\pm$0.282	&	0.30$\pm ^{0.01}_{0.01}$	&	0.007$\pm^{0.001}_{0.001}$	&	0.031$\pm^{0.002}_{0.002}$	&	-0.126$\pm^{0.034}_{0.036}$	&	0.029$\pm$0.012	\\ 
J045809.37+390538.7	&	49$\pm$3	&	8$\pm$1	&	-28.613$\pm$0.018	&	3.499$\pm$0.338	&	0.61$\pm ^{0.11}_{0.08}$	&	0.010$\pm^{0.002}_{0.001}$	&	0.035$\pm^{0.007}_{0.006}$	&	-0.280$\pm^{0.186}_{0.154}$	&	0.213$\pm$0.005	\\ 
\textbf{J045841.57+364234.2}	&	102$\pm$5	&	7$\pm$1	&	-30.248$\pm$0.002	&	61.904$\pm$17.079	&	\,.\,.\,.	&	\,.\,.\,.	&	\,.\,.\,.	&	\,.\,.\,.	&	0.070$\pm$0.024	\\ 
\textbf{J045854.82+482831.0}	&	59$\pm$4	&	8$\pm$1	&	-29.829$\pm$0.010	&	10.439$\pm$4.919	&	\,.\,.\,.	&	\,.\,.\,.	&	\,.\,.\,.	&	\,.\,.\,.	&	0.001$\pm$0.025	\\ 
\textbf{J050405.54+383324.9}	&	118$\pm$21	&	8$\pm$1	&	-28.479$\pm$0.012	&	40.139$\pm$13.803	&	5.47$\pm ^{1.02}_{0.88}$	&	0.104$\pm^{0.019}_{0.017}$	&	4.175$\pm^{1.634}_{1.587}$	&	3.271$\pm^{0.353}_{0.343}$	&	0.650$\pm$0.006	\\ 
\textbf{J050933.75+345104.9}	&	144$\pm$24	&	7$\pm$1	&	-29.643$\pm$0.011	&	190.016$\pm$37.862	&	1.31$\pm ^{0.06}_{0.05}$	&	0.008$\pm^{0.001}_{0.001}$	&	1.475$\pm^{0.302}_{0.299}$	&	1.366$\pm^{0.289}_{0.288}$	&	0.360$\pm$0.010	\\ 
\textbf{J052925.68+402700.9}	&	66$\pm$7	&	8$\pm$1	&	-29.997$\pm$0.006	&	20.727$\pm$10.184	&	\,.\,.\,.	&	\,.\,.\,.	&	\,.\,.\,.	&	\,.\,.\,.	&	0.090$\pm$0.050	\\ 
\textbf{J053207.61+404445.3}	&	161$\pm$36	&	9$\pm$1	&	-30.367$\pm$0.015	&	20.873$\pm$16.554	&	\,.\,.\,.	&	\,.\,.\,.	&	\,.\,.\,.	&	\,.\,.\,.	&	0.202$\pm$0.100	\\ 
\bottomrule 
\end{tabular} 

}
\end{table*}

\begin{table}
\centering
\caption{WD stars classification according to the regions defined by \citet{Wellhouse2005}. The objects in bold are the hot WDs with $T_{\rm eff}\geq 50\,000$~K.}
\label{tab:wd_regions}
\begin{tabular}{lcc}
\hline
                IGAPS Name & Region$_{\rm 2MASS}$ & Region$_{\rm UKIDSS}$ \\
\hline
{\bf J023215.29+632115.9} &   III &        \\
{\bf J023223.07+632040.9} &   III &        \\
 J023527.80+632738.4 &     I &        \\
 J025319.52+544414.1 &    II &        \\
 J031048.65+530022.7 &       &    III \\
 J033300.81+524748.0 &       &     IV \\
 J034015.23+615204.1 &     I &        \\
{\bf J035737.37+570303.8} &     I &      I \\
{\bf J035934.41+571348.5} &     I &      I \\
{\bf J040522.62+555339.4} &     I &      I \\
 J041531.64+450209.2 &       &      I \\
 J042641.65+444404.7 &    II &     II \\
 J042703.35+414349.2 &       &     IV \\
{\bf J043137.79+415249.2} &     I &      I \\
{\bf J043143.43+415247.0} &     I &      I \\
{\bf J043145.95+445458.7} &     I &      I \\
{\bf J043326.01+413116.1} &    II &    III \\
 J043519.17+543932.7 &    II &        \\
 J044326.63+464520.2 &   III &    III \\
{\bf J044554.05+492824.2} &     I &      I \\
 J044822.93+483429.1 &     I &      I \\
 J044839.59+483606.0 &     I &      I \\
 J045739.49+505757.4 &       &      I \\
 J045809.37+390538.7 &       &    III \\
{\bf J045841.57+364234.2} &    II &      I \\
 J045843.35+390046.5 &       &      I \\
 J045930.01+485819.9 &       &     II \\
 J050117.70+485213.5 &       &     II \\
{\bf J050405.54+383324.9} &     I &    III \\
{\bf J050933.75+345104.9} &     I &      I \\
 J051112.03+344722.5 &    II &     II \\
{\bf J052925.68+402700.9} &       &     IV \\
 J190347.51+170140.6 &   III &      I \\
\hline
\end{tabular}
\end{table}

\section{SED fitting}\label{sec:SED_analysis}

We estimated the physical parameters ($T_{\rm eff}$, radius, and interstellar reddening) of the candidates by performing fits of their spectral energy distributions (SED).

Theoretically, the equation that describes the magnitude observed of any source in the
AB magnitude system, is
\begin{equation}
\label{eq:pogson}
    m_{\rm AB} = -2.5\log\left(F_\nu \left[e^{\alpha}\right]^{2}\right) - 48.6
\end{equation}
with
\begin{equation*}
        \alpha \equiv \ln\left(\frac{R_{1}}{D}\right)
\end{equation*}
where $R$ is the radius of the star, D is the distance to it, and
$F_\nu=F_\nu\left(T_{\rm eff}, \log(g), Z\right)$ the flux of the star at frequency $\nu$. The total
number of free parameters are five: $R$, $D$, $T_{\rm eff}$, $\log(g)$,
and the metallicity, $Z$.

Equation~\ref{eq:pogson} would be enough to fit any WD star. However, it is difficult to
achieve accurate values of stellar parameters ($T_{\rm WD}$, $R_{\rm WD}$) with
few observable variables (2 points from \textit{GALEX}, when available, and 4 points from IGAPS).
To reduce the number of free parameters, we fitted colour indexes instead of magnitudes as follows,
\begin{equation}
    \label{eq:color_indexes}
    m_{\rm A} - m_{\rm B} = -2.5 \log\left(F_{\rm A} / F_{\rm B}\right)
\end{equation}
which remove the stellar radius,$R$, and the distance to the object. The A and B
subscripts indicate the magnitude difference between filter A and B.

In the case of binary systems, equation~\ref{eq:color_indexes} transforms to
\begin{equation}
    \label{eq:colour_index_binary}
    m_{\rm A} - m_{\rm B} = -2.5 \log\left(\frac{F^{1}_{\rm A} + F^{2}_{\rm A}\beta^{2}}{F^{1}_{\rm B} + F^{2}_{\rm B}\beta^{2}}\right)
\end{equation}
where
\begin{equation*}
    \beta \equiv R_{2}/R_{1}.
\end{equation*}

Note that in all the equations the flux is accounting for the extinction in the following form,
\begin{equation}
    \label{eq:extinction}
    F_{\nu} = f_{\nu}10^{0.4E(B-V)X_\lambda}
\end{equation}
where $f_{\nu}$ is the observed flux, and $X_\lambda$ is the extinction coefficient according to \citet{Cardelli1989} extinction law.

A Nelder-Mead algorithm (NMA) was employed to find the best fit to the equation~\ref{eq:color_indexes}
(and equation~\ref{eq:colour_index_binary} in the case of binary systems) and then equation~\ref{eq:pogson}.
Appropriate atmosphere
models for WD stars and for MS stars, for possible cool companions, were selected. 
The H-He non-LTE TLUSTY \citep{Hubeny1995} solar atmosphere models for the hot-WD stars \citep{Bianchi2009apss,Bianchi2011} were used, which covers a range of effective temperature, $T_{\rm eff}$, from  20\,000~K$<T_{\rm eff}<$200\,000~K and a range of gravities of 4.0$<\log g<$9.0.
For the MS
stars the solar LTE BT-Settl stellar atmosphere models from Phoenix/NextGen group \citep{Hauschildt1999,Allard2012}
which covers a range from 2600~K$<T_{\rm eff}<$15\,000~K and $-0.5<\log g<5.5$ were used.
Each atmosphere model was convolved with the \textit{GALEX} and IGAPS transmission curves in order
to create a grid of synthetic flux models. It is important to mention that we fixed the value
of $\log g $= 5 and 7 for MS and WD atmosphere models, respectively, as have been proven that
$\log g$ is not very sensitive to colour indexes  \citep[see][]{Barker2018}.
After NMA fitting, we sampled the posterior probability of the stellar parameters using the Markov-Chain Monte-Carlo (MCMC) through the \emph{emcee} \citep{ForemanMackey2013} python package.

The fitting procedure of each object was as follows.
First, we run the NMA fitting procedure, taking the equation~\ref{eq:color_indexes} as a kernel (or equation~\ref{eq:colour_index_binary} if a binary is guessed), to first estimate the stellar parameters and $E(B-V)$, and calculate their posterior probability with the MCMC algorithm.
For this, we reddened the theoretical spectra and convolved with \textit{GALEX} and IGAPS bands on the fly, varying the $E(B-V)$ according to equation~\ref{eq:extinction}.
 For objects with reliable Gaia distances,
 we set the $E(B-V)$ and the error \citep[as obtained using Bayestar2019;][]{Green2019} as priors to the  MCMC run.
 For Bayestar2019, we only selected $E(B-V)$ values and distances that contained both 'reliable\_dist' and 'converged' flags set to
 \texttt{True}\footnote{According to the
 `dustmaps' package \citep{Green2018}, the reliability is based on the distance modulus method in which a distance is not
 reliable if is smaller or greater than the minimum or maximum distance modulus, respectively. See \url{https://dustmaps.readthedocs.io/en/latest/maps.html}.}.
 For cases where the value of $E(B-V)$ could not be obtained using Bayestar2019, we selected the reddening
 value from GUVcat catalogue \citep[which is calculated using the extinction map of][]{Schlegel1998} and varied
 $E(B-V)$ from 0 to GUVcat $E(B-V)$ value.
 
A second run was used to obtain the marginalised posterior
distribution of the stellar parameters of the star (or binary system) by first using
equation~\ref{eq:color_indexes}, to obtain $T_{\rm WD}$ (and $\beta$ and $T_{\rm MS}$ if a binary
system is previously fitted), and then equation~\ref{eq:pogson} to obtain the scaling factor, $\alpha$. Note
that for the second run we fixed the value of $E(B-V)$ as resulted from the first run; the observed magnitudes were previously unreddened with this value to be compared with the grid of
theoretical flux models.

Table~\ref{tab:wd_simple} shows the stellar parameters resulted from the MCMC method of single
WD stars along with the $E(B-V)$ value. Most of the WD stars have an $T_{\rm eff}<50\,000$~K with
only four greater than 50\,000~K. Table~\ref{tab:wd_binary} shows the objects for which a binary
system was fitted. In this case, a total of 27 binary systems contain WD stars with
$T_{\rm WD}>50\,000$~K. Note that sources with more than one
missing magnitude measurement were not fitted because of MCMC
convergence problems (i.e., MCMC did not converge due to a few
observable values). Stellar radius was calculated using Gaia distances and $\alpha$ whereas the luminosity
was calculated using the Stefan-Boltzmann luminosity relation.

\begin{figure*}
    \centering
    \includegraphics[width=0.7\textwidth]{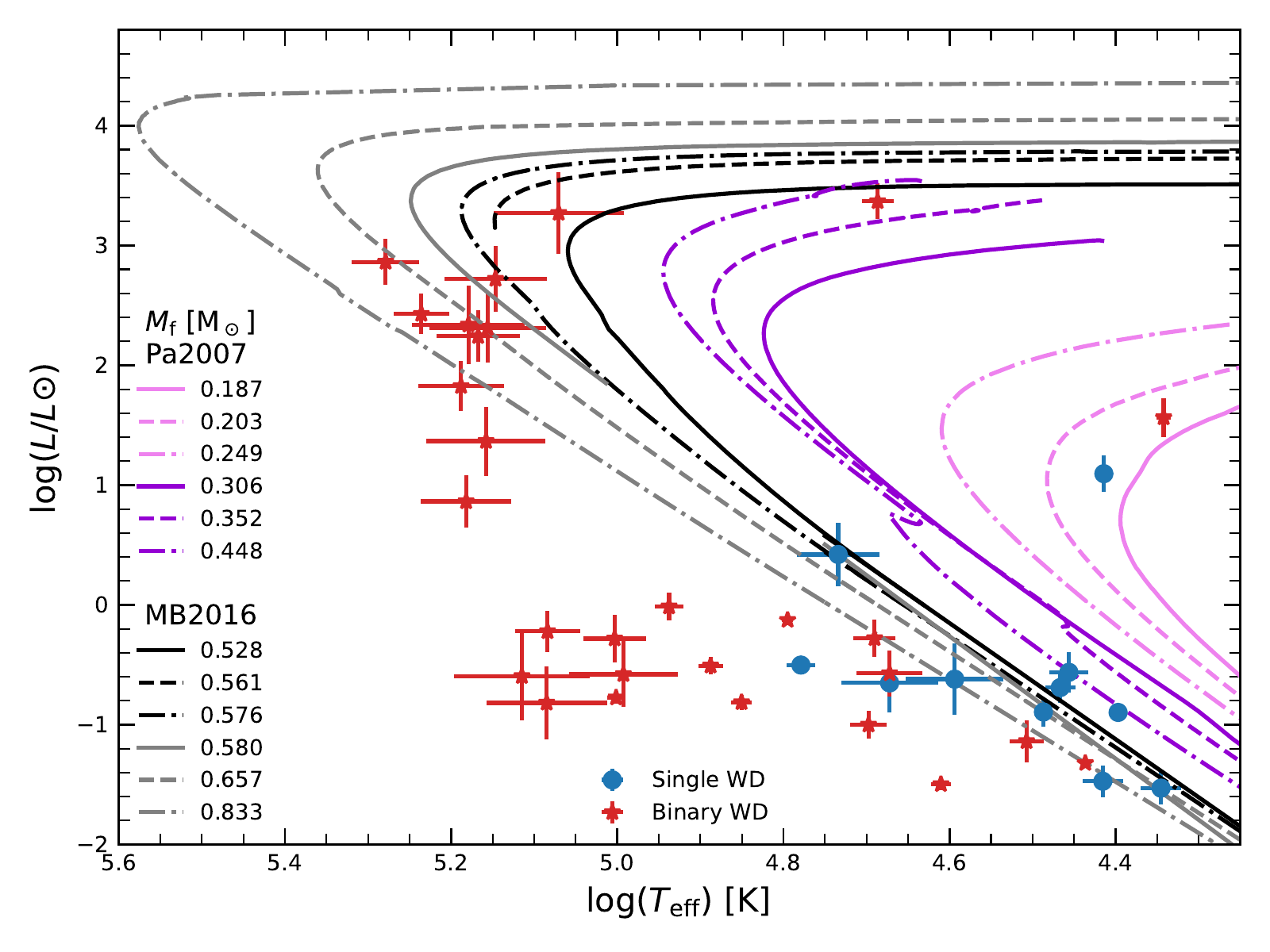}
    \caption{H-R diagram showing the single (blue) and binary (red) WD candidates in conjunction with the low- and intermediate-mass stars theoretical tracks of \citetalias{Miller2016} (gray lines) and the He- and O-core WD
    tracks of \citetalias{Panei2007} (purple lines). Masses of the evolutionary tracks are labelled in the figure.}
    \label{fig:teff_logg_track}
\end{figure*}

Gaia EDR3 provides a direct measurement of distance for the majority of the WD stars candidates of our sample. Most of them with distances $<$1~kpc and only ten with distances $>$1~kpc.
The combination of Gaia EDR3 distances allowed us to 
concurrently determine the stellar radius of the WD and binary WD candidates, using the distance to solve
the $\alpha$ and $\beta$ parameters, and its luminosity, $L_{\rm WD}$.
Figure~\ref{fig:teff_logg_track} shows the H-R diagram of the WD (blue) and binary WD candidates (red)
along with the low- and intermediate-mass stars theoretically evolutionary tracks of
\citet[][MB2016]{Miller2016}\defcitealias{Miller2016}{MB2016} and the low-mass He- and O-core WD tracks of \citet[][Pa2007]{Panei2007}\defcitealias{Panei2007}{Pa2007}.
Note that for \citetalias{Miller2016} models we linearly extended the theoretical tracks to account for luminosities below
of $\log(L/L_{\odot})<0.$ instead of the limit of $\log(L/L_{\odot})=1$ as the original ones.
The evolutionary stage of the majority of the objects corresponds to the WD cooling track except for four objects whose evolutionary stage corresponds to the constant luminosity phase for low mass stars (J044839.59+483606.0, J050405.54+383324.9, J044822.93+483429.1, and J014150.56+650546.9).

\begin{figure*}
    \centering
    \includegraphics[width=0.7\textwidth]{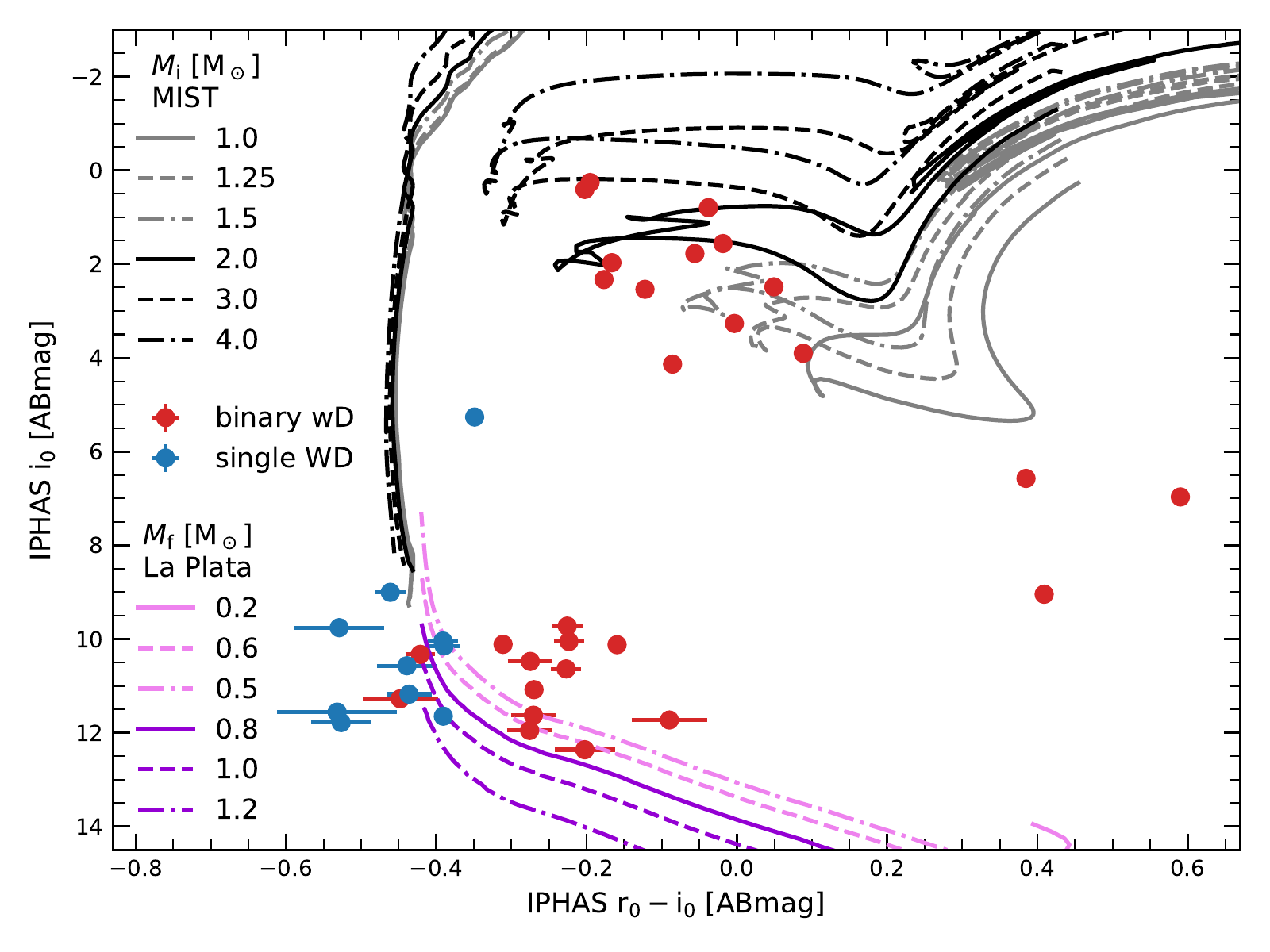}
    \caption{Colour-magnitude diagram using the MIST theoretical tracks (gray/black lines) and La Plata WD cooling tracks (purple lines) for different initial masses (labelled in the figure). Single (blue dots) and binary (red dots) WD candidates are also shown.}
    \label{fig:cmd_iphas}
\end{figure*}

In Figure~\ref{fig:cmd_iphas}, we show the IGAPS colour-magnitude (CMD) diagram of the candidate WDs. For reference, we plot the MESA/MIST
\citep[][and references therein]{Choi2016} evolutionary tracks for stars of 1--4\,M$_{\odot}$ and the La Plata cooling tracks for WDs of 0.2--1.2\,M$_{\odot}$ with hydrogen-rich atmospheres \citep{althaus2013,camisassa2016,camisassa2019}. The absolute magnitudes of WDs were computed by means of appropriate synthetic spectra \citep[][]{koester2010}.

We converted the IGAPS magnitudes of the candidate WDs into absolute magnitudes by using the $E(V-B)$ and distances from
Tables~\ref{tab:wd_simple} and \ref{tab:wd_binary}.
The single WD stars from the GaPHAS catalogue appear in the WD cooling phase, as expected, in the CMD
whereas the colours of the binary WD stars are scattered around in the CMD due to the blending of the two components; a few of them overlap with the
main-sequence (MS)
while others are close to the WD cooling sequence. Binary objects that are located in the
MS loci are dominated by the flux of the companion star in the optical range (e.g., J023215.29+632115.9,
J0333.07+632040.9, and J040522.62+555339.4)
whereas the magnitude of the binary objects located close to the WD loci is the composition of 
both the WD and the binary companion flux (e.g., J045739.49+505757.4, J025931.05+550806.3, and
J045809.37+390538.7).
One additional problem is that for these hot and very hot stars, a small $T_{\rm eff}$ uncertainty
may actually mean a large uncertainty in the reddening/extinction and hence affecting the position
in the CMD; the latter could be solved by spectroscopic observations.

\section{Discussion}\label{discussion}

In the following subsections, we present the discussion of the SED fitting in terms of the estimated stellar parameters and their comparison with different theoretical evolutionary tracks. We also discuss the WD
candidates classification on infrared (IR) colour-colour maps as well as the nature of the binary companions found in our sample. Finally, a discussion related to the nebulae around WD star candidates is also included.

\subsection{Physical properties of WD candidates}

The combination of \textit{GALEX}
\textit{FUV} and \textit{NUV}, and IGAPS g r H$\alpha$ i, analysed with synthetic spectra, enable a preliminary
characterisation of the WD and unresolved binary WD candidates
\citep[see e.g.,][]{bianchi2018}.
Figures~\ref{fig:single_wd1} to \ref{fig:binary_wd3} show the best fitted synthetic spectrum to the
UV-optical photometry, as extracted from GaPHAS catalogue, for
single and binary WD candidates, respectively.
The majority of the fitted $T_{\rm eff}$ correspond to the
$T_{\rm eff}$ expected from the \textit{GALEX} \textit{FUV}$-$\textit{NUV} colour\footnote{For a MilkyWay type dust, with $R_{\rm V}$=3.1, the \textit{GALEX} \textit{FUV}$-$\textit{NUV} colour is almost reddening free \citep[see Table~1 of][]{Bianchi2017}. The absorption in each UV band, however, is higher than in optical bands. Other colour indexes between \textit{GALEX} and optical bands could be severely affected by extinction.}; the \textit{GALEX} \textit{FUV}$-$\textit{NUV}$<-$0.53
corresponds to $T_{\rm eff}\geq$30\,000~K \citep[see Figure~7 of][]{Bianchi2011}.
However, large errors in the photometry could strongly affect the position of the WD stars in the colour-colour
diagrams presented in Figure~\ref{fig:cc_plots_1}. Objects with \textit{GALEX} colours \textit{FUV}$-$\textit{NUV}$\sim-$0.53
but large errors could result in  smaller or higher $T_{\rm eff}$ during the fitting procedure;
this could be the case for objects resulting with $T_{\rm eff}<$30\,000~K. This can also occur when the
binary companion contributes to the overall flux in the \textit{GALEX} \textit{NUV} band (e.g., J023527.80+632738.4 
and J044822.93+483429.1).

As mentioned in section~\ref{sec:SED_analysis} the stellar parameters were obtained by fitting the
different combinations of colour indexes between \textit{GALEX} and IGAPS catalogues by the implementation
of the MCMC algorithm. Most of the fitted parameters are very well constrained between
20\,000~K$\leq T_{\rm eff} \leq$100\,000~K, however, for $T_{\rm eff}>$100\,000~K the fitting is
not accurate because colour indexes, specially \textit{GALEX} \textit{FUV}$-$\textit{NUV} \citep[see Figure~7][]{Bianchi2011},
are not very sensitive to higher effective
temperatures. Objects resulting with $T_{\rm eff}>$100\,000~K must be taken as an upper limit and
must be combined with spectral data in order to corroborate its nature.

\begin{figure}
    \centering
    \includegraphics[width=\columnwidth]{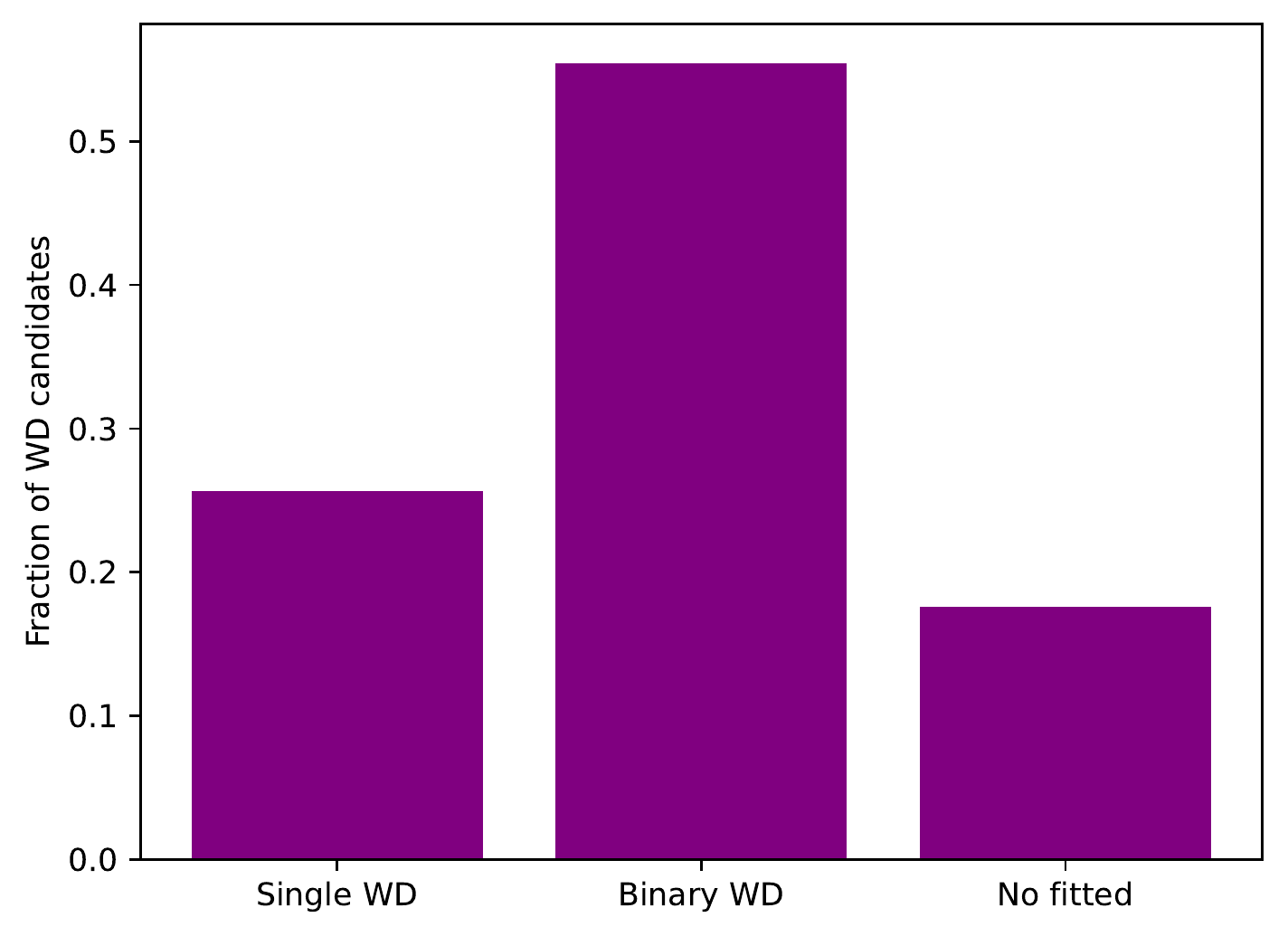}
    \caption{Fraction of single and binary WD star candidates from the GaPHAS catalogue resulted from the fitting procedure. The fraction of WD stars that were not fitted is also shown. A binary fraction of 55\% was estimated.}
    \label{fig:barplot_wds}
\end{figure}

Figure~\ref{fig:barplot_wds} show the total number of single and binary WD candidates obtained from the SED fitting. Out of the 74 WD candidates, a total of 41 resulted in probably binary systems in which a binary fraction of 55.4\% is estimated, similar to the binary fractions expected for solar-like and M-type MS stars \citep[e.g.,][]{Raghavan2010,Winters2019} and lower than the fraction of binary central stars of PNe \citep[$\sim$70\%;][]{deMarco2013,Jones2017}.

\subsection{WD classification}
\begin{figure}
    \centering
    \includegraphics[width=\columnwidth]{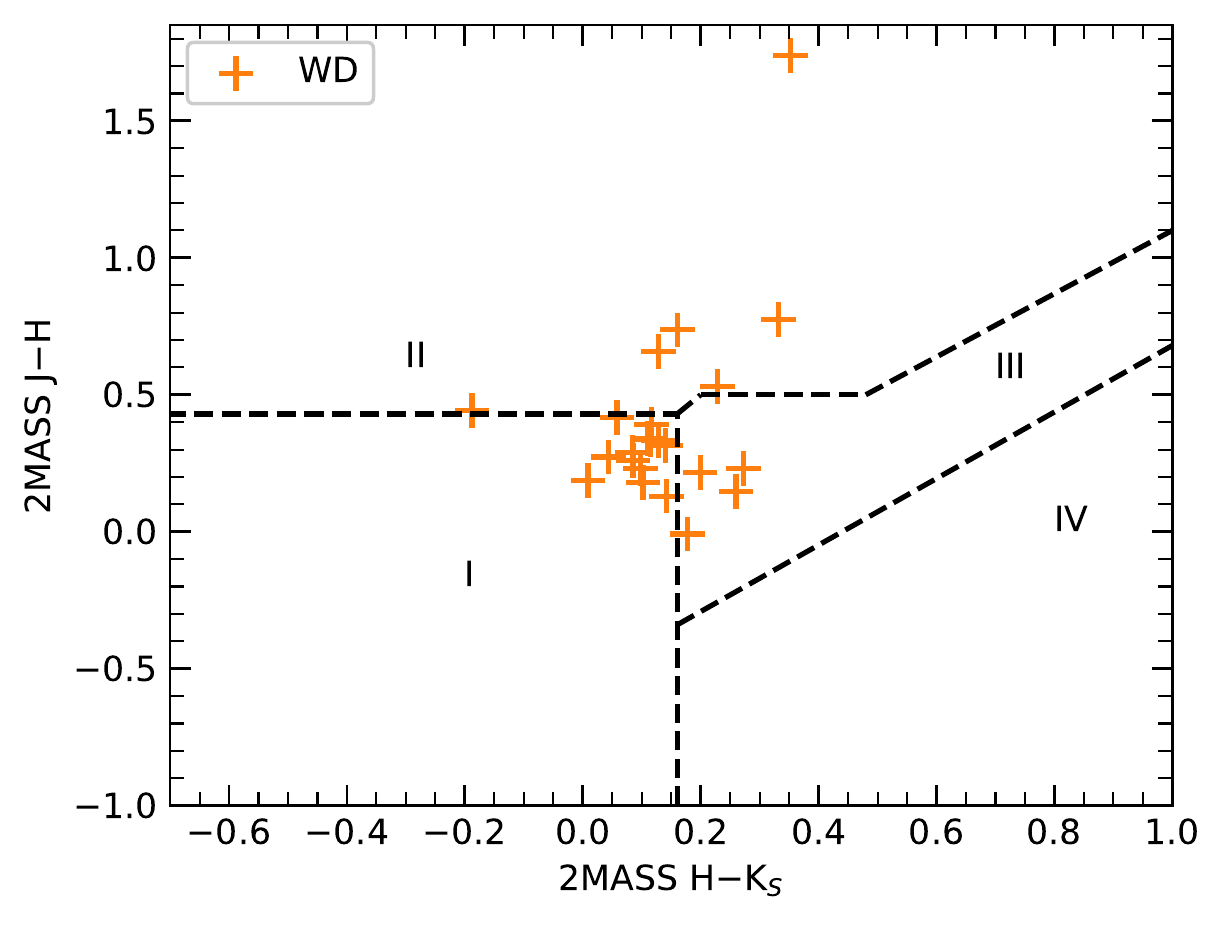}
    \includegraphics[width=\columnwidth]{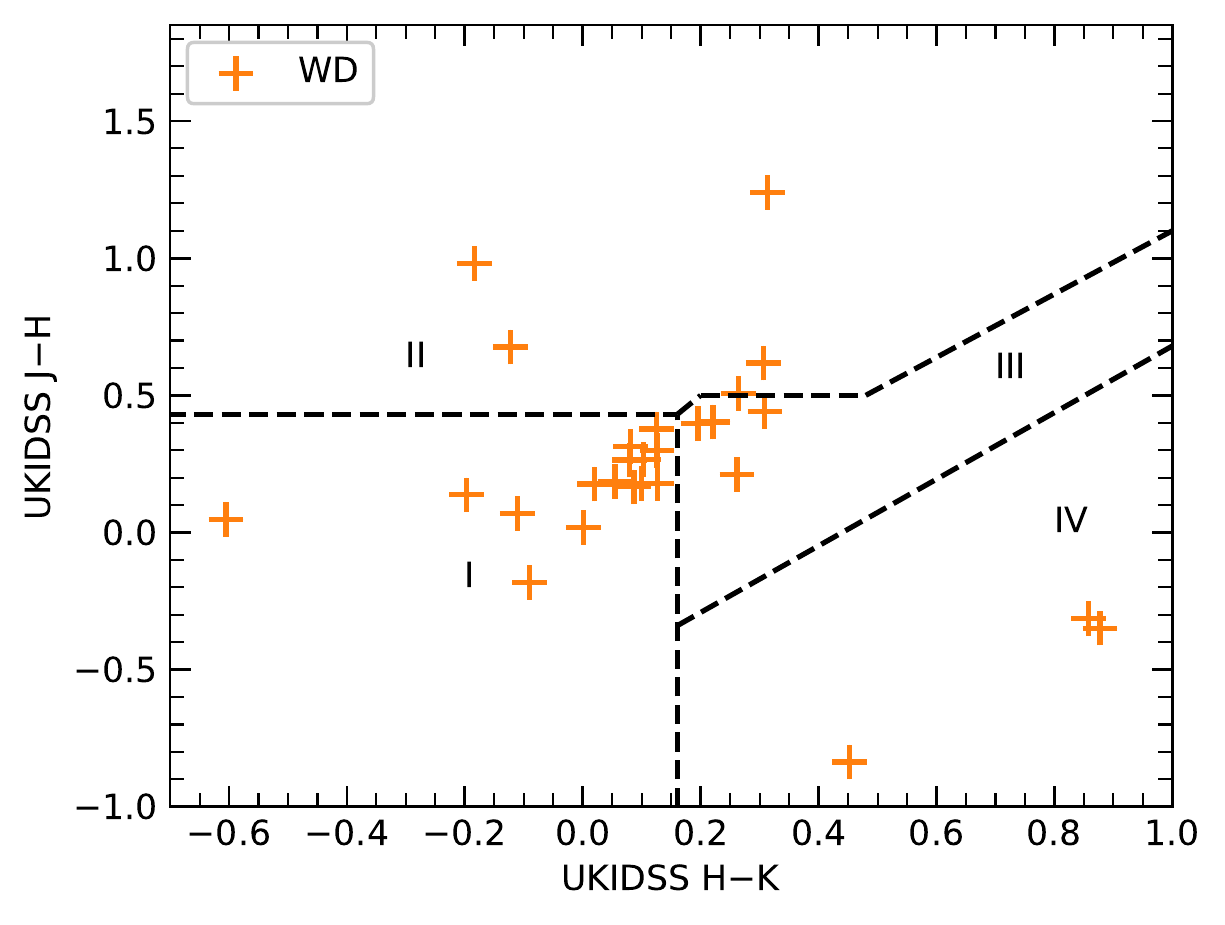}
    \caption{Near-IR (2MASS-top and UKIDSS -bottom) colour-colour diagrams of the
    hot-WDs selection (orange crosses) are shown.
    The four regions delimited
    by a dashed line represent single WD (I), WDs with low-mass, late-type MS star (or L-type star) as a companion (II), WDs with low-mass
    brown dwarf (L-type star) companion (III), and WDs that are
    probably contaminated by the presence of circumstellar material (IV)
    \citep[see][]{Wellhouse2005}.}
    \label{fig:wd_2mass}
\end{figure}

According to \citet{Wellhouse2005}, the infrared colour-colour diagrams of WDs can be classified into four different regions that are occupied by: \texttt{I}) single WDs; \texttt{II})  binary WDs with colours dominated by a low-mass late MS companions or a very low-mass L-type companions; \texttt{III}) binary WDs that have very low-mass brown dwarf companions; \texttt{IV}) objects with colours that are probably contaminated by the presence of circumstellar material \citep[see Figure~1 from][]{Wellhouse2005}.
In this context, an analysis of the IR 2MASS and UKIDSS colour indexes of the WD stars that are shown in Table~\ref{tab:example_wd} was performed. Figure~\ref{fig:wd_2mass} shows the colour-colour diagrams for 2MASS (top) and UKIDSS (bottom) colour indexes.
In the 2MASS diagram we identified 13 objects in region \texttt{I}, 6 in region \texttt{II}, 4 in region \texttt{III}, and none in region \texttt{IV}. In the UKIDSS diagram we found 15 objects in region \texttt{I}, 4 in region \texttt{II}, 5 in region \texttt{III}, and 3 in region \texttt{IV}. Only 13 WD candidates (ten of which are part of the region \texttt{I}) coincide in terms of identification in both diagrams.\\
We notice that objects selected as single in Fig.\ref{fig:wd_2mass} are also described as such based on their SED analysis (\ref{tab:wd_simple}). The same applies to those classified as binary WDs (\ref{tab:wd_binary}).
This is underlying the strength of the method used in Section~\ref{sec:SED_analysis}.

The objects present in region III are of particular interest as they might be a case of a WD with a planetary companion, although a low mass brown dwarf is not discarded. 
We, therefore, performed a search in the Transiting Exoplanet Survey Satellite (TESS) database \citep{Ricker2015} for all the objects in regions III and IV. We expect this investigation to be completed by follow-up observations using ground-based facilities in order to obtain more accurate information on the binary parameters.

\subsection{Nature of the binary companion}
Due to these targets mostly being faint and located in crowded fields, they received low priority in the TESS Candidate Target List (CTL; \citealt{2018AJ....156..102S}) and were therefore not selected for short-cadence observations. 
We created light curves from the TESS full-frame images (FFIs) using the \texttt{eleanor} Python package \citep{2019PASP..131i4502F}, by specifying the target coordinates. 
We then detrended these raw light curves by two different methods; 1) by regressing against background level and co-trending basis vectors from \texttt{eleanor}\footnote{We used the \texttt{linea} Python package available at \url{https://github.com/bmorris3/linea}}; and 2) by a 10-hour biweight window slider \citep{2019AJ....158..143H}, which can account better for unknown systematics, such as coming from other stars within the aperture.
We found the light curves to be negatively impacted by momentum dumps (every ${\sim} 5$ days), so we applied the detrending separately in regions bounded by the times of these dumps.
We then searched these corrected lightcurves for periodic signals using the Box-fitting Least Squares algorithm (BLS; \citealt{2002A&A...391..369K})\footnote{\url{https://github.com/hpparvi/PyBLS}}.

% limitations:
While transits of any planet around a white dwarf would be very large compared to the photometric precision of TESS, there are a number of effects that act to diminish the detection sensitivity of this search.
The vast majority of the targets were observed in sectors 18 and 19 and therefore were observed at a cadence of 30 minutes. Transits of short-period objects around white dwarfs are much shorter than this; consequently, their transit depths will be significantly reduced when sampled with 30-minute integrations.
For example, the 8-minute, 56\% deep signal of the planet candidate around WD\,1856+534 \citep{2020Natur.585..363V} would be reduced to ${\sim}$10\% at the 30-minute cadence. 
The crowdedness of the fields also has a significant impact. Due to the large pixel size of TESS (${\sim} 20\arcsec$), the light from many stars is combined in the aperture around the target. 
This dilutes any signals on the white dwarf by an unknown amount and includes stellar signals, diminishing their significance. 
Our search resulted in a handful of low-SNR candidates, which we hope to follow up with ground-based photometric observations.

\subsection{Nebulae around the WDs}
Finally, we searched for the presence of H$\alpha$ emitting nebulae around the 74 hottest new WD candidates using the IPHAS/IGAPS imagery, and we found no obvious sign of such nebulosities in the close vicinity of the objects in our sample (up to 15$\arcmin$). We note however that the exposure time used for the H$\alpha$ images is relatively short (120s), and as it was shown by \citet{Sabin2021}, deeper observations of these IPHAS/IGAPS objects can reveal new faint external structures. Then, in very few cases, we found the WDs candidates embedded in an H$\alpha$ cloud. We also searched for H$\alpha$ excess in these sources using the data by \citep{Fratta2021}. In Table \ref{tab:wd_simple}, the last column indicates that only 17 out of 74 objects present such excess with a 3$\sigma$ significance. We identified 6 of them as single and 11 in a binary system. In the latter case, the H$\alpha$-excess could be due to accretion onto the WD.
Therefore, deeper optical imaging is needed to investigate if the WDs are (still) surrounded by an ionised (planetary) nebula.

\section{Conclusions}\label{conclusion}

\textit{GALEX} UV in combination with optical databases offers a unique opportunity to investigate and characterise WD stars that are usually elusive in the optical range. We combined \textit{GALEX} UV with the optical IGAPS catalogue and found 74 WD star candidates in the footprint of both catalogues. The selection of the WD candidates was made by analysing the colour-colour diagrams using different combinations of colours
(see Figure~\ref{fig:cc_plots_1}).

An analysis of the SED was done by implementing an MCMC algorithm to fit the \textit{GALEX} UV and IGAPS
photometry resulting in 19 and 31 single and binary WD candidates, respectively (see apendix~\ref{ap:single_wd} and \ref{ap:binary_wd}). The combination
of Gaia EDR3 distances enables us to determine the stellar parameters which in turn enable the
characterisation of different evolutionary stages. However, for the very hot stars, a small
$T_{\rm eff}$ uncertainty, may increase the uncertainty of the reddening calculation and hence
affecting the derived evolutionary stage (see Figures~\ref{fig:teff_logg_track} and \ref{fig:cmd_iphas}).
A spectral analysis must be done in order to
obtain more accurate values of the hot star stellar parameters and to confirm its single
and/or binary nature.

We classified the WD star candidates by analysing the different IR colours using photometric data
from 2MASS and UKIDSS catalogues. Objects found to be single and binary WD candidates in
the IR colour-colour diagrams (see Figure~\ref{fig:wd_2mass}) also resulted as such in the SED analysis.
This supports the
strength of the method used in Section~\ref{sec:SED_analysis} and the advantages of the combination
of \textit{GALEX} UV with optical photometry. The IR colour-colour diagrams enable us to also identify WD stars that probably contain a brown dwarf or planetary companion. Follow-up TESS observations with higher cadence as well as follow-up using ground-based facilities are required in order to obtain more accurate information
on the binary parameters. Finally, no nebulae were detected around the hot-WDs by exploiting the short exposure IGAPS imagery, further deeper observations are needed to unveil the very evolved PNe.

\section*{Acknowledgements}

MAGM and LS acknowledge support from UNAM PAPIIT projects IN101819 and IN110122.
MAGM also acknowledges the postdoctoral fellowship granted by the Instituto de Astronom{\'i}a of the Universidad Nacional Aut{\'o}noma de M{\'e}xico.
RR has received funding from the postdoctoral fellowship programme Beatriu de Pin\'os, funded by the Secretary of Universities and Research (Government of Catalonia) and by the Horizon 2020 programme of research and innovation of the European Union under the Maria Sk\l{}odowska-Curie grant agreement No 801370.

We thank Dra. Luciana Bianchi for sharing with us the theoretical atmosphere models for WD stars that were used in the colour-colour diagram analysis and for the SED fitting of our sample of hot WD stars.
We also thank the referee, Prof. Martin Barstow, for his careful reading of the paper and his valuable comments.
This research has made use of the SIMBAD database operated at CDS (Strasbourg, France). 
This paper makes use of data obtained as part of the IGAPS merger of the IPHAS and UVEX surveys (www.igapsimages.org) carried out at the Isaac Newton Telescope (INT). The INT is operated on the island of La Palma by the Isaac Newton Group in the Spanish Observatorio del Roque de los Muchachos of the Instituto de Astrofisica de Canarias. All IGAPS data were processed by the Cambridge Astronomical Survey Unit, at the Institute of Astronomy in Cambridge. The uniformly-calibrated bandmerged IGAPS catalogue was assembled using the high performance computing cluster via the Centre for Astrophysics Research, University of Hertfordshire.
This paper is based on data products from observations made with ESO Telescopes at the La Silla Paranal Observatory under programme ID 177.D-3023, as part of the VST Photometric H$\alpha$ Survey of the Southern Galactic Plane and Bulge (VPHAS+, www.vphas.eu).

This research is also based on observations made with \textit{GALEX}, obtained from the MAST data archive at the Space Telescope Science Institute, which is operated by the Association of Universities for Research in  Astronomy, Inc., under NASA contract NAS 5$-$26555. This work also presents results from the European Space Agency (ESA) space mission Gaia. Gaia data are being processed by the Gaia Data Processing and Analysis Consortium (DPAC). Funding for the DPAC is provided by national institutions, in particular the institutions participating in the Gaia MultiLateral Agreement (MLA). The Gaia mission website is \url{https://www.cosmos.esa.int/gaia}. The Gaia archive website is \url{https://archives.esac.esa.int/gaia.}

\section*{Data Availability}

All data used in this manuscript are available through public archives.



\bibliographystyle{mnras}
\bibliography{main} % if your bibtex file is called example.bib

\begin{thebibliography}{}
\makeatletter
\relax
\def\mn@urlcharsother{\let\do\@makeother \do\$\do\&\do\#\do\^\do\_\do\%\do\~}
\def\mn@doi{\begingroup\mn@urlcharsother \@ifnextchar [ {\mn@doi@}
  {\mn@doi@[]}}
\def\mn@doi@[#1]#2{\def\@tempa{#1}\ifx\@tempa\@empty \href
  {http://dx.doi.org/#2} {doi:#2}\else \href {http://dx.doi.org/#2} {#1}\fi
  \endgroup}
\def\mn@eprint#1#2{\mn@eprint@#1:#2::\@nil}
\def\mn@eprint@arXiv#1{\href {http://arxiv.org/abs/#1} {{\tt arXiv:#1}}}
\def\mn@eprint@dblp#1{\href {http://dblp.uni-trier.de/rec/bibtex/#1.xml}
  {dblp:#1}}
\def\mn@eprint@#1:#2:#3:#4\@nil{\def\@tempa {#1}\def\@tempb {#2}\def\@tempc
  {#3}\ifx \@tempc \@empty \let \@tempc \@tempb \let \@tempb \@tempa \fi \ifx
  \@tempb \@empty \def\@tempb {arXiv}\fi \@ifundefined
  {mn@eprint@\@tempb}{\@tempb:\@tempc}{\expandafter \expandafter \csname
  mn@eprint@\@tempb\endcsname \expandafter{\@tempc}}}

\bibitem[\protect\citeauthoryear{{Allard}, {Homeier}  \& {Freytag}}{{Allard}
  et~al.}{2012}]{Allard2012}
{Allard} F.,  {Homeier} D.,   {Freytag} B.,  2012, \mn@doi [Philosophical
  Transactions of the Royal Society of London Series A]
  {10.1098/rsta.2011.0269}, \href
  {https://ui.adsabs.harvard.edu/abs/2012RSPTA.370.2765A} {370, 2765}

\bibitem[\protect\citeauthoryear{{Almaini} et~al.,}{{Almaini}
  et~al.}{2007}]{Almaini2007}
{Almaini} O.,  et~al., 2007, in {Metcalfe} N.,  {Shanks} T.,  eds,
  Astronomical Society of the Pacific Conference Series Vol. 379, Cosmic
  Frontiers. p.~163

\bibitem[\protect\citeauthoryear{{Althaus}, {Miller Bertolami}  \&
  {C{\'o}rsico}}{{Althaus} et~al.}{2013}]{althaus2013}
{Althaus} L.~G.,  {Miller Bertolami} M.~M.,   {C{\'o}rsico} A.~H.,  2013,
  \mn@doi [\aap] {10.1051/0004-6361/201321868}, \href
  {https://ui.adsabs.harvard.edu/abs/2013A&A...557A..19A} {557, A19}

\bibitem[\protect\citeauthoryear{{Bailer-Jones}, {Rybizki}, {Fouesneau},
  {Demleitner}  \& {Andrae}}{{Bailer-Jones} et~al.}{2021}]{BailerJones2021}
{Bailer-Jones} C.~A.~L.,  {Rybizki} J.,  {Fouesneau} M.,  {Demleitner} M.,
  {Andrae} R.,  2021, \mn@doi [\aj] {10.3847/1538-3881/abd806}, \href
  {https://ui.adsabs.harvard.edu/abs/2021AJ....161..147B} {161, 147}

\bibitem[\protect\citeauthoryear{{Barentsen} et~al.,}{{Barentsen}
  et~al.}{2014}]{Barentsen2014}
{Barentsen} G.,  et~al., 2014, \mn@doi [\mnras] {10.1093/mnras/stu1651}, \href
  {https://ui.adsabs.harvard.edu/abs/2014MNRAS.444.3230B} {444, 3230}

\bibitem[\protect\citeauthoryear{{Barker}, {Zijlstra}, {De Marco}, {Frew},
  {Drew}, {Corradi}, {Eisl{\"o}ffel}  \& {Parker}}{{Barker}
  et~al.}{2018}]{Barker2018}
{Barker} H.,  {Zijlstra} A.,  {De Marco} O.,  {Frew} D.~J.,  {Drew} J.~E.,
  {Corradi} R. L.~M.,  {Eisl{\"o}ffel} J.,   {Parker} Q.~A.,  2018, \mn@doi
  [\mnras] {10.1093/mnras/stx3240}, \href
  {https://ui.adsabs.harvard.edu/abs/2018MNRAS.475.4504B} {475, 4504}

\bibitem[\protect\citeauthoryear{{Bianchi}}{{Bianchi}}{2009}]{Bianchi2009apss}
{Bianchi} L.,  2009, \mn@doi [\apss] {10.1007/s10509-008-9761-3}, \href
  {https://ui.adsabs.harvard.edu/abs/2009Ap&SS.320...11B} {320, 11}

\bibitem[\protect\citeauthoryear{{Bianchi} \& {Shiao}}{{Bianchi} \&
  {Shiao}}{2020}]{bianchi2020}
{Bianchi} L.,  {Shiao} B.,  2020, \mn@doi [\apjs] {10.3847/1538-4365/aba2d7},
  \href {https://ui.adsabs.harvard.edu/abs/2020ApJS..250...36B} {250, 36}

\bibitem[\protect\citeauthoryear{{Bianchi}, {Hutchings}, {Efremova}, {Herald},
  {Bressan}  \& {Martin}}{{Bianchi} et~al.}{2009}]{Bianchi2009}
{Bianchi} L.,  {Hutchings} J.~B.,  {Efremova} B.,  {Herald} J.~E.,  {Bressan}
  A.,   {Martin} C.,  2009, \mn@doi [\aj] {10.1088/0004-6256/137/4/3761}, \href
  {https://ui.adsabs.harvard.edu/abs/2009AJ....137.3761B} {137, 3761}

\bibitem[\protect\citeauthoryear{{Bianchi}, {Efremova}, {Herald}, {Girardi},
  {Zabot}, {Marigo}  \& {Martin}}{{Bianchi} et~al.}{2011}]{Bianchi2011}
{Bianchi} L.,  {Efremova} B.,  {Herald} J.,  {Girardi} L.,  {Zabot} A.,
  {Marigo} P.,   {Martin} C.,  2011, \mn@doi [\mnras]
  {10.1111/j.1365-2966.2010.17890.x}, \href
  {https://ui.adsabs.harvard.edu/abs/2011MNRAS.411.2770B} {411, 2770}

\bibitem[\protect\citeauthoryear{{Bianchi}, {Shiao}  \& {Thilker}}{{Bianchi}
  et~al.}{2017}]{Bianchi2017}
{Bianchi} L.,  {Shiao} B.,   {Thilker} D.,  2017, \mn@doi [\apjs]
  {10.3847/1538-4365/aa7053}, \href
  {https://ui.adsabs.harvard.edu/abs/2017ApJS..230...24B} {230, 24}

\bibitem[\protect\citeauthoryear{{Bianchi}, {Keller}, {Bohlin}, {Barstow}  \&
  {Casewell}}{{Bianchi} et~al.}{2018}]{bianchi2018}
{Bianchi} L.,  {Keller} G.~R.,  {Bohlin} R.,  {Barstow} M.,   {Casewell} S.,
  2018, \mn@doi [\apss] {10.1007/s10509-018-3369-z}, \href
  {https://ui.adsabs.harvard.edu/abs/2018Ap&SS.363..166B} {363, 166}

\bibitem[\protect\citeauthoryear{{Camisassa}, {Althaus}, {C{\'o}rsico},
  {Vinyoles}, {Serenelli}, {Isern}, {Miller Bertolami}  \&
  {Garc{\'\i}a-Berro}}{{Camisassa} et~al.}{2016}]{camisassa2016}
{Camisassa} M.~E.,  {Althaus} L.~G.,  {C{\'o}rsico} A.~H.,  {Vinyoles} N.,
  {Serenelli} A.~M.,  {Isern} J.,  {Miller Bertolami} M.~M.,
  {Garc{\'\i}a-Berro} E.,  2016, \mn@doi [\apj] {10.3847/0004-637X/823/2/158},
  \href {https://ui.adsabs.harvard.edu/abs/2016ApJ...823..158C} {823, 158}

\bibitem[\protect\citeauthoryear{{Camisassa} et~al.,}{{Camisassa}
  et~al.}{2019}]{camisassa2019}
{Camisassa} M.~E.,  et~al., 2019, \mn@doi [\aap] {10.1051/0004-6361/201833822},
  \href {https://ui.adsabs.harvard.edu/abs/2019A&A...625A..87C} {625, A87}

\bibitem[\protect\citeauthoryear{{Capaccioli} et~al.,}{{Capaccioli}
  et~al.}{2012}]{Capaccioli2012}
{Capaccioli} M.,  et~al., 2012, in Science from the Next Generation Imaging and
  Spectroscopic Surveys. p.~1

\bibitem[\protect\citeauthoryear{{Cardelli}, {Clayton}  \& {Mathis}}{{Cardelli}
  et~al.}{1989}]{Cardelli1989}
{Cardelli} J.~A.,  {Clayton} G.~C.,   {Mathis} J.~S.,  1989, \mn@doi [\apj]
  {10.1086/167900}, \href
  {https://ui.adsabs.harvard.edu/abs/1989ApJ...345..245C} {345, 245}

\bibitem[\protect\citeauthoryear{{Castelli} \& {Kurucz}}{{Castelli} \&
  {Kurucz}}{2003}]{Castelli2003}
{Castelli} F.,  {Kurucz} R.~L.,  2003, in {Piskunov} N.,  {Weiss} W.~W.,
  {Gray} D.~F.,  eds, ~ Vol. 210, Modelling of Stellar Atmospheres. p.~A20
  (\mn@eprint {arXiv} {astro-ph/0405087})

\bibitem[\protect\citeauthoryear{{Choi}, {Dotter}, {Conroy}, {Cantiello},
  {Paxton}  \& {Johnson}}{{Choi} et~al.}{2016}]{Choi2016}
{Choi} J.,  {Dotter} A.,  {Conroy} C.,  {Cantiello} M.,  {Paxton} B.,
  {Johnson} B.~D.,  2016, \mn@doi [\apj] {10.3847/0004-637X/823/2/102}, \href
  {https://ui.adsabs.harvard.edu/abs/2016ApJ...823..102C} {823, 102}

\bibitem[\protect\citeauthoryear{{De Marco}, {Passy}, {Frew}, {Moe}  \&
  {Jacoby}}{{De Marco} et~al.}{2013}]{deMarco2013}
{De Marco} O.,  {Passy} J.-C.,  {Frew} D.~J.,  {Moe} M.,   {Jacoby} G.~H.,
  2013, \mn@doi [\mnras] {10.1093/mnras/sts180}, \href
  {https://ui.adsabs.harvard.edu/abs/2013MNRAS.428.2118D} {428, 2118}

\bibitem[\protect\citeauthoryear{{Drew} et~al.,}{{Drew}
  et~al.}{2014}]{Drew2014}
{Drew} J.~E.,  et~al., 2014, \mn@doi [\mnras] {10.1093/mnras/stu394}, \href
  {https://ui.adsabs.harvard.edu/abs/2014MNRAS.440.2036D} {440, 2036}

\bibitem[\protect\citeauthoryear{{Drew} et~al.,}{{Drew}
  et~al.}{2015}]{Drew2005}
{Drew} J.~E.,  et~al., 2015, VizieR Online Data Catalog, \href
  {https://ui.adsabs.harvard.edu/abs/2015yCat..74402036D} {p. J/MNRAS/440/2036}

\bibitem[\protect\citeauthoryear{{Feinstein} et~al.,}{{Feinstein}
  et~al.}{2019}]{2019PASP..131i4502F}
{Feinstein} A.~D.,  et~al., 2019, \mn@doi [\pasp] {10.1088/1538-3873/ab291c},
  \href {https://ui.adsabs.harvard.edu/abs/2019PASP..131i4502F} {131, 094502}

\bibitem[\protect\citeauthoryear{{Foreman-Mackey}, {Hogg}, {Lang}  \&
  {Goodman}}{{Foreman-Mackey} et~al.}{2013}]{ForemanMackey2013}
{Foreman-Mackey} D.,  {Hogg} D.~W.,  {Lang} D.,   {Goodman} J.,  2013, \mn@doi
  [\pasp] {10.1086/670067}, \href
  {https://ui.adsabs.harvard.edu/abs/2013PASP..125..306F} {125, 306}

\bibitem[\protect\citeauthoryear{{Fratta} et~al.,}{{Fratta}
  et~al.}{2021}]{Fratta2021}
{Fratta} M.,  et~al., 2021, \mn@doi [\mnras] {10.1093/mnras/stab1258}, \href
  {https://ui.adsabs.harvard.edu/abs/2021MNRAS.505.1135F} {505, 1135}

\bibitem[\protect\citeauthoryear{{Gaia Collaboration}}{{Gaia
  Collaboration}}{2020}]{GaiaColab2020}
{Gaia Collaboration} 2020, VizieR Online Data Catalog, \href
  {https://ui.adsabs.harvard.edu/abs/2020yCat.1350....0G} {p. I/350}

\bibitem[\protect\citeauthoryear{{Gaia Collaboration} \& et al.}{{Gaia
  Collaboration} \& et~al.}{2018}]{Gaia2018}
{Gaia Collaboration} et al. 2018, \mn@doi [\aap] {10.1051/0004-6361/201833051},
  \href {https://ui.adsabs.harvard.edu/abs/2018A&A...616A...1G} {616, A1}

\bibitem[\protect\citeauthoryear{{Gaia Collaboration} et~al.,}{{Gaia
  Collaboration} et~al.}{2021}]{EDR32021}
{Gaia Collaboration} et~al., 2021, \mn@doi [\aap]
  {10.1051/0004-6361/202039657}, \href
  {https://ui.adsabs.harvard.edu/abs/2021A&A...649A...1G} {649, A1}

\bibitem[\protect\citeauthoryear{{Gentile Fusillo} et~al.,}{{Gentile Fusillo}
  et~al.}{2019}]{Gentile2019}
{Gentile Fusillo} N.~P.,  et~al., 2019, \mn@doi [\mnras]
  {10.1093/mnras/sty3016}, \href
  {https://ui.adsabs.harvard.edu/abs/2019MNRAS.482.4570G} {482, 4570}

\bibitem[\protect\citeauthoryear{{Gentile Fusillo} et~al.,}{{Gentile Fusillo}
  et~al.}{2021}]{Gentile2021}
{Gentile Fusillo} N.~P.,  et~al., 2021, \mn@doi [\mnras]
  {10.1093/mnras/stab2672}, \href
  {https://ui.adsabs.harvard.edu/abs/2021MNRAS.508.3877G} {508, 3877}

\bibitem[\protect\citeauthoryear{{Green}}{{Green}}{2018}]{Green2018}
{Green} G.,  2018, \mn@doi [The Journal of Open Source Software]
  {10.21105/joss.00695}, \href
  {https://ui.adsabs.harvard.edu/abs/2018JOSS....3..695G} {3, 695}

\bibitem[\protect\citeauthoryear{{Green}, {Schlafly}, {Zucker}, {Speagle}  \&
  {Finkbeiner}}{{Green} et~al.}{2019}]{Green2019}
{Green} G.~M.,  {Schlafly} E.,  {Zucker} C.,  {Speagle} J.~S.,   {Finkbeiner}
  D.,  2019, \mn@doi [\apj] {10.3847/1538-4357/ab5362}, \href
  {https://ui.adsabs.harvard.edu/abs/2019ApJ...887...93G} {887, 93}

\bibitem[\protect\citeauthoryear{{Groot} et~al.,}{{Groot}
  et~al.}{2009}]{Groot2009}
{Groot} P.~J.,  et~al., 2009, \mn@doi [\mnras]
  {10.1111/j.1365-2966.2009.15273.x}, \href
  {https://ui.adsabs.harvard.edu/abs/2009MNRAS.399..323G} {399, 323}

\bibitem[\protect\citeauthoryear{{Hauschildt}, {Allard}  \&
  {Baron}}{{Hauschildt} et~al.}{1999}]{Hauschildt1999}
{Hauschildt} P.~H.,  {Allard} F.,   {Baron} E.,  1999, \mn@doi [\apj]
  {10.1086/306745}, \href
  {https://ui.adsabs.harvard.edu/abs/1999ApJ...512..377H} {512, 377}

\bibitem[\protect\citeauthoryear{{Hippke}, {David}, {Mulders}  \&
  {Heller}}{{Hippke} et~al.}{2019}]{2019AJ....158..143H}
{Hippke} M.,  {David} T.~J.,  {Mulders} G.~D.,   {Heller} R.,  2019, \mn@doi
  [\aj] {10.3847/1538-3881/ab3984}, \href
  {https://ui.adsabs.harvard.edu/abs/2019AJ....158..143H} {158, 143}

\bibitem[\protect\citeauthoryear{{Hubeny} \& {Lanz}}{{Hubeny} \&
  {Lanz}}{1995}]{Hubeny1995}
{Hubeny} I.,  {Lanz} T.,  1995, \mn@doi [\apj] {10.1086/175226}, \href
  {https://ui.adsabs.harvard.edu/abs/1995ApJ...439..875H} {439, 875}

\bibitem[\protect\citeauthoryear{{Jones}}{{Jones}}{2020}]{Jones2020}
{Jones} D.,  2020, \mn@doi [Galaxies] {10.3390/galaxies8020028}, \href
  {https://ui.adsabs.harvard.edu/abs/2020Galax...8...28J} {8, 28}

\bibitem[\protect\citeauthoryear{{Jones} \& {Boffin}}{{Jones} \&
  {Boffin}}{2017}]{Jones2017}
{Jones} D.,  {Boffin} H. M.~J.,  2017, \mn@doi [Nature Astronomy]
  {10.1038/s41550-017-0117}, \href
  {https://ui.adsabs.harvard.edu/abs/2017NatAs...1E.117J} {1, 0117}

\bibitem[\protect\citeauthoryear{{Kepler} et~al.,}{{Kepler}
  et~al.}{2016}]{Kepler2016}
{Kepler} S.~O.,  et~al., 2016, \mn@doi [\mnras] {10.1093/mnras/stv2526}, \href
  {https://ui.adsabs.harvard.edu/abs/2016MNRAS.455.3413K} {455, 3413}

\bibitem[\protect\citeauthoryear{{Kepler} et~al.,}{{Kepler}
  et~al.}{2019}]{Kepler2019}
{Kepler} S.~O.,  et~al., 2019, \mn@doi [\mnras] {10.1093/mnras/stz960}, \href
  {https://ui.adsabs.harvard.edu/abs/2019MNRAS.486.2169K} {486, 2169}

\bibitem[\protect\citeauthoryear{{Koester}}{{Koester}}{2010}]{koester2010}
{Koester} D.,  2010, \memsai, \href
  {https://ui.adsabs.harvard.edu/abs/2010MmSAI..81..921K} {81, 921}

\bibitem[\protect\citeauthoryear{{Koester} \& {Chanmugam}}{{Koester} \&
  {Chanmugam}}{1990}]{Koester1990}
{Koester} D.,  {Chanmugam} G.,  1990, \mn@doi [Reports on Progress in Physics]
  {10.1088/0034-4885/53/7/001}, \href
  {https://ui.adsabs.harvard.edu/abs/1990RPPh...53..837K} {53, 837}

\bibitem[\protect\citeauthoryear{{Kov{\'a}cs}, {Zucker}  \&
  {Mazeh}}{{Kov{\'a}cs} et~al.}{2002}]{2002A&A...391..369K}
{Kov{\'a}cs} G.,  {Zucker} S.,   {Mazeh} T.,  2002, \mn@doi [\aap]
  {10.1051/0004-6361:20020802}, \href
  {https://ui.adsabs.harvard.edu/abs/2002A&A...391..369K} {391, 369}

\bibitem[\protect\citeauthoryear{{Kuijken} et~al.,}{{Kuijken}
  et~al.}{2002}]{Kuijken2002}
{Kuijken} K.,  et~al., 2002, The Messenger, \href
  {https://ui.adsabs.harvard.edu/abs/2002Msngr.110...15K} {110, 15}

\bibitem[\protect\citeauthoryear{{Kwitter} \& {Henry}}{{Kwitter} \&
  {Henry}}{2022}]{Kwitter2022}
{Kwitter} K.~B.,  {Henry} R.~B.~C.,  2022, \mn@doi [\pasp]
  {10.1088/1538-3873/ac32b1}, \href
  {https://ui.adsabs.harvard.edu/abs/2022PASP..134b2001K} {134, 022001}

\bibitem[\protect\citeauthoryear{{Lawrence} et~al.,}{{Lawrence}
  et~al.}{2007}]{Lawrence2007}
{Lawrence} A.,  et~al., 2007, \mn@doi [\mnras]
  {10.1111/j.1365-2966.2007.12040.x}, \href
  {https://ui.adsabs.harvard.edu/abs/2007MNRAS.379.1599L} {379, 1599}

\bibitem[\protect\citeauthoryear{{Martin} et~al.,}{{Martin}
  et~al.}{2005}]{Martin2005}
{Martin} D.~C.,  et~al., 2005, \mn@doi [\apjl] {10.1086/426387}, \href
  {https://ui.adsabs.harvard.edu/abs/2005ApJ...619L...1M} {619, L1}

\bibitem[\protect\citeauthoryear{{Miller Bertolami}}{{Miller
  Bertolami}}{2016}]{Miller2016}
{Miller Bertolami} M.~M.,  2016, \mn@doi [\aap] {10.1051/0004-6361/201526577},
  \href {https://ui.adsabs.harvard.edu/abs/2016A&A...588A..25M} {588, A25}

\bibitem[\protect\citeauthoryear{{Mongui{\'o}} et~al.,}{{Mongui{\'o}}
  et~al.}{2020}]{Monguio2020}
{Mongui{\'o}} M.,  et~al., 2020, \mn@doi [\aap] {10.1051/0004-6361/201937333},
  \href {https://ui.adsabs.harvard.edu/abs/2020A&A...638A..18M} {638, A18}

\bibitem[\protect\citeauthoryear{{Morrissey} et~al.,}{{Morrissey}
  et~al.}{2007}]{Morrissey2007}
{Morrissey} P.,  et~al., 2007, \mn@doi [\apjs] {10.1086/520512}, \href
  {https://ui.adsabs.harvard.edu/abs/2007ApJS..173..682M} {173, 682}

\bibitem[\protect\citeauthoryear{{Osuna} et~al.,}{{Osuna} et~al.}{2008}]{adql}
{Osuna} P.,  et~al., 2008, {IVOA Astronomical Data Query Language Version
  2.00}, IVOA Recommendation 30 October 2008 (\mn@eprint {arXiv} {1110.0503}),
  \mn@doi{10.5479/ADS/bib/2008ivoa.spec.1030O}

\bibitem[\protect\citeauthoryear{{Panei}, {Althaus}, {Chen}  \& {Han}}{{Panei}
  et~al.}{2007}]{Panei2007}
{Panei} J.~A.,  {Althaus} L.~G.,  {Chen} X.,   {Han} Z.,  2007, \mn@doi
  [\mnras] {10.1111/j.1365-2966.2007.12400.x}, \href
  {https://ui.adsabs.harvard.edu/abs/2007MNRAS.382..779P} {382, 779}

\bibitem[\protect\citeauthoryear{{Parker} et~al.,}{{Parker}
  et~al.}{2005}]{Parker2005}
{Parker} Q.~A.,  et~al., 2005, \mn@doi [\mnras]
  {10.1111/j.1365-2966.2005.09350.x}, \href
  {https://ui.adsabs.harvard.edu/abs/2005MNRAS.362..689P} {362, 689}

\bibitem[\protect\citeauthoryear{{Parker} et~al.,}{{Parker}
  et~al.}{2006}]{Parker2006}
{Parker} Q.~A.,  et~al., 2006, \mn@doi [\mnras]
  {10.1111/j.1365-2966.2006.10950.x}, \href
  {https://ui.adsabs.harvard.edu/abs/2006MNRAS.373...79P} {373, 79}

\bibitem[\protect\citeauthoryear{{Raddi} et~al.,}{{Raddi}
  et~al.}{2017}]{Raddi2017}
{Raddi} R.,  et~al., 2017, \mn@doi [\mnras] {10.1093/mnras/stx2243}, \href
  {https://ui.adsabs.harvard.edu/abs/2017MNRAS.472.4173R} {472, 4173}

\bibitem[\protect\citeauthoryear{{Raghavan} et~al.,}{{Raghavan}
  et~al.}{2010}]{Raghavan2010}
{Raghavan} D.,  et~al., 2010, \mn@doi [\apjs] {10.1088/0067-0049/190/1/1},
  \href {https://ui.adsabs.harvard.edu/abs/2010ApJS..190....1R} {190, 1}

\bibitem[\protect\citeauthoryear{{Ricker} et~al.,}{{Ricker}
  et~al.}{2015}]{Ricker2015}
{Ricker} G.~R.,  et~al., 2015, \mn@doi [Journal of Astronomical Telescopes,
  Instruments, and Systems] {10.1117/1.JATIS.1.1.014003}, \href
  {https://ui.adsabs.harvard.edu/abs/2015JATIS...1a4003R} {1, 014003}

\bibitem[\protect\citeauthoryear{{Sabin} et~al.,}{{Sabin}
  et~al.}{2014}]{Sabin2014}
{Sabin} L.,  et~al., 2014, \mn@doi [\mnras] {10.1093/mnras/stu1404}, \href
  {https://ui.adsabs.harvard.edu/abs/2014MNRAS.443.3388S} {443, 3388}

\bibitem[\protect\citeauthoryear{{Sabin} et~al.,}{{Sabin}
  et~al.}{2021}]{Sabin2021}
{Sabin} L.,  et~al., 2021, arXiv e-prints, \href
  {https://ui.adsabs.harvard.edu/abs/2021arXiv210813612S} {p. arXiv:2108.13612}

\bibitem[\protect\citeauthoryear{{Schlegel}, {Finkbeiner}  \&
  {Davis}}{{Schlegel} et~al.}{1998}]{Schlegel1998}
{Schlegel} D.~J.,  {Finkbeiner} D.~P.,   {Davis} M.,  1998, \mn@doi [\apj]
  {10.1086/305772}, \href
  {https://ui.adsabs.harvard.edu/abs/1998ApJ...500..525S} {500, 525}

\bibitem[\protect\citeauthoryear{{Skrutskie} et~al.,}{{Skrutskie}
  et~al.}{2006}]{Skrustskie2006}
{Skrutskie} M.~F.,  et~al., 2006, \mn@doi [\aj] {10.1086/498708}, \href
  {https://ui.adsabs.harvard.edu/abs/2006AJ....131.1163S} {131, 1163}

\bibitem[\protect\citeauthoryear{{Stassun} et~al.,}{{Stassun}
  et~al.}{2018}]{2018AJ....156..102S}
{Stassun} K.~G.,  et~al., 2018, \mn@doi [\aj] {10.3847/1538-3881/aad050}, \href
  {https://ui.adsabs.harvard.edu/abs/2018AJ....156..102S} {156, 102}

\bibitem[\protect\citeauthoryear{{Vanderburg} et~al.,}{{Vanderburg}
  et~al.}{2020}]{2020Natur.585..363V}
{Vanderburg} A.,  et~al., 2020, \mn@doi [\nat] {10.1038/s41586-020-2713-y},
  \href {https://ui.adsabs.harvard.edu/abs/2020Natur.585..363V} {585, 363}

\bibitem[\protect\citeauthoryear{{Weidmann} et~al.,}{{Weidmann}
  et~al.}{2020}]{Weidmann2020}
{Weidmann} W.~A.,  et~al., 2020, \mn@doi [\aap] {10.1051/0004-6361/202037998},
  \href {https://ui.adsabs.harvard.edu/abs/2020A&A...640A..10W} {640, A10}

\bibitem[\protect\citeauthoryear{{Wellhouse}, {Hoard}, {Howell}, {Wachter}  \&
  {Esin}}{{Wellhouse} et~al.}{2005}]{Wellhouse2005}
{Wellhouse} J.~W.,  {Hoard} D.~W.,  {Howell} S.~B.,  {Wachter} S.,   {Esin}
  A.~A.,  2005, \mn@doi [\pasp] {10.1086/497084}, \href
  {https://ui.adsabs.harvard.edu/abs/2005PASP..117.1378W} {117, 1378}

\bibitem[\protect\citeauthoryear{{Winters} et~al.,}{{Winters}
  et~al.}{2019}]{Winters2019}
{Winters} J.~G.,  et~al., 2019, \mn@doi [\aj] {10.3847/1538-3881/ab05dc}, \href
  {https://ui.adsabs.harvard.edu/abs/2019AJ....157..216W} {157, 216}

\bibitem[\protect\citeauthoryear{{York} et~al.,}{{York}
  et~al.}{2000}]{York2000}
{York} D.~G.,  et~al., 2000, \mn@doi [\aj] {10.1086/301513}, \href
  {https://ui.adsabs.harvard.edu/abs/2000AJ....120.1579Y} {120, 1579}

\makeatother
\end{thebibliography}

\appendix

\section{Single WD fitting}
\label{ap:single_wd}

\begin{figure*}
    \centering
\includegraphics[width=0.33\textwidth]{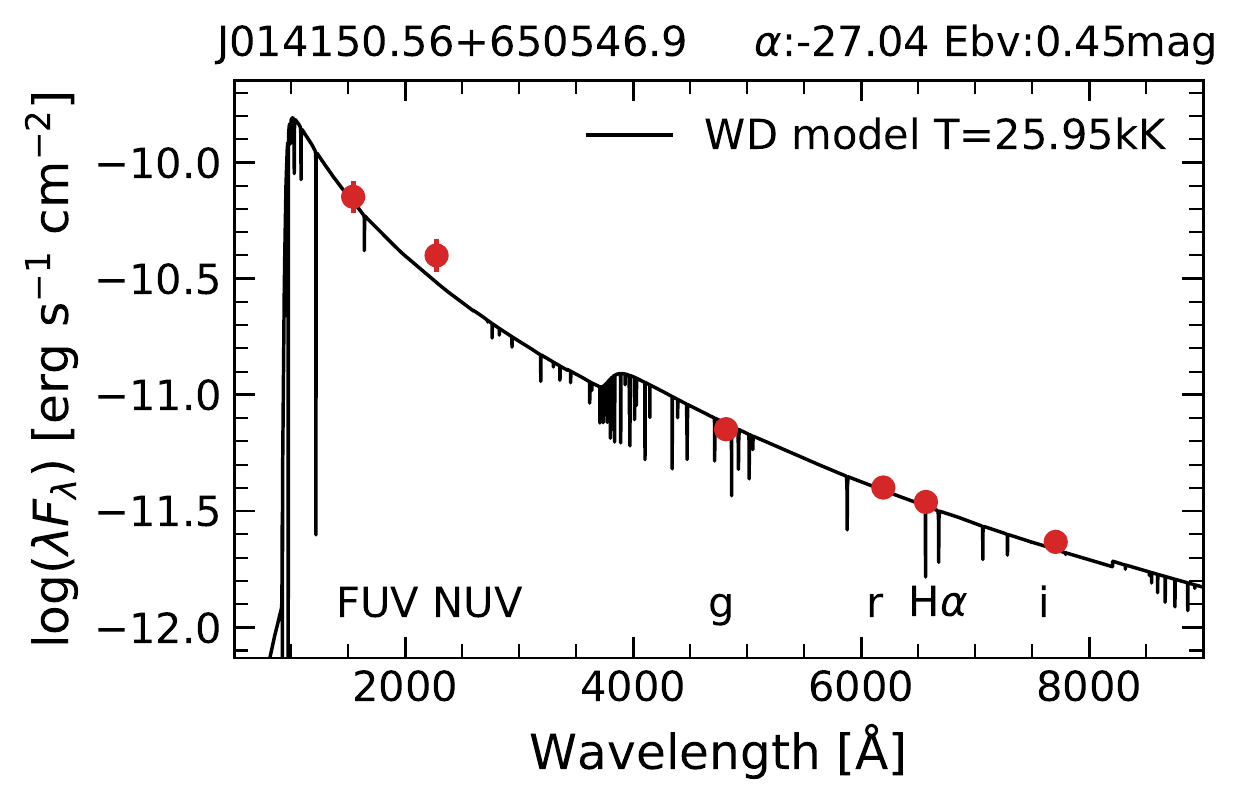}
\includegraphics[width=0.33\textwidth]{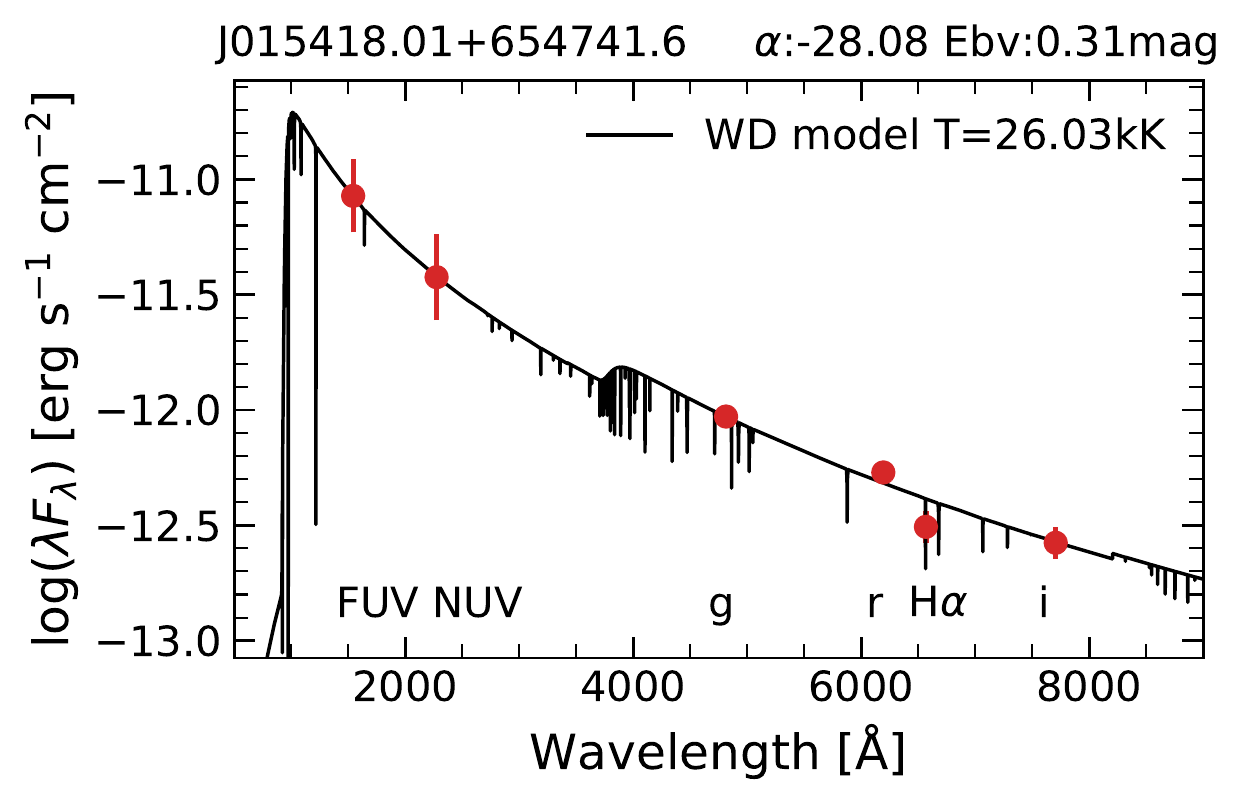}
\includegraphics[width=0.33\textwidth]{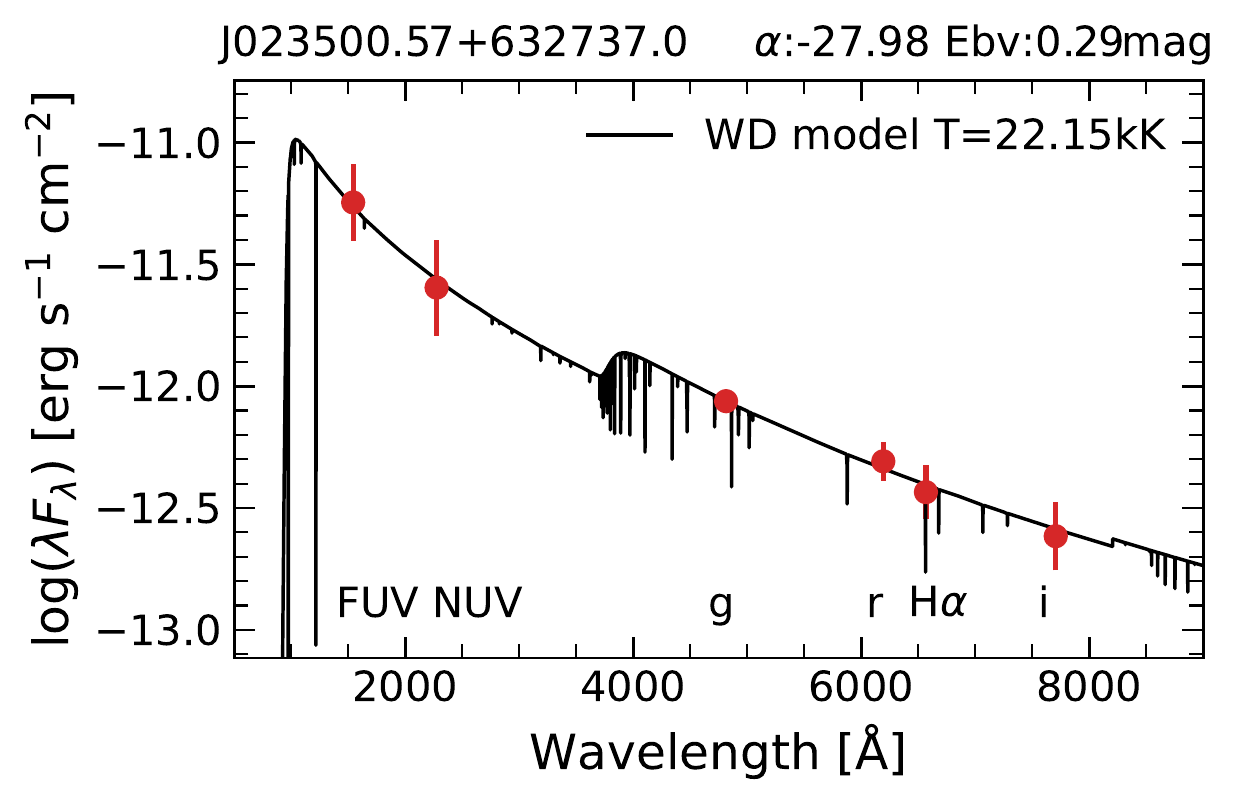}
\includegraphics[width=0.33\textwidth]{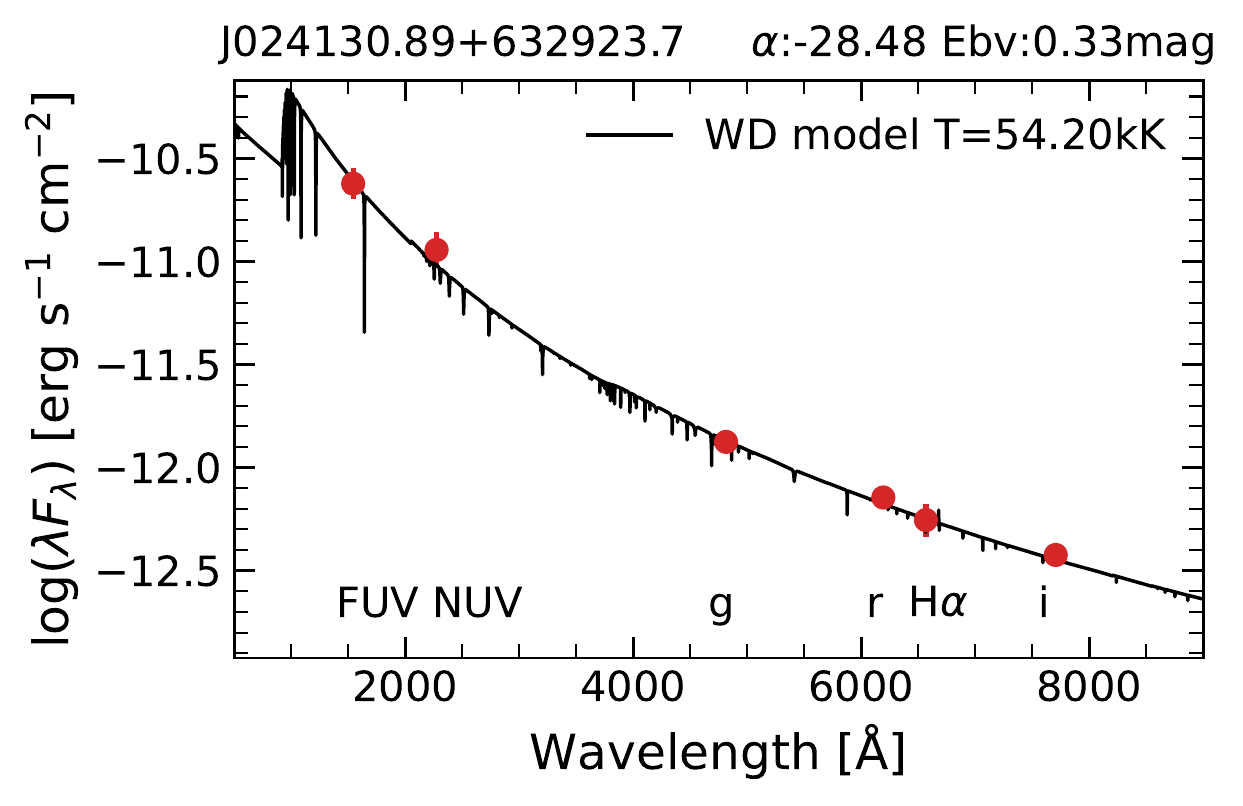}
\includegraphics[width=0.33\textwidth]{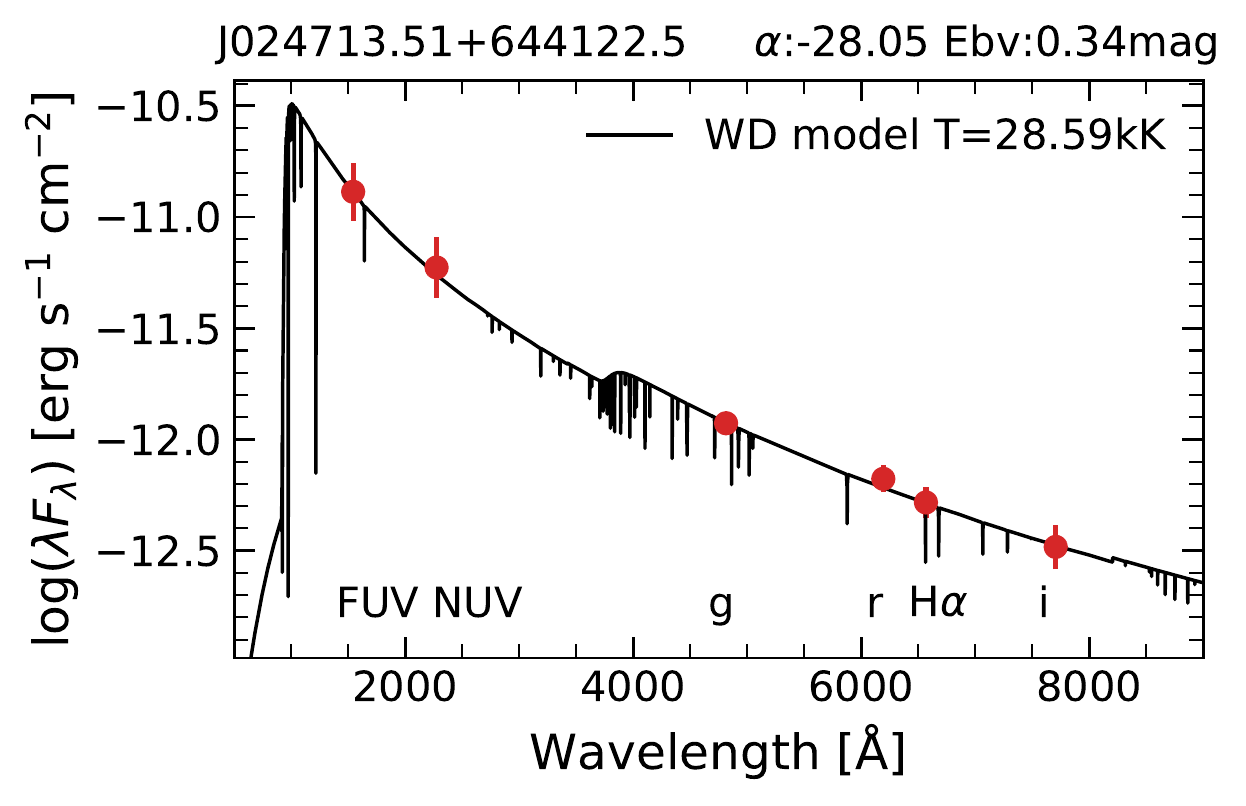}
\includegraphics[width=0.33\textwidth]{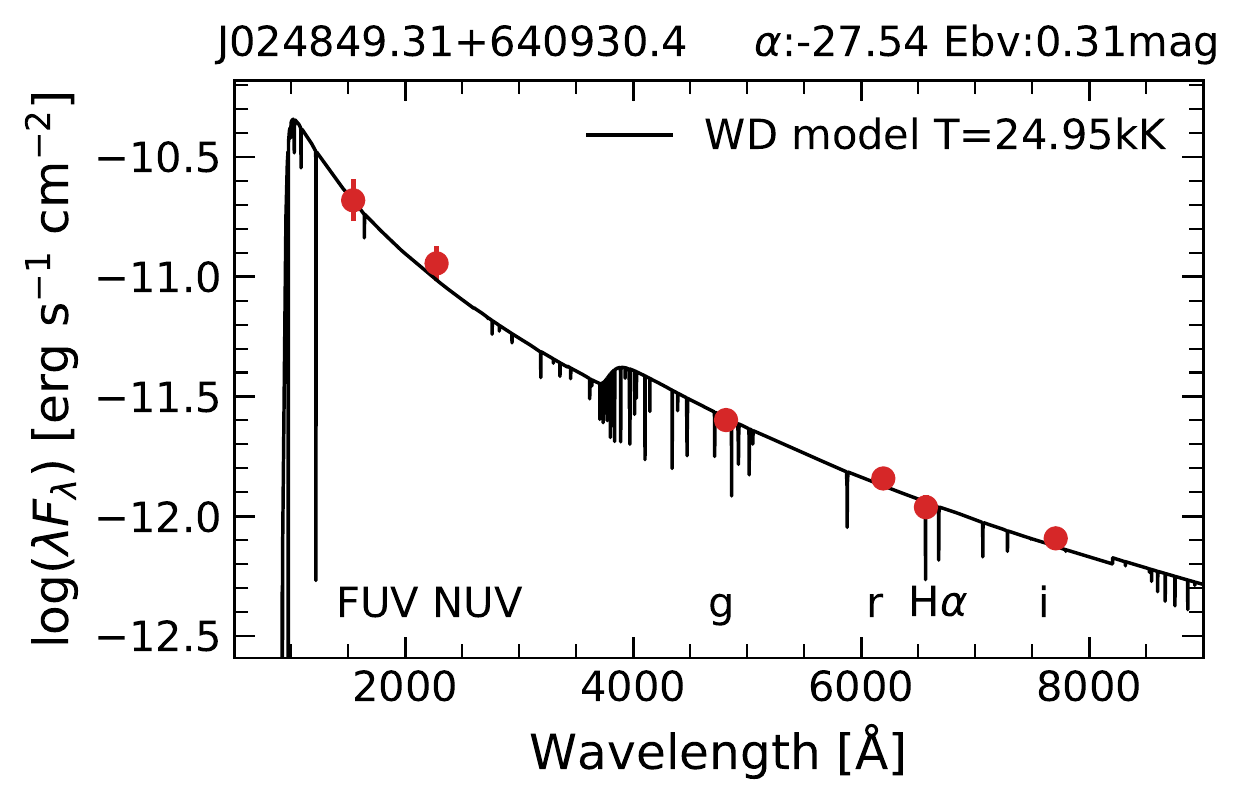}
\includegraphics[width=0.33\textwidth]{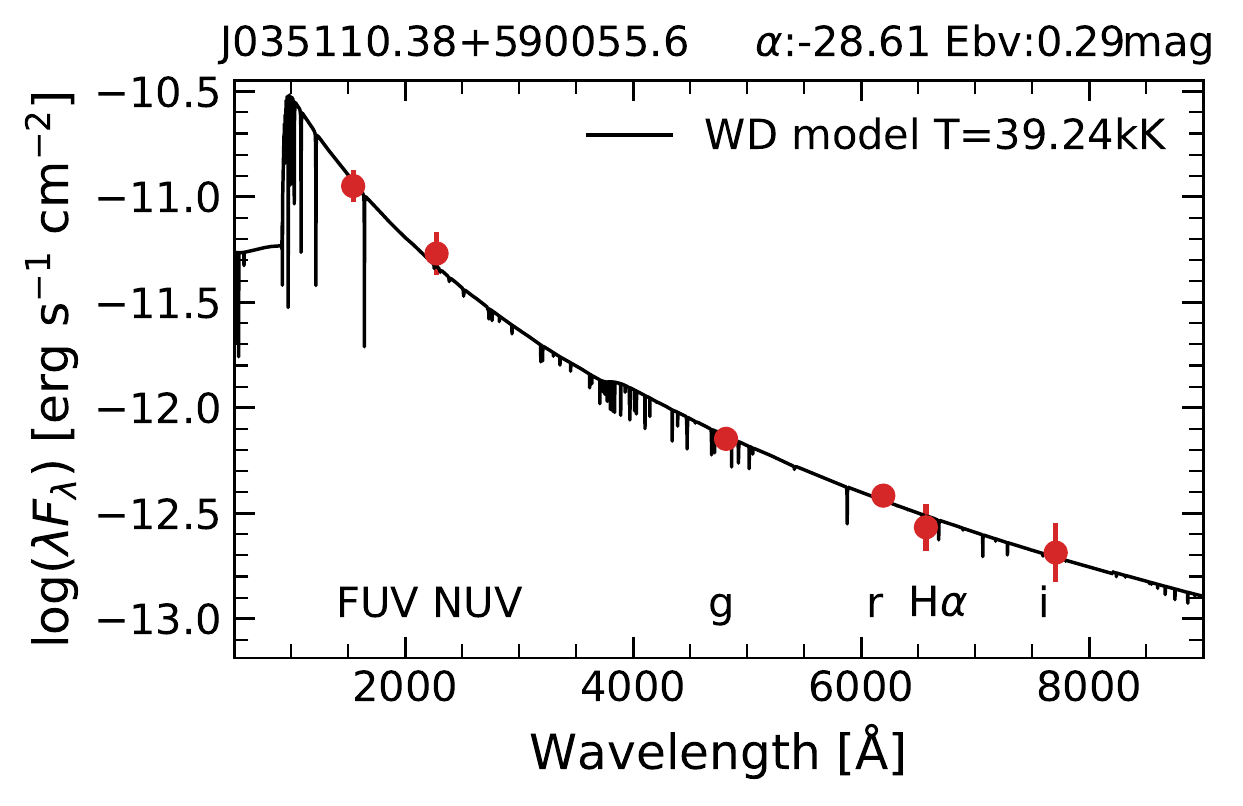}
\includegraphics[width=0.33\textwidth]{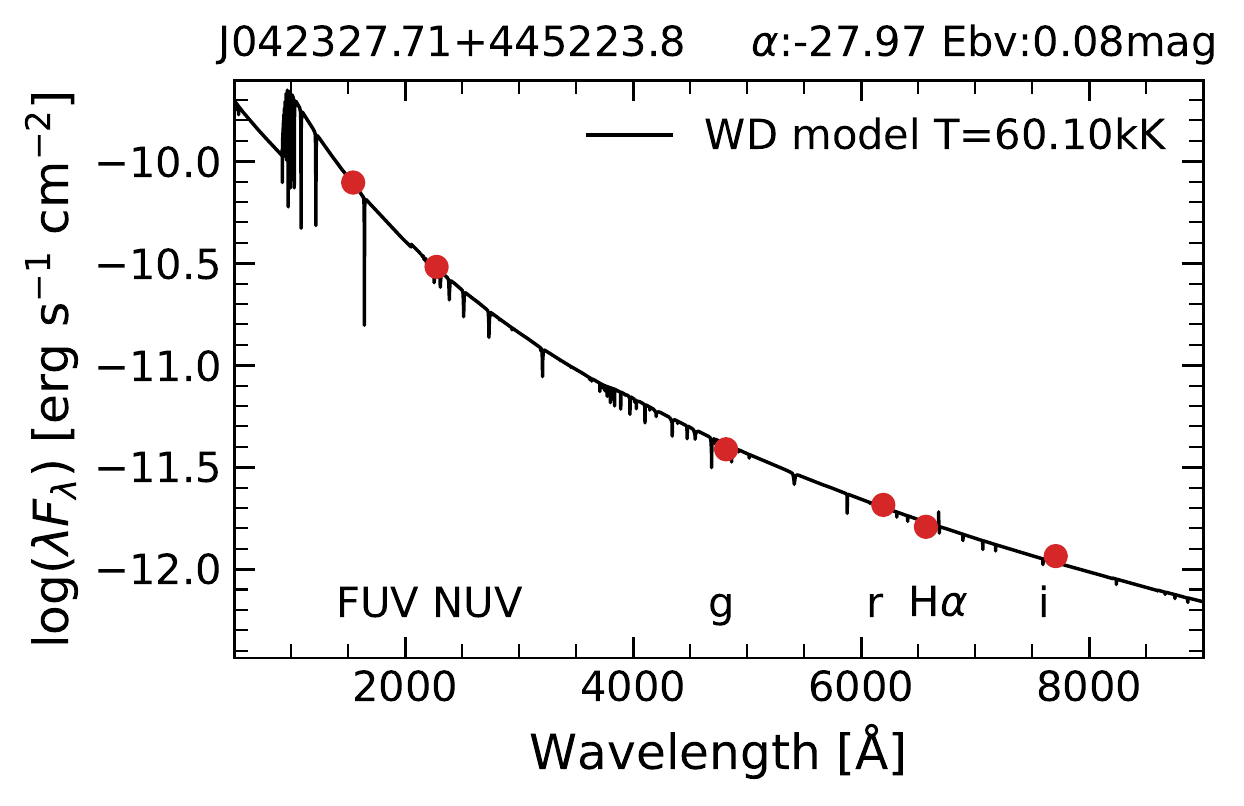}
\includegraphics[width=0.33\textwidth]{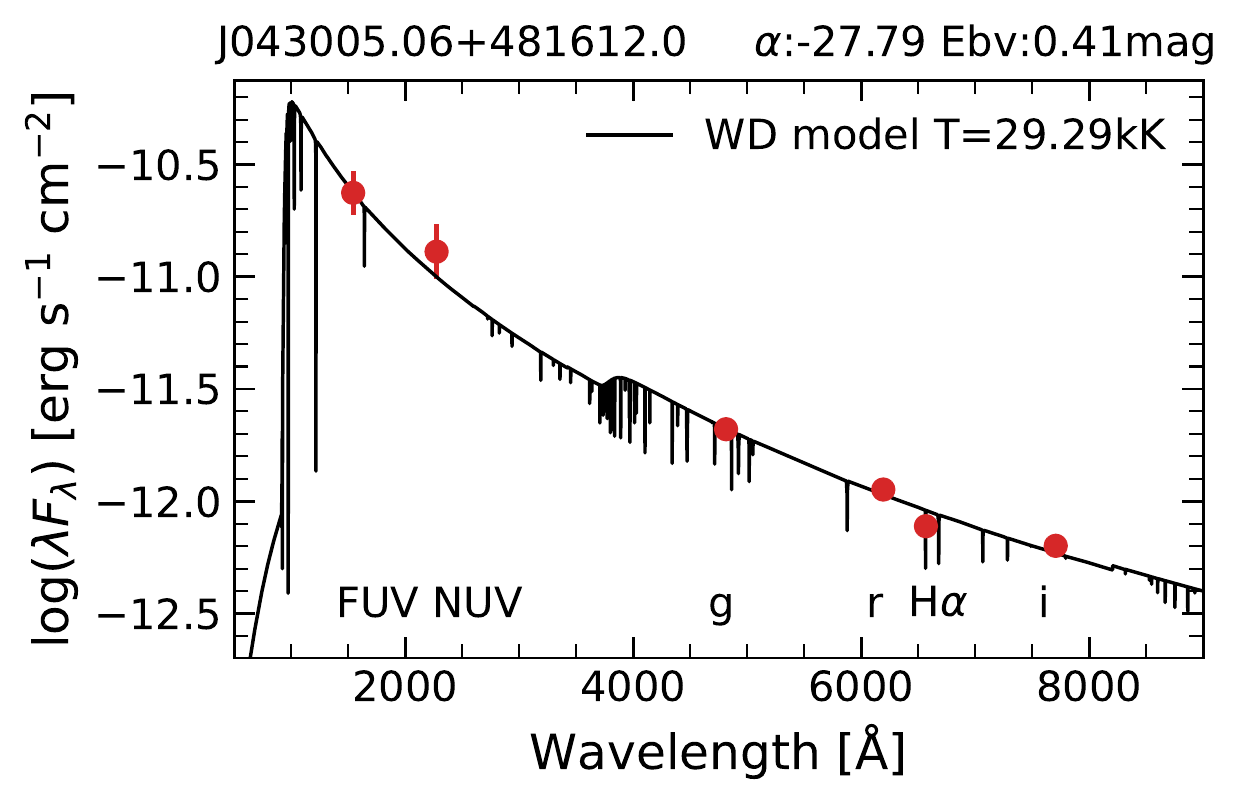}
\includegraphics[width=0.33\textwidth]{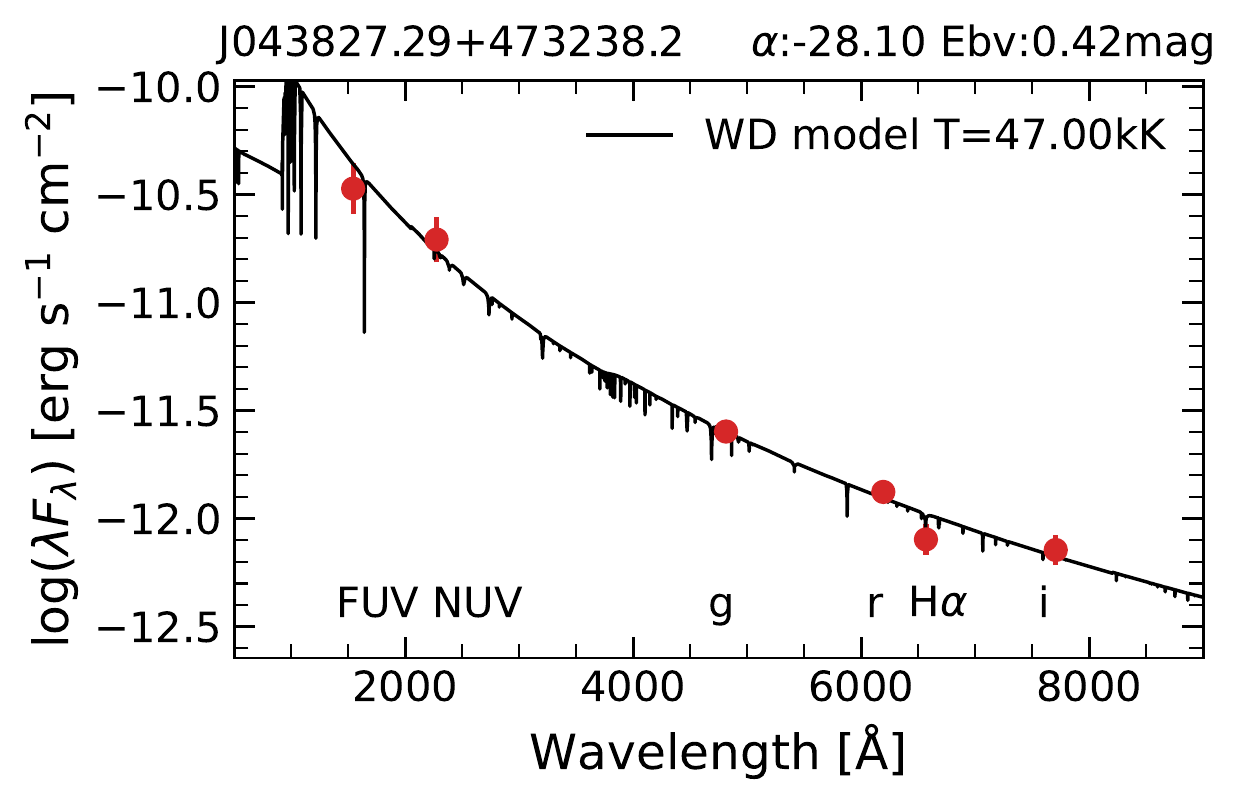}
\includegraphics[width=0.33\textwidth]{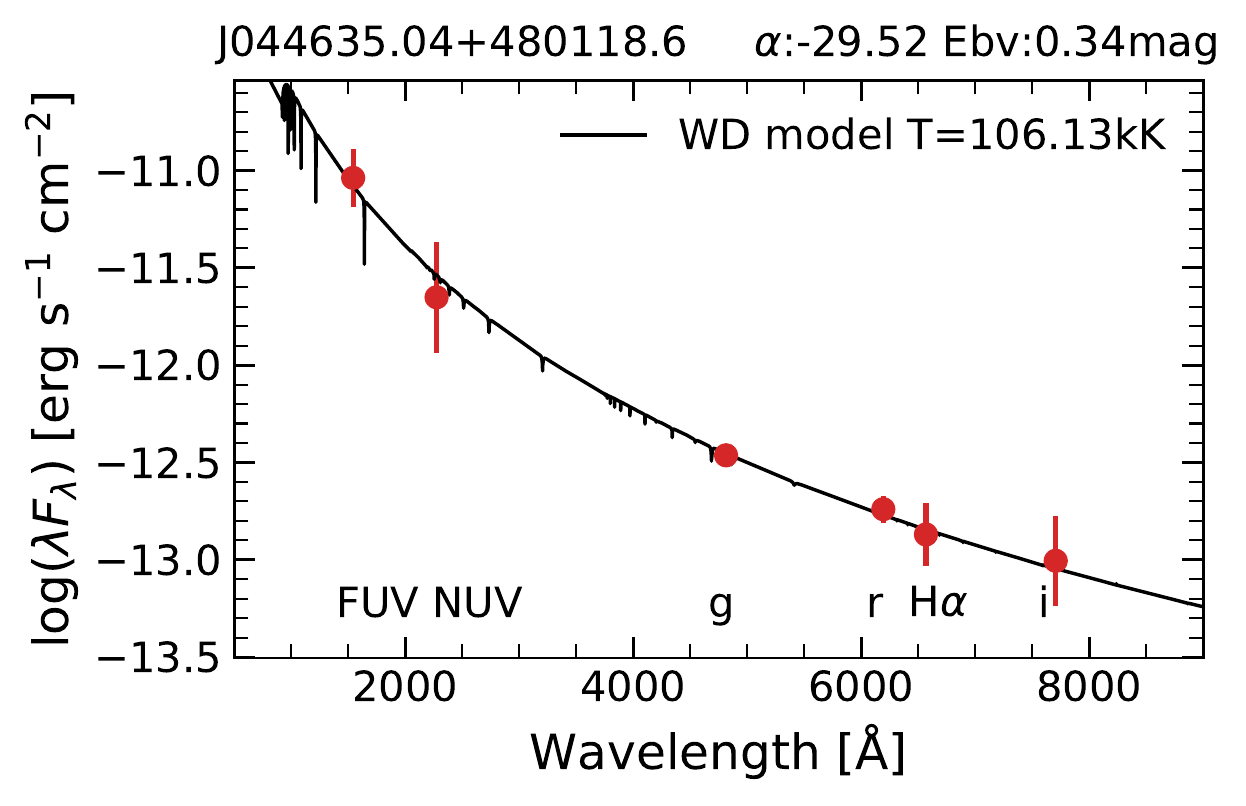}
\includegraphics[width=0.33\textwidth]{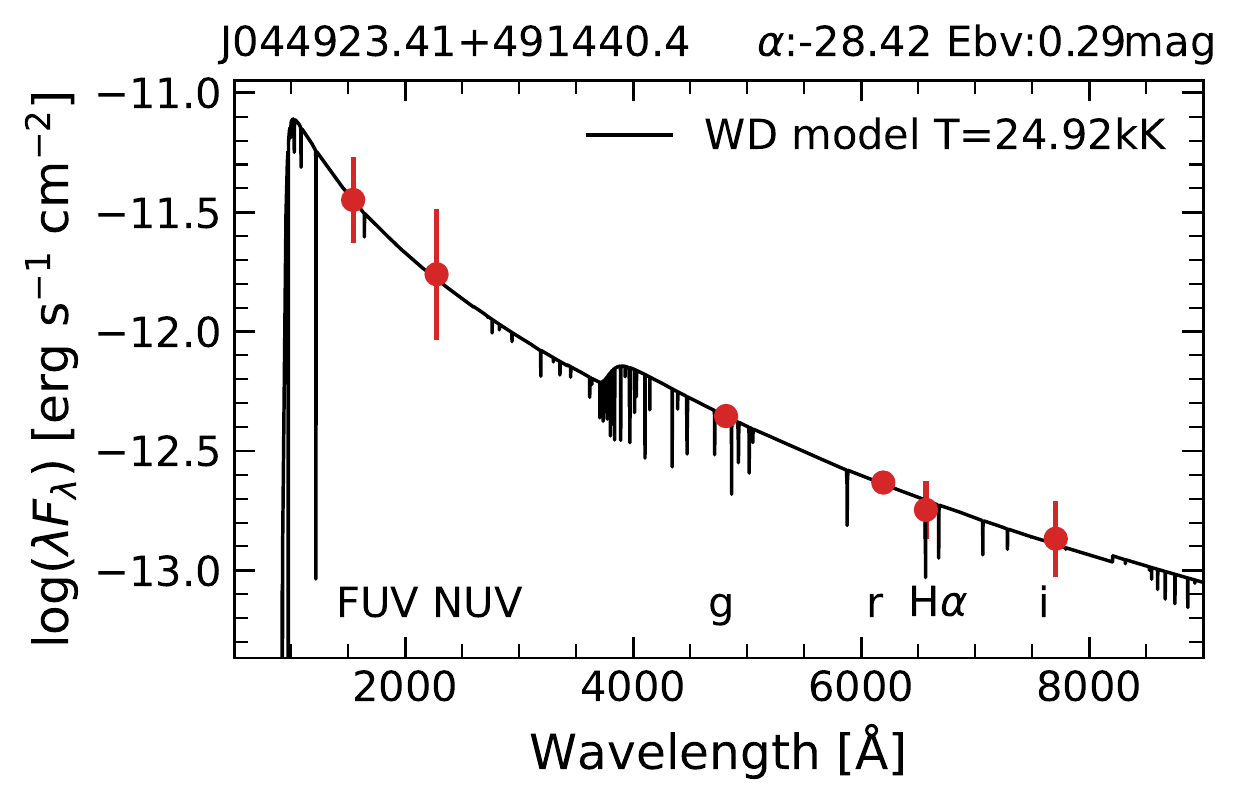}
\includegraphics[width=0.33\textwidth]{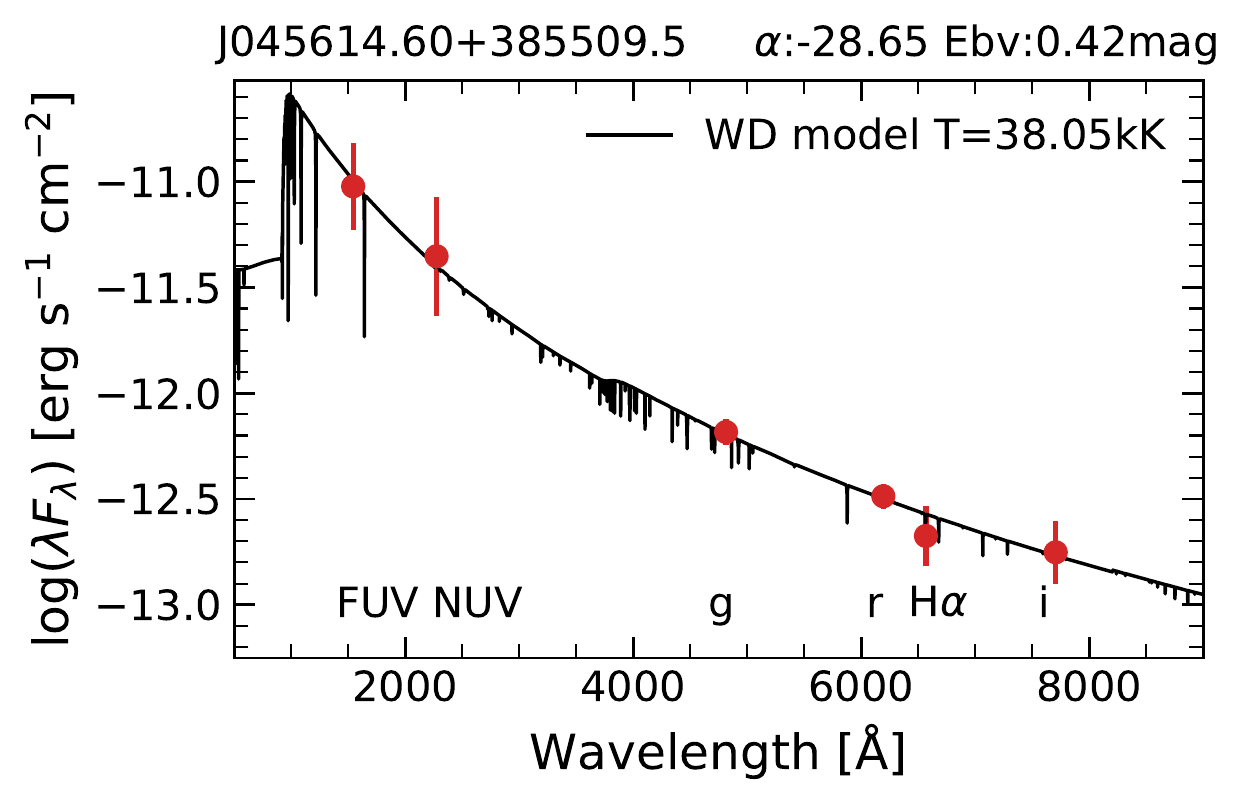}
\includegraphics[width=0.33\textwidth]{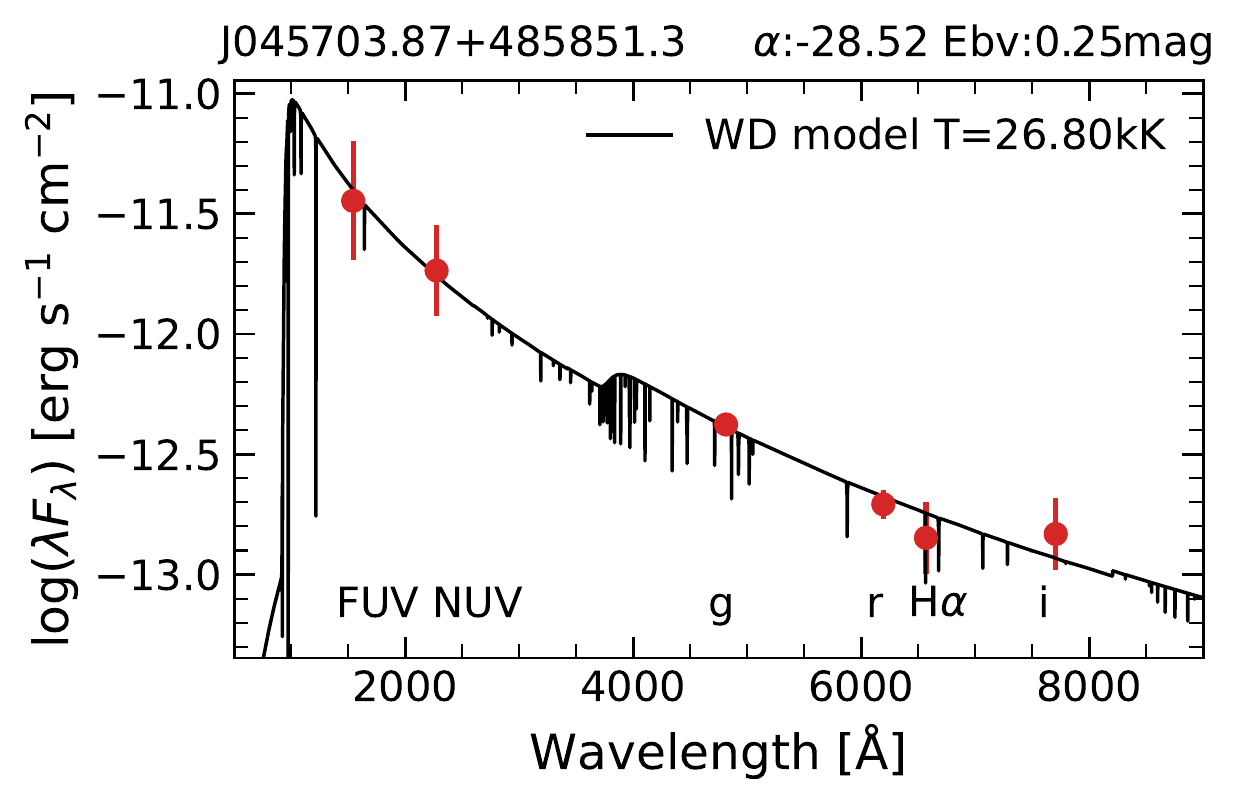}
\includegraphics[width=0.33\textwidth]{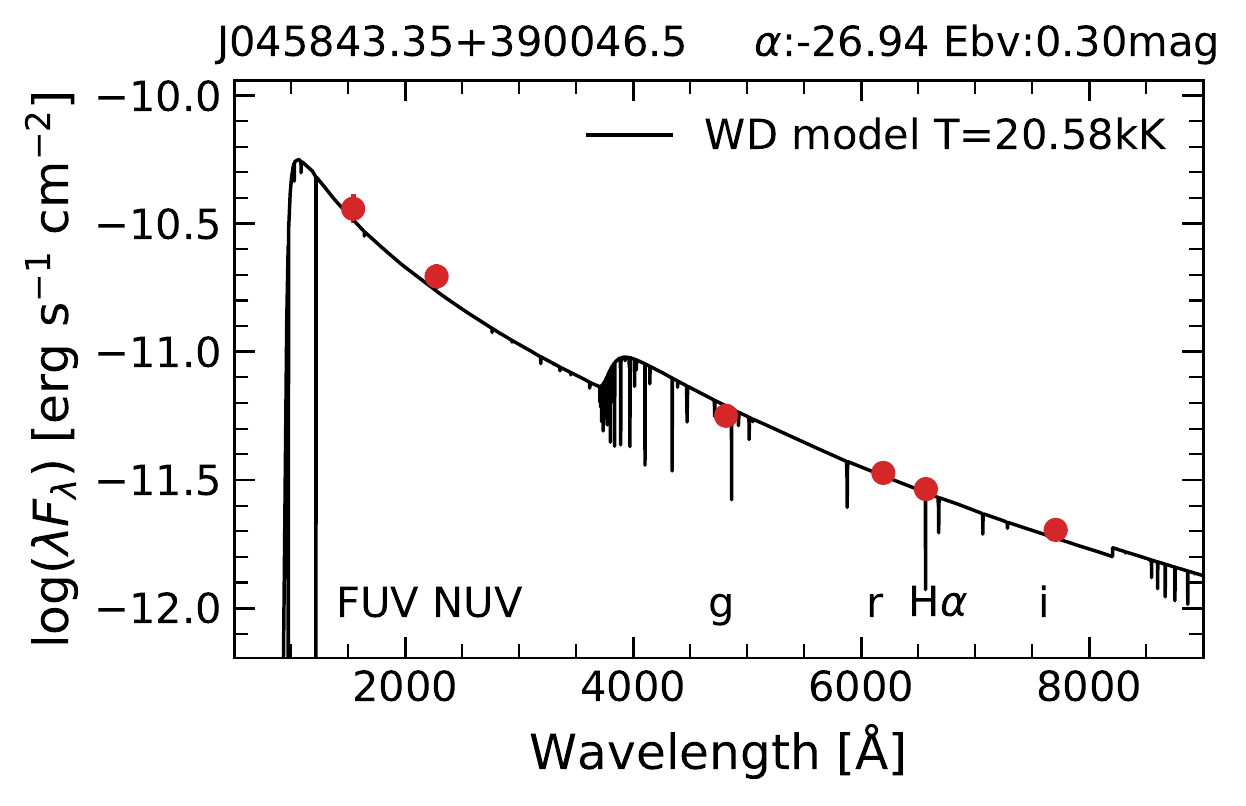}
    \caption{SED fit of the \textit{GALEX} \textit{FUV} and \textit{NUV} + IGAPS g, r, i, and H$\alpha$ photometry (red dots) from which
    a single WD stars is infered. Synthetic spectrum of the best fit is also shown (black solid line).}
    \label{fig:single_wd1}
\end{figure*}    
    
\begin{figure*}
\centering
\includegraphics[width=0.33\textwidth]{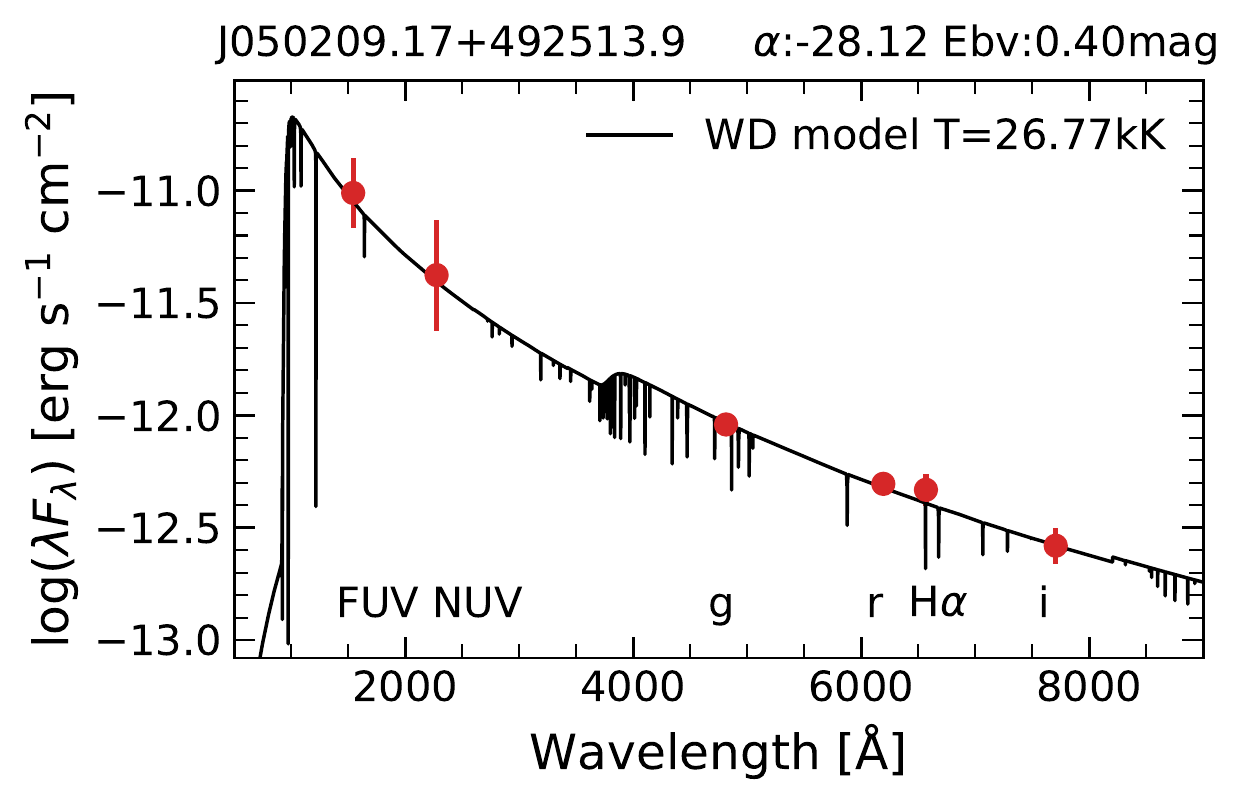}
\includegraphics[width=0.33\textwidth]{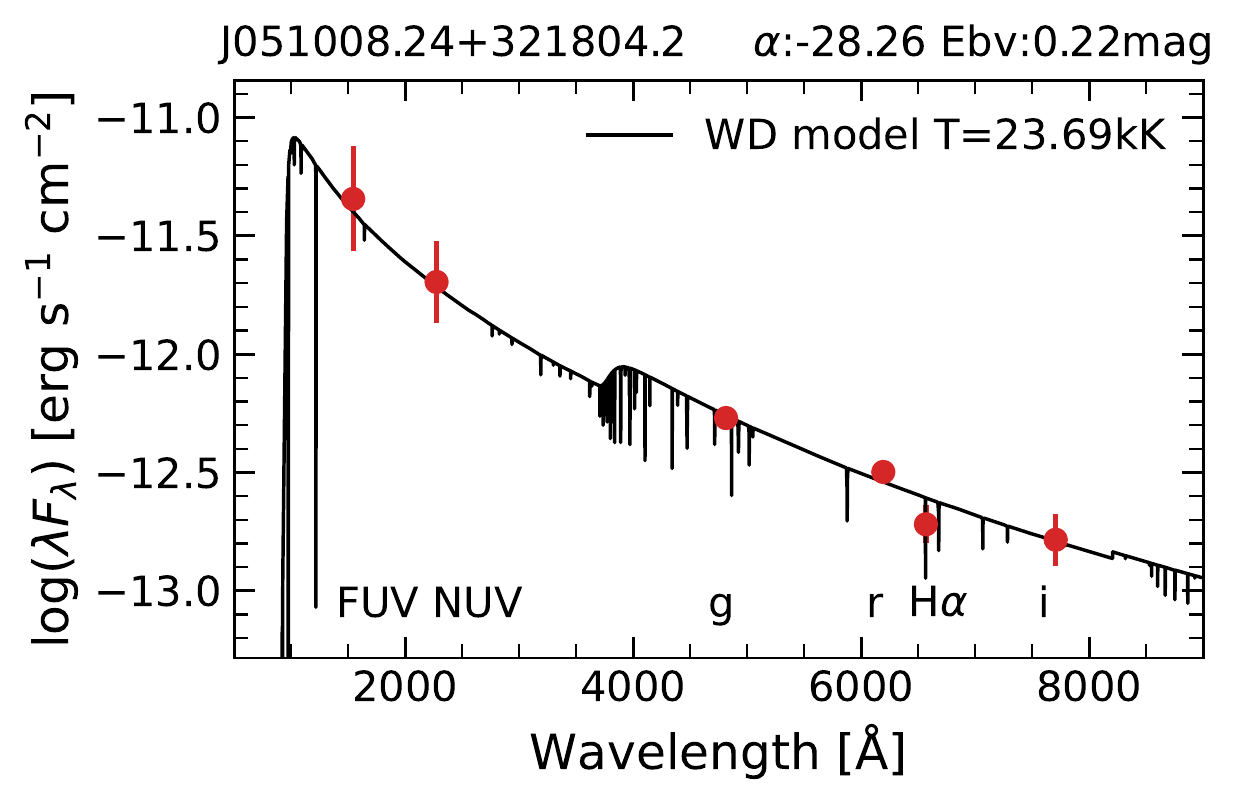}
\includegraphics[width=0.33\textwidth]{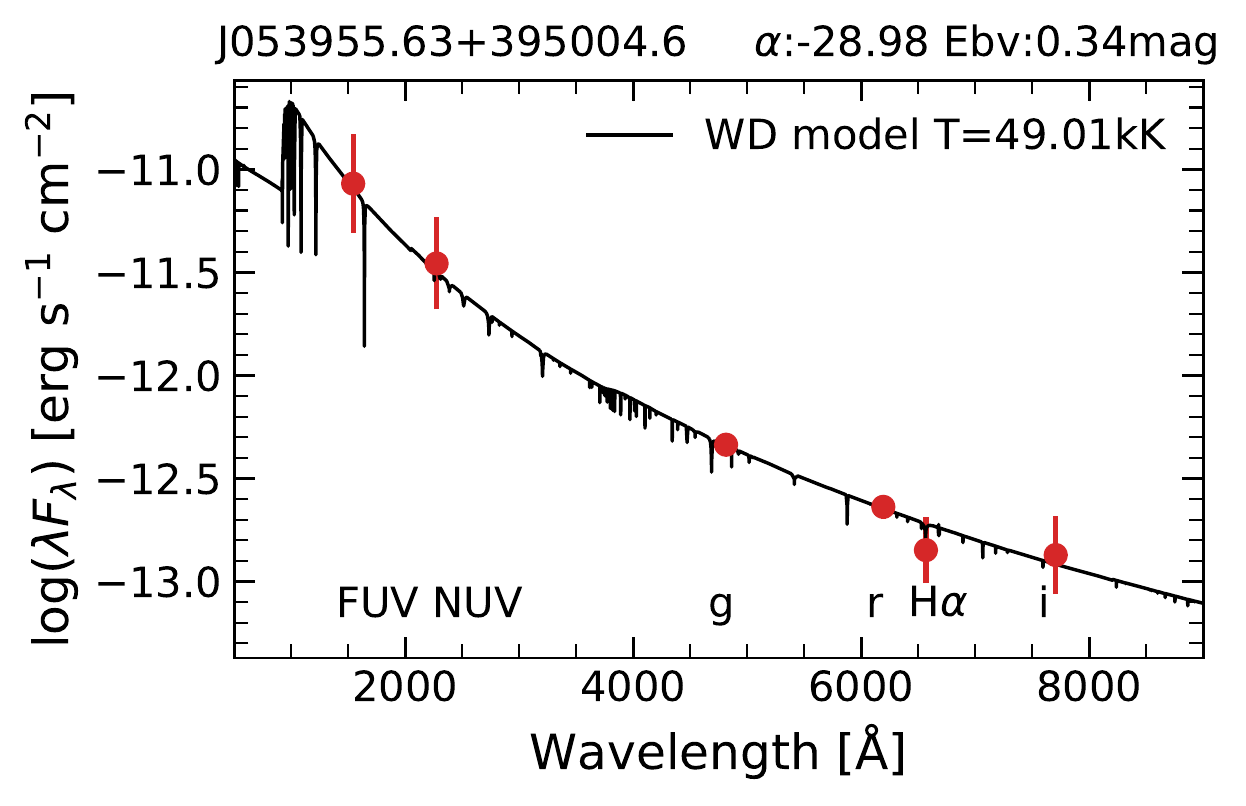}
\includegraphics[width=0.33\textwidth]{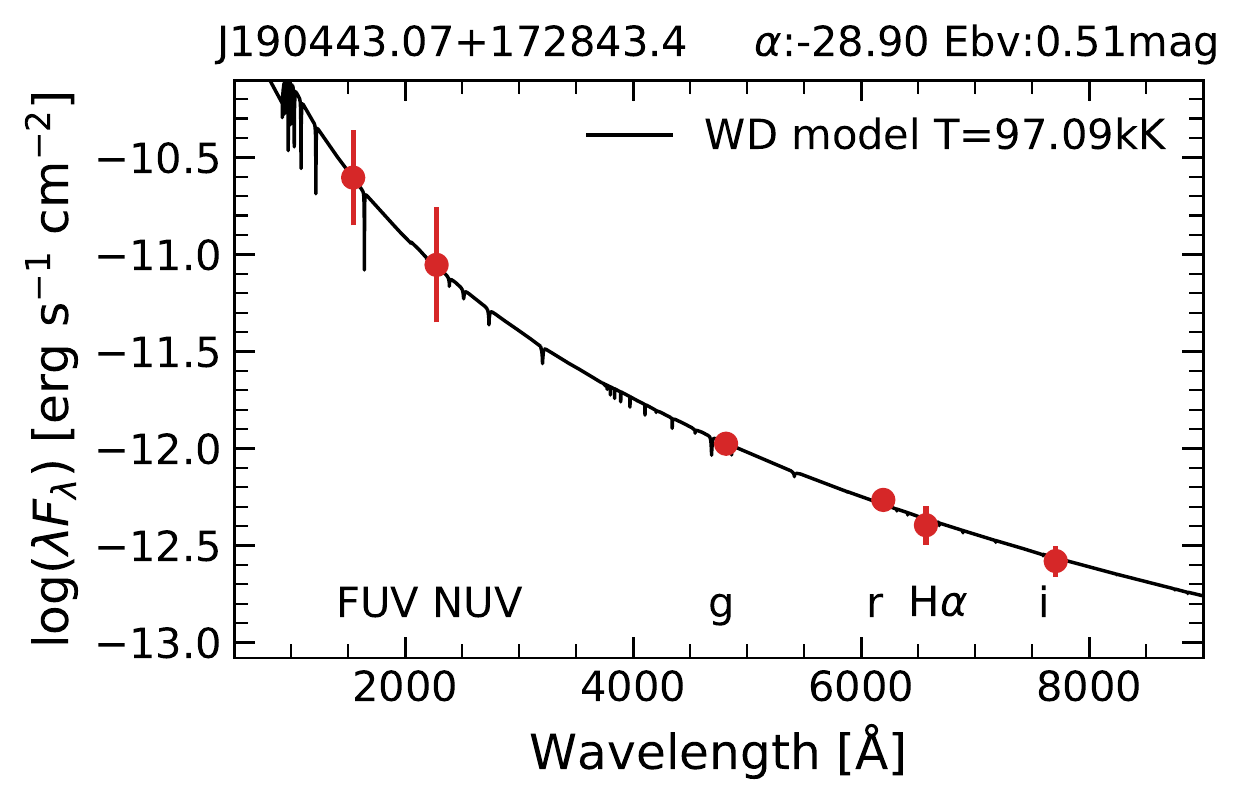}
    \caption{Similar to Figure~\ref{fig:single_wd1}.}
    \label{fig:single_wd2}
\end{figure*}

\section{Binary WD SED fitting}
\label{ap:binary_wd}

\begin{figure*}
    \centering
\includegraphics[width=0.33\textwidth]{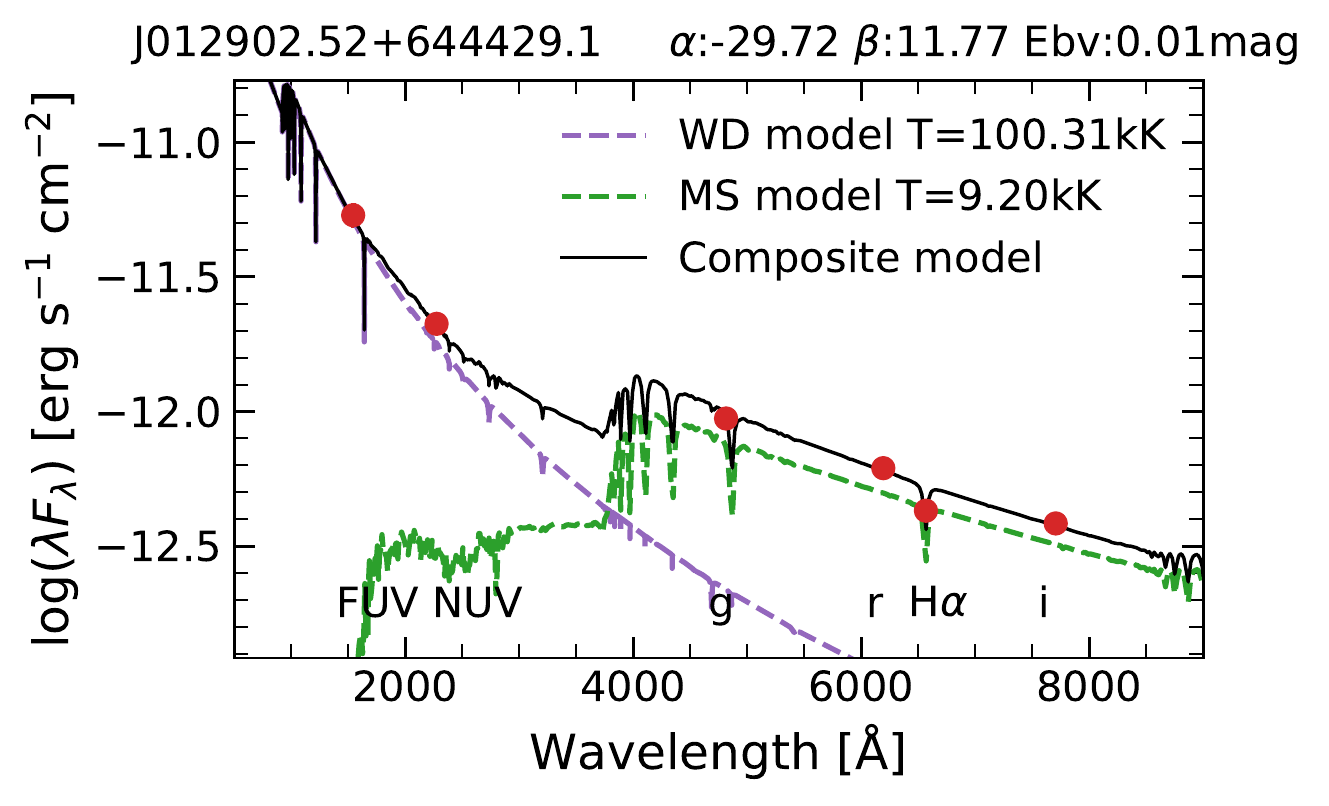}
\includegraphics[width=0.33\textwidth]{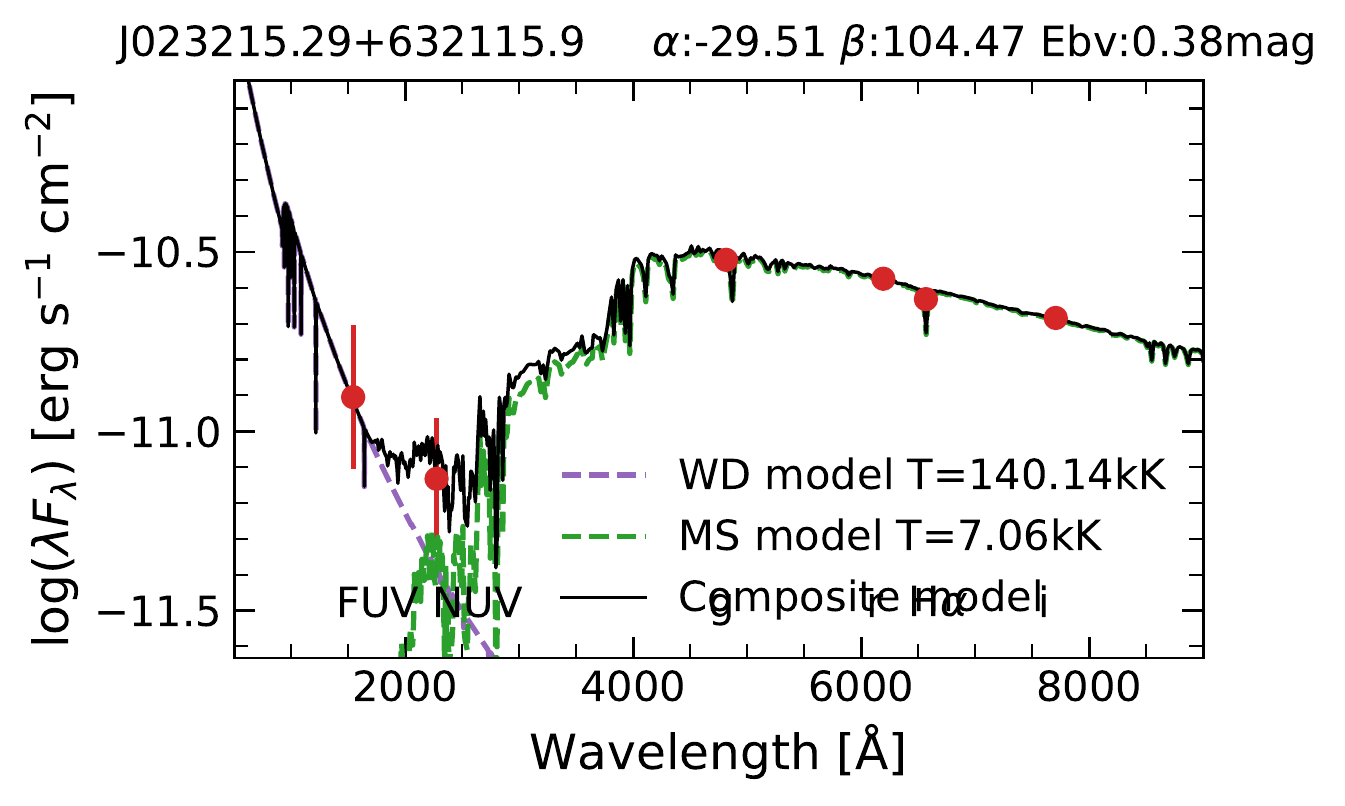}
\includegraphics[width=0.33\textwidth]{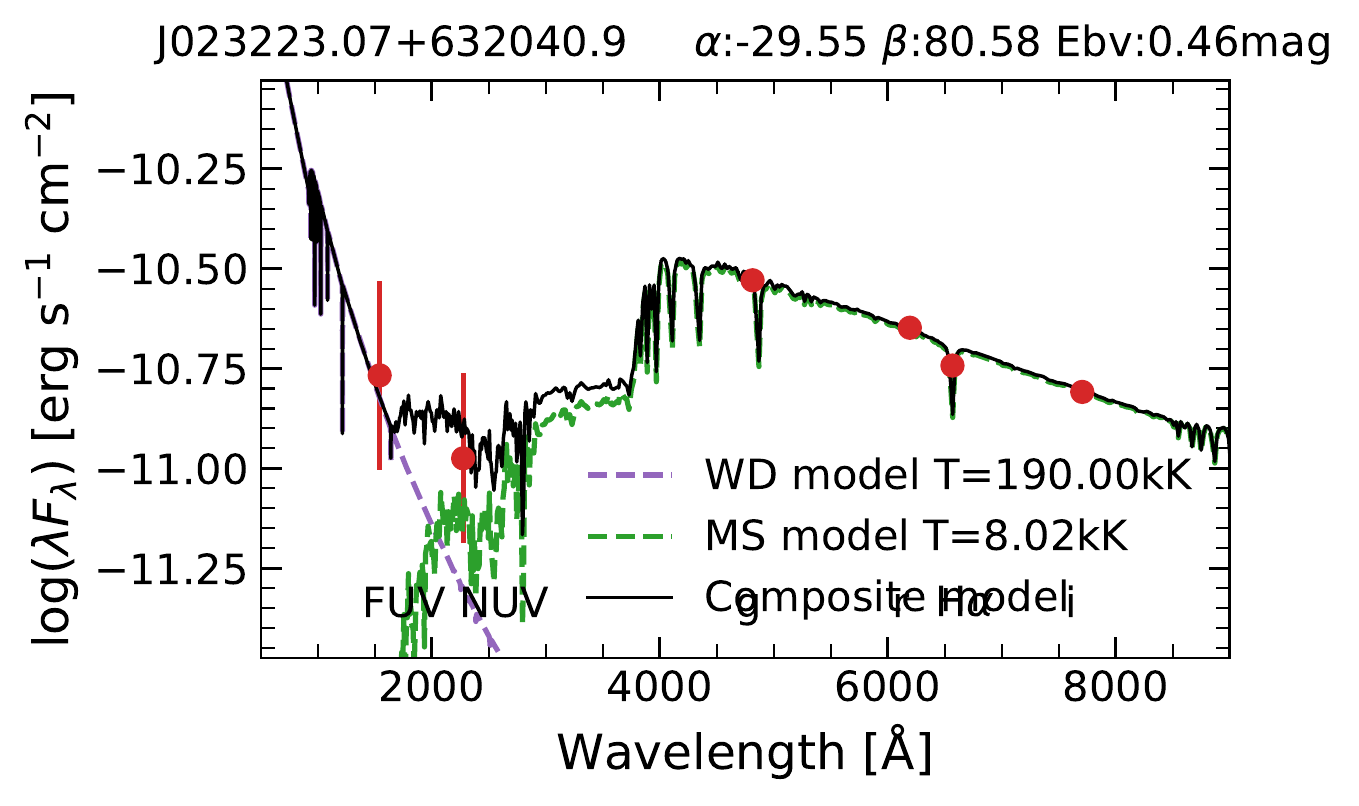}
\includegraphics[width=0.33\textwidth]{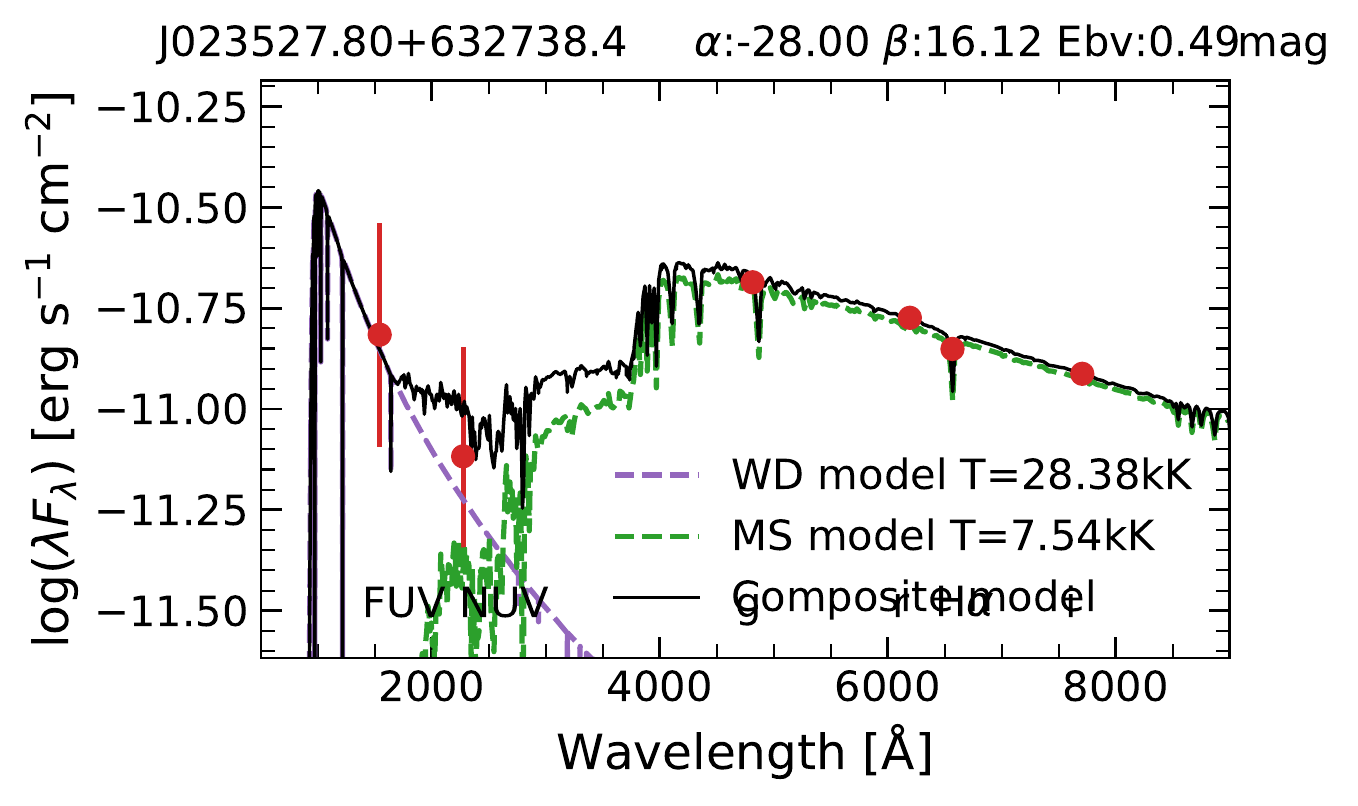}
\includegraphics[width=0.33\textwidth]{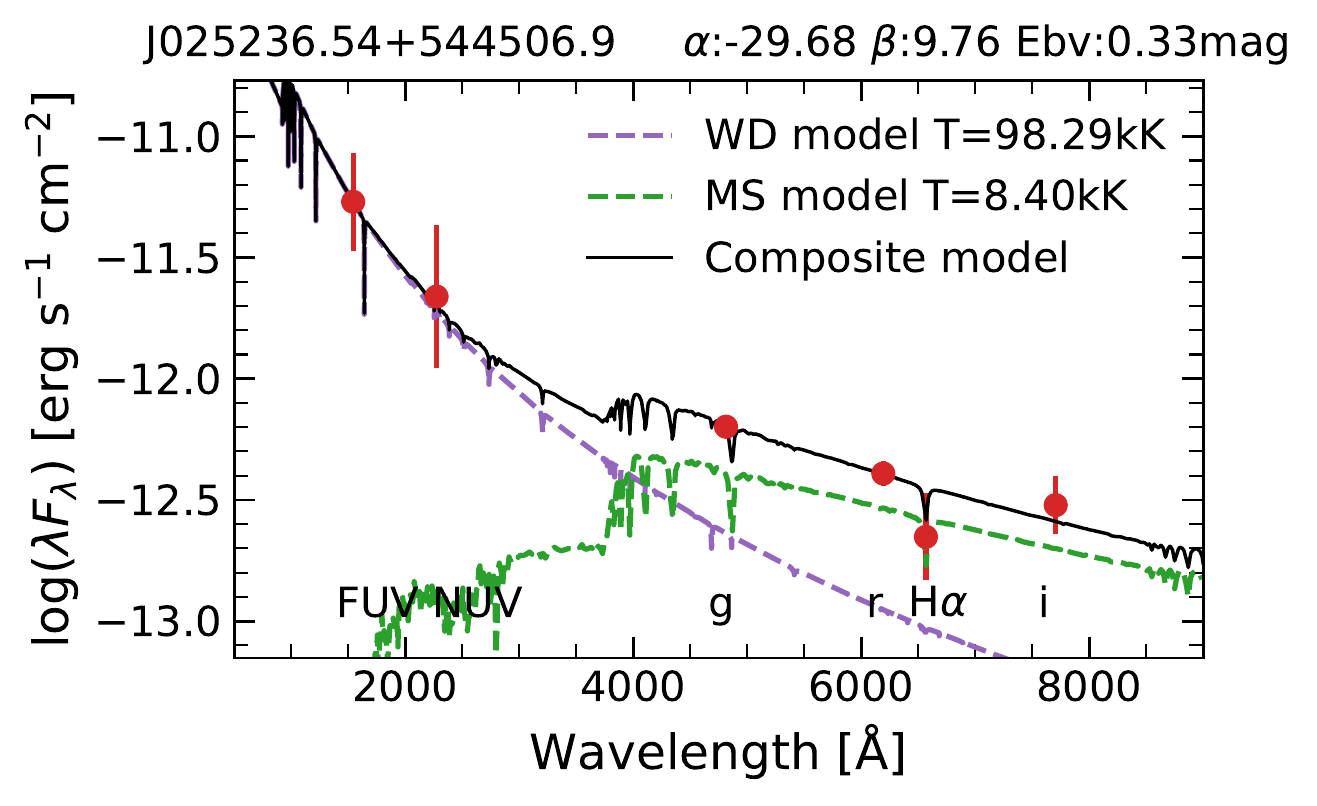}
\includegraphics[width=0.33\textwidth]{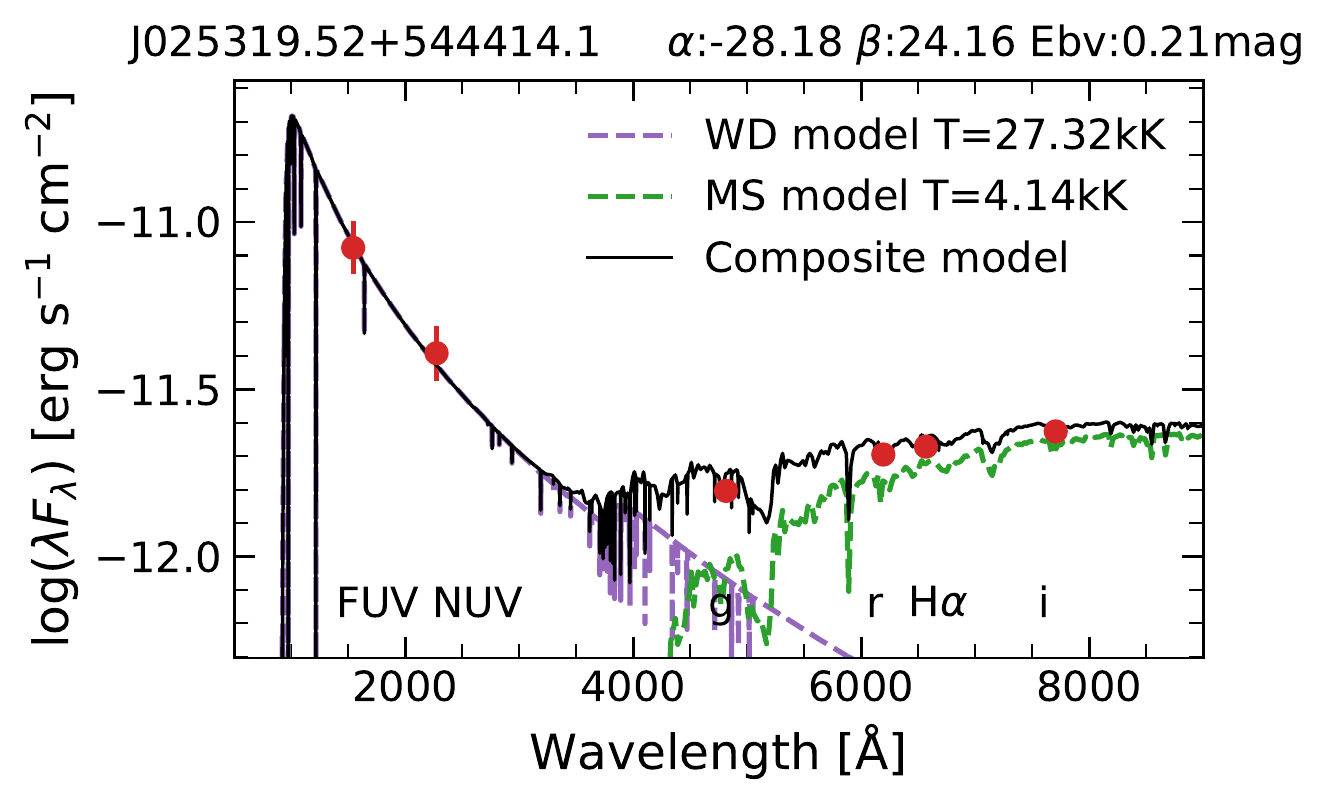}
\includegraphics[width=0.33\textwidth]{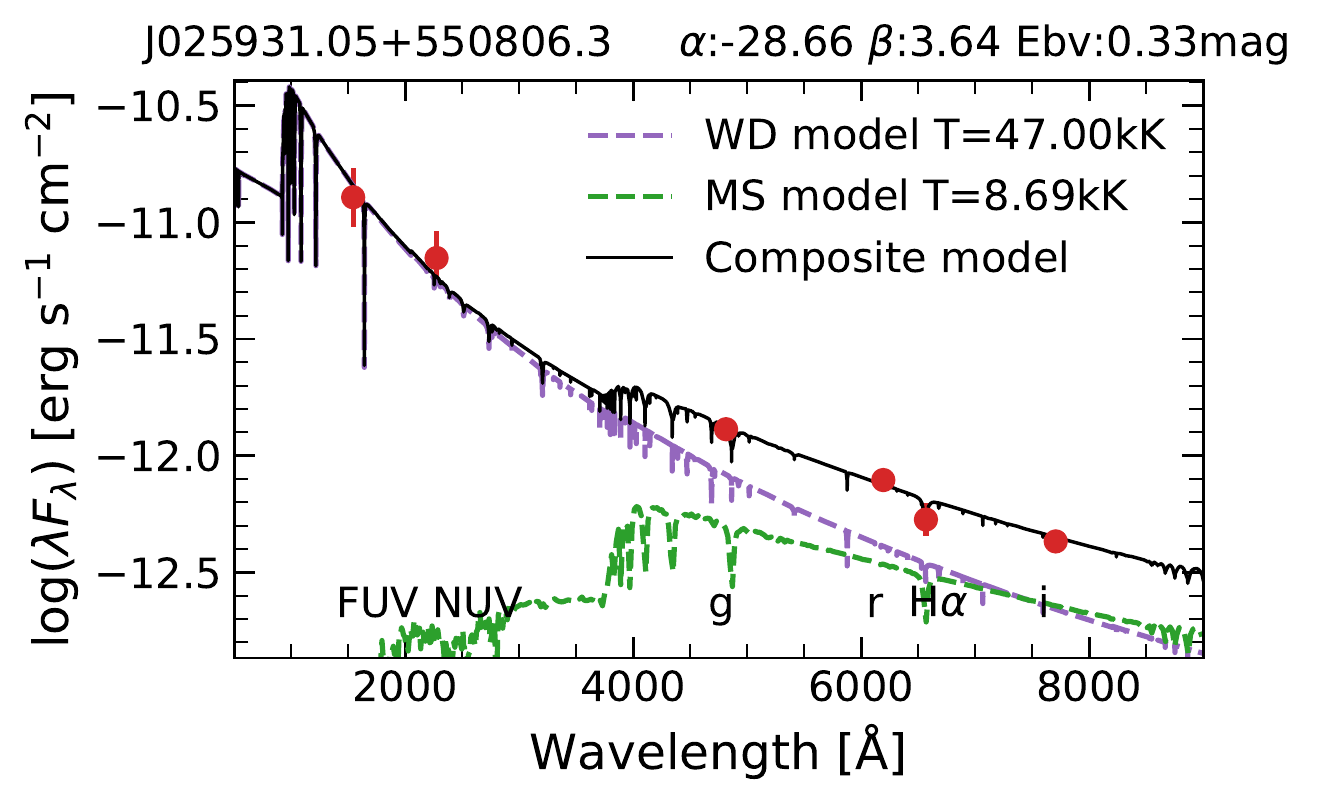}
\includegraphics[width=0.33\textwidth]{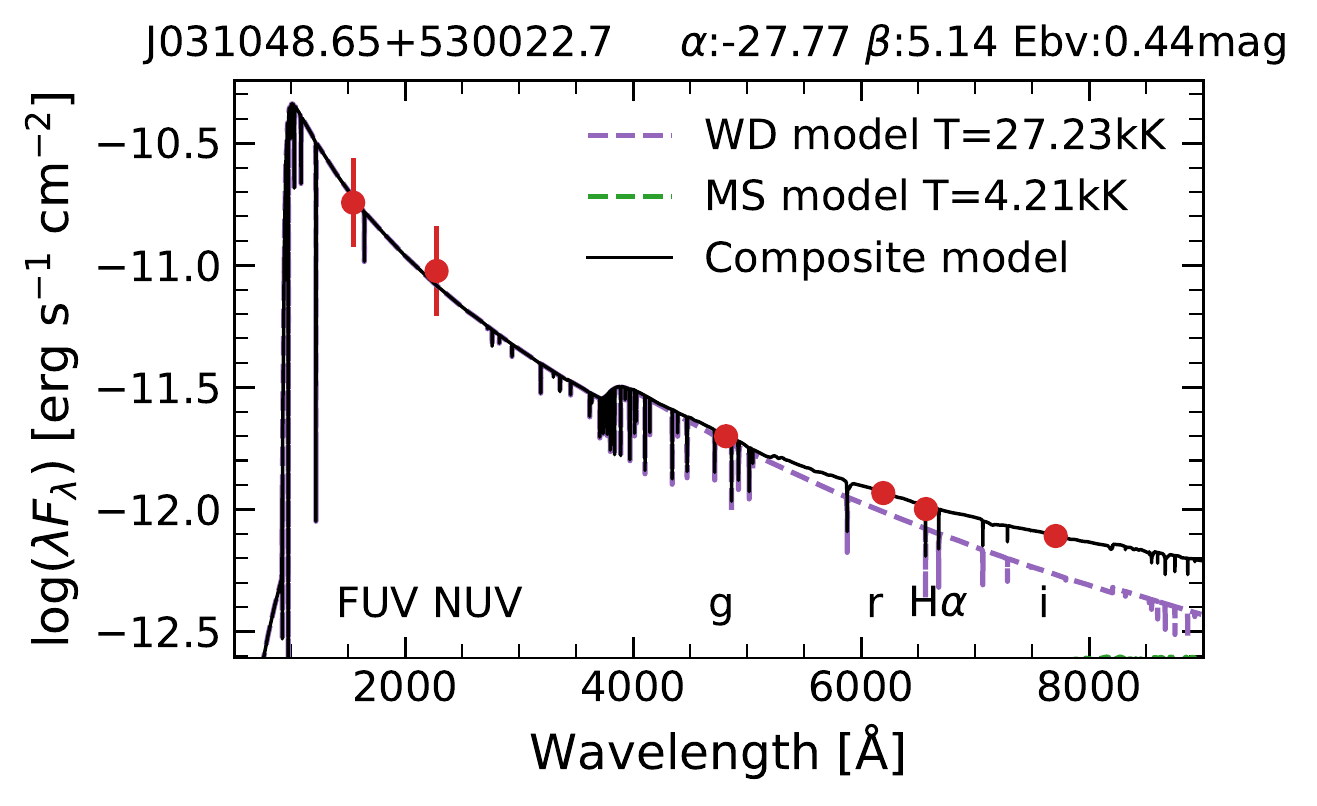}
\includegraphics[width=0.33\textwidth]{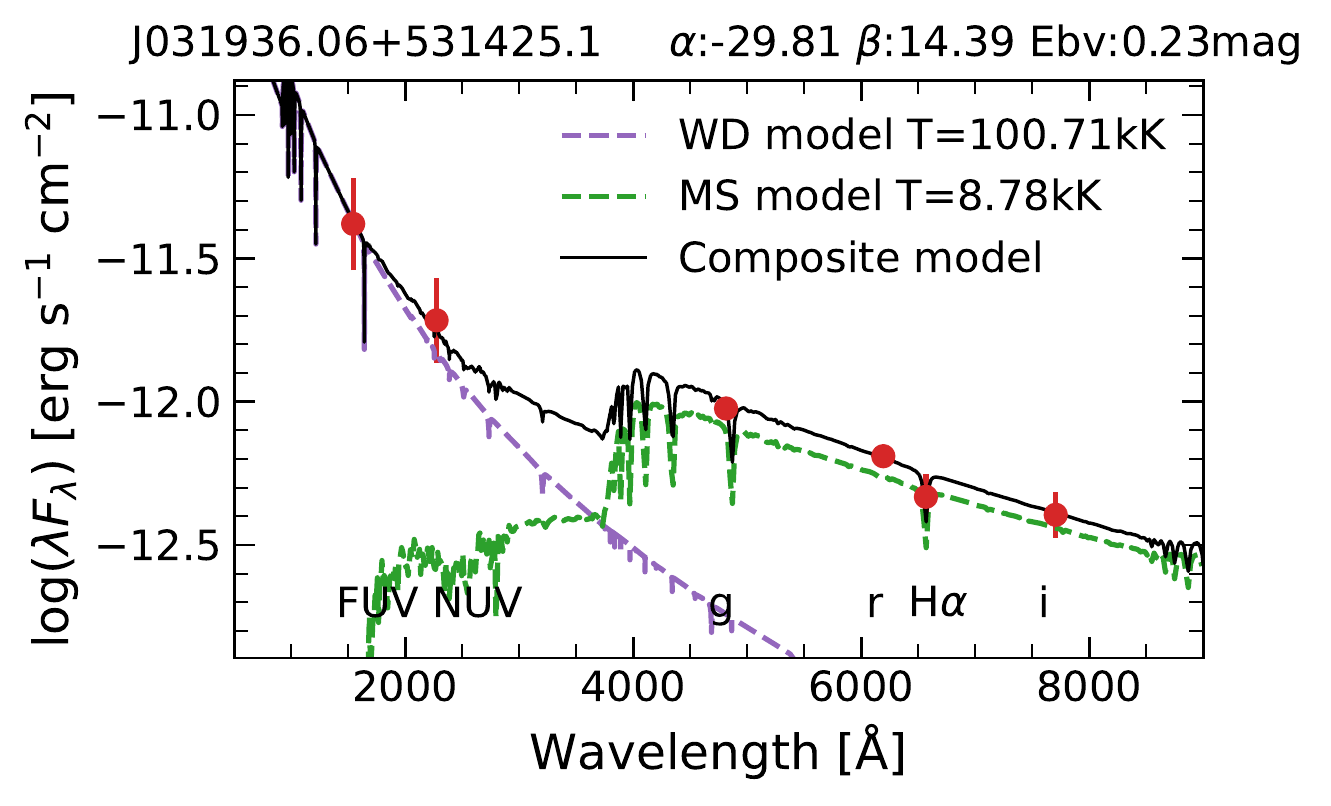}
\includegraphics[width=0.33\textwidth]{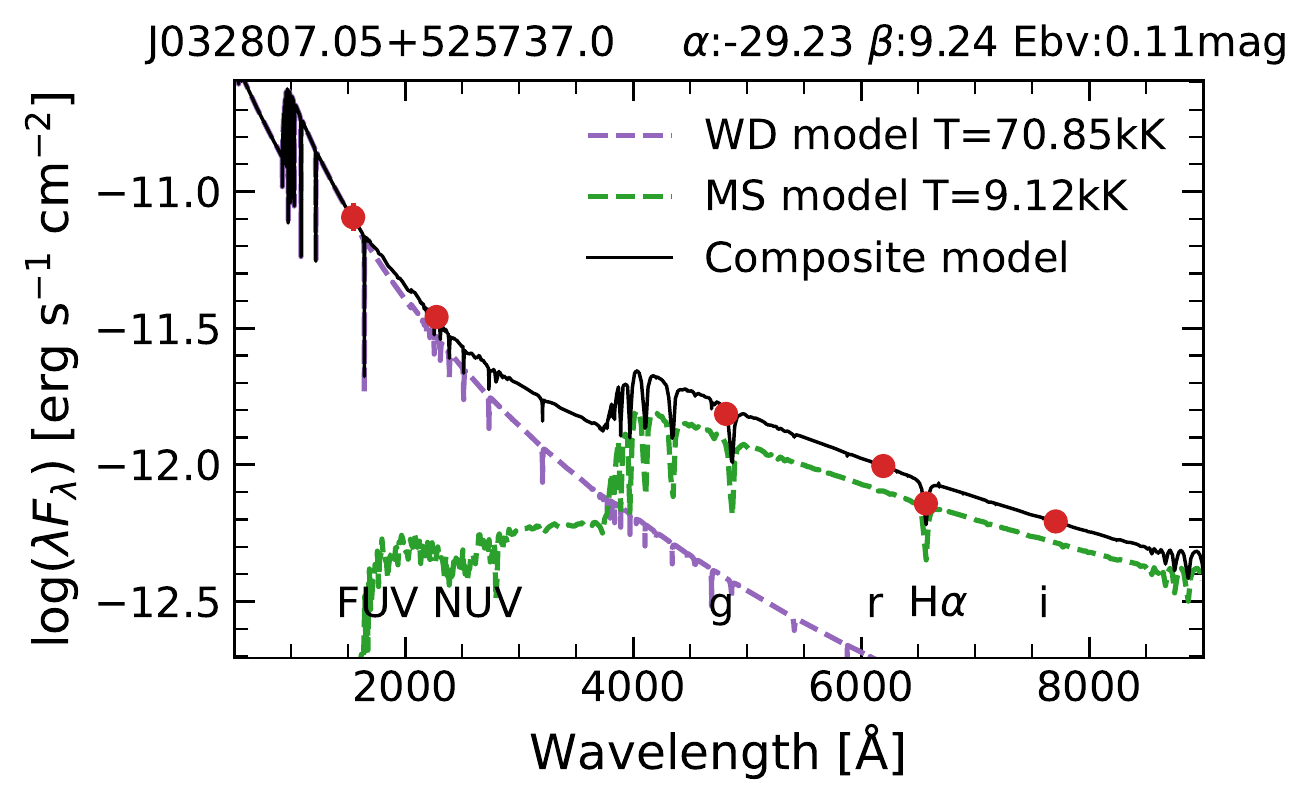}
\includegraphics[width=0.33\textwidth]{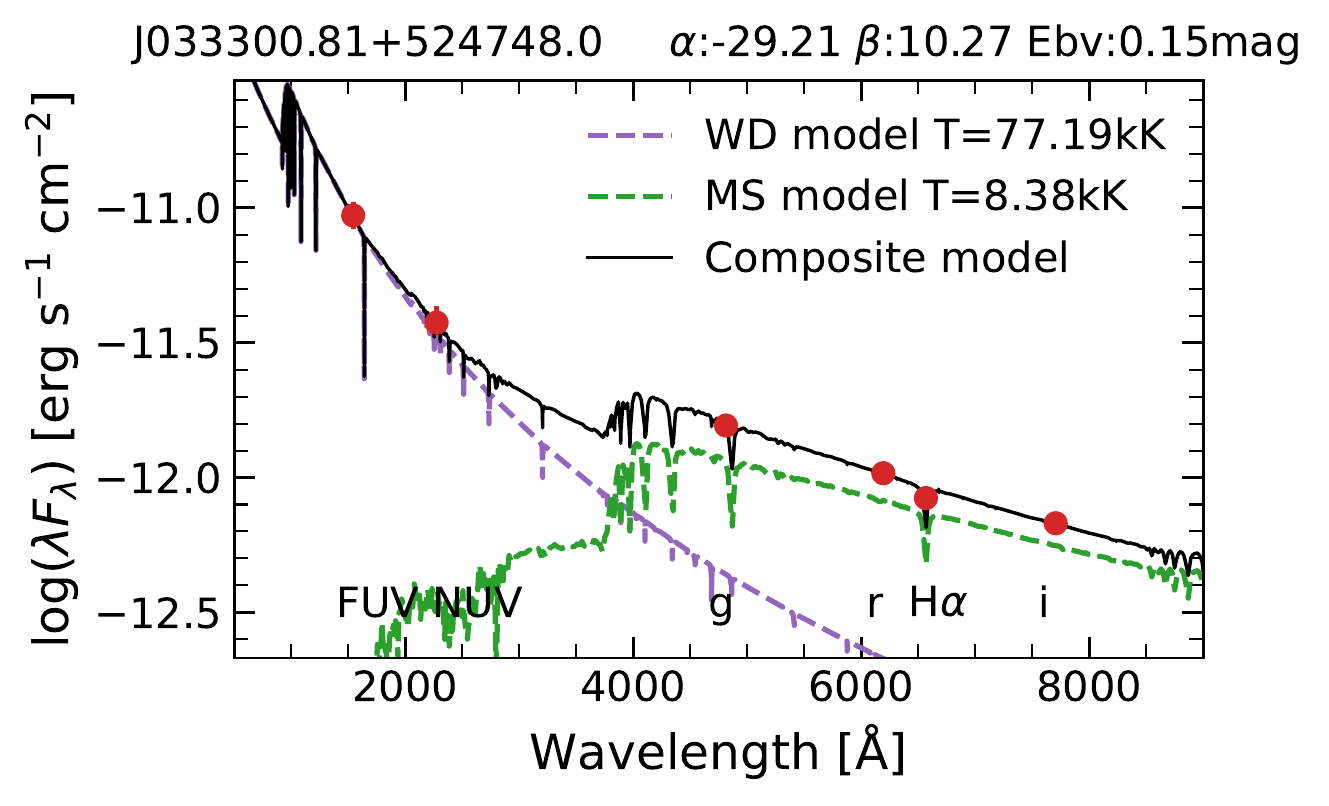}
\includegraphics[width=0.33\textwidth]{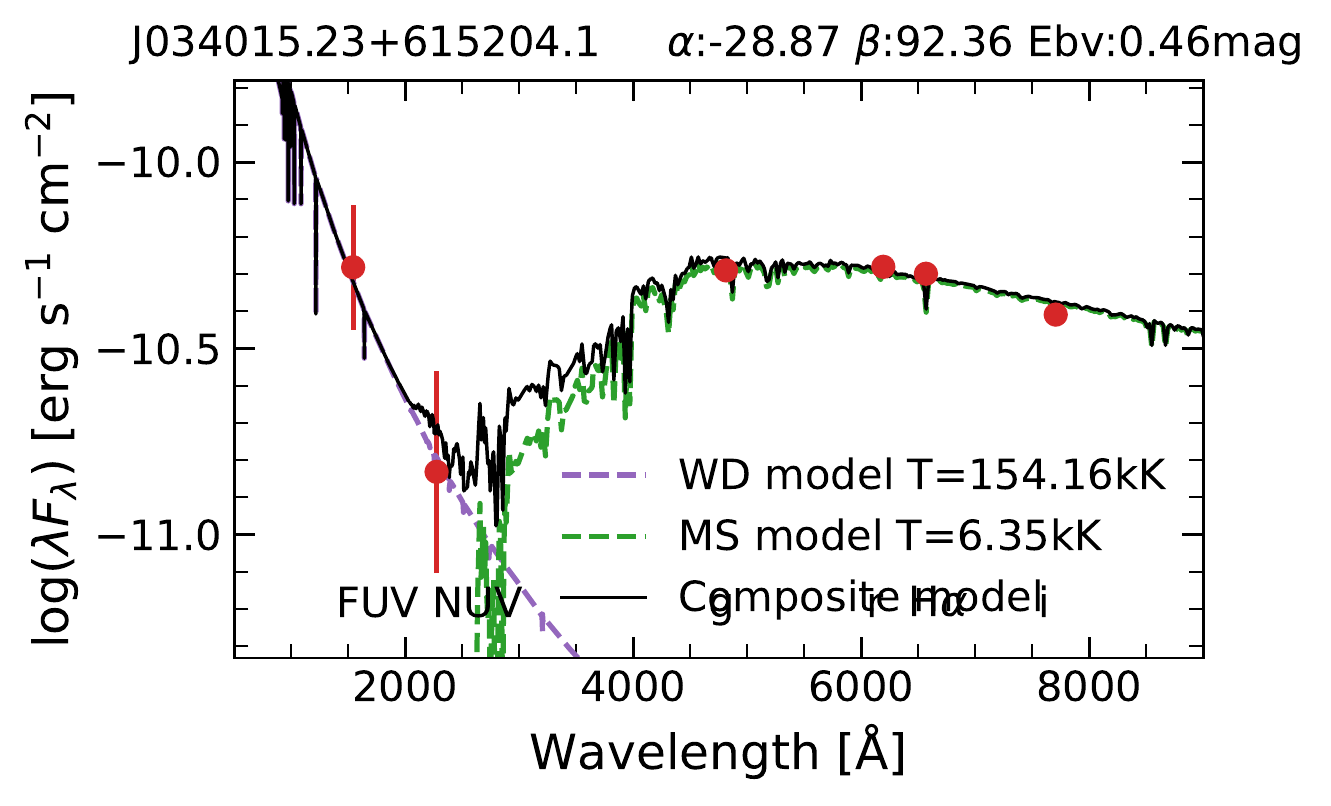}
\includegraphics[width=0.33\textwidth]{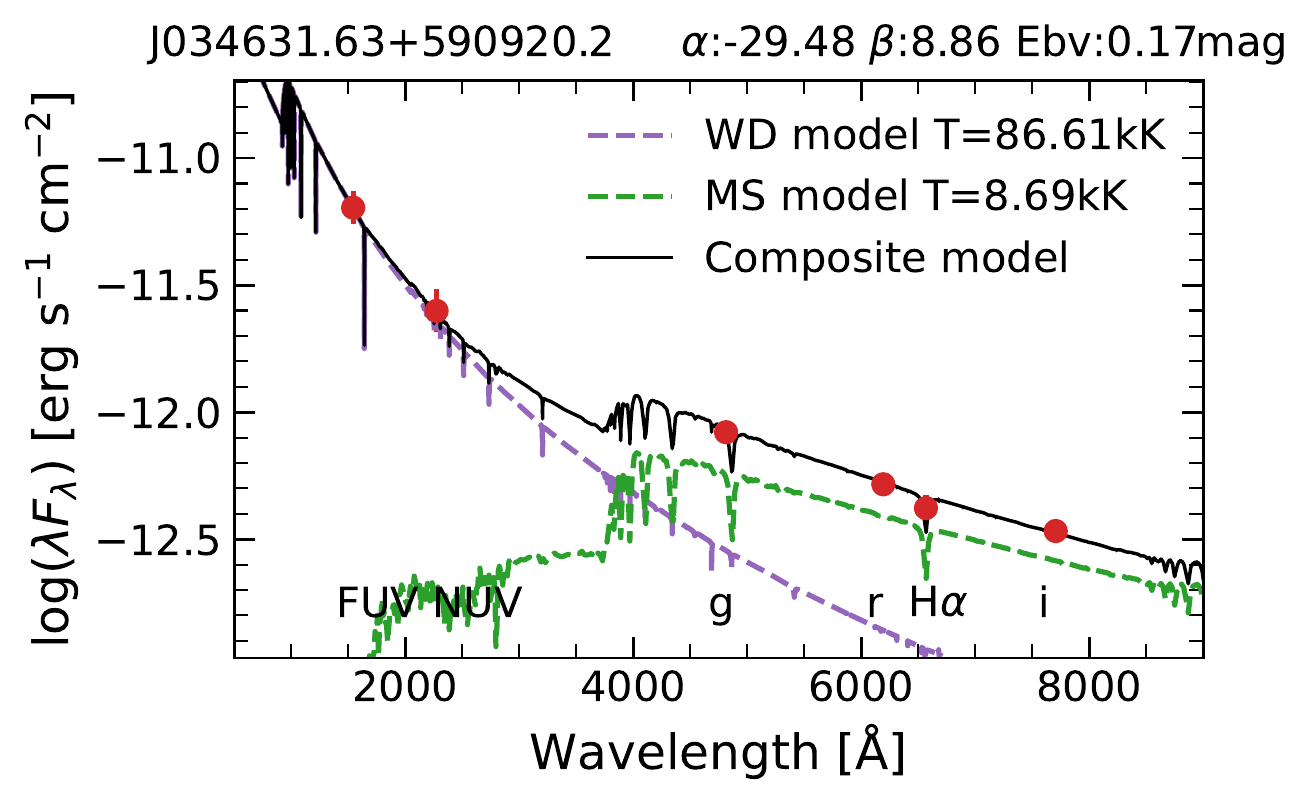}
\includegraphics[width=0.33\textwidth]{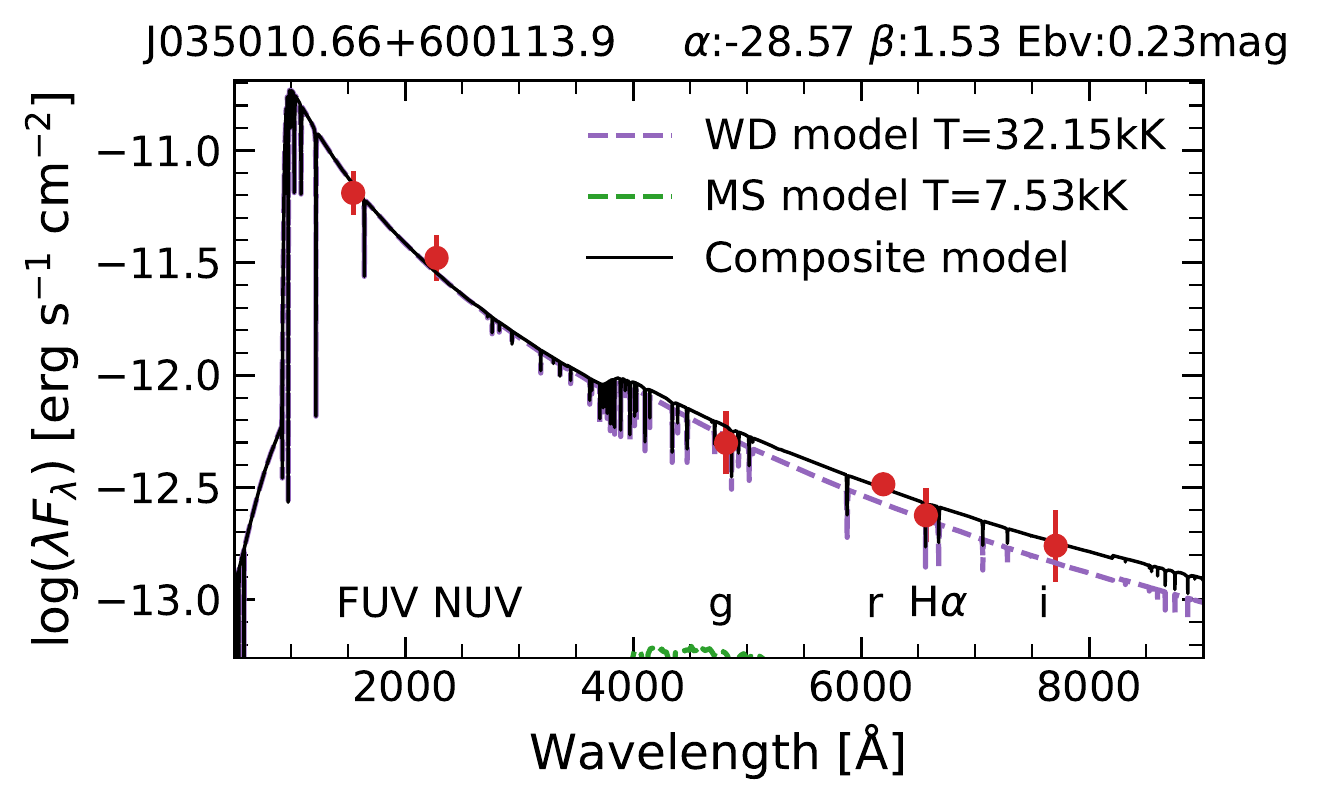}
\includegraphics[width=0.33\textwidth]{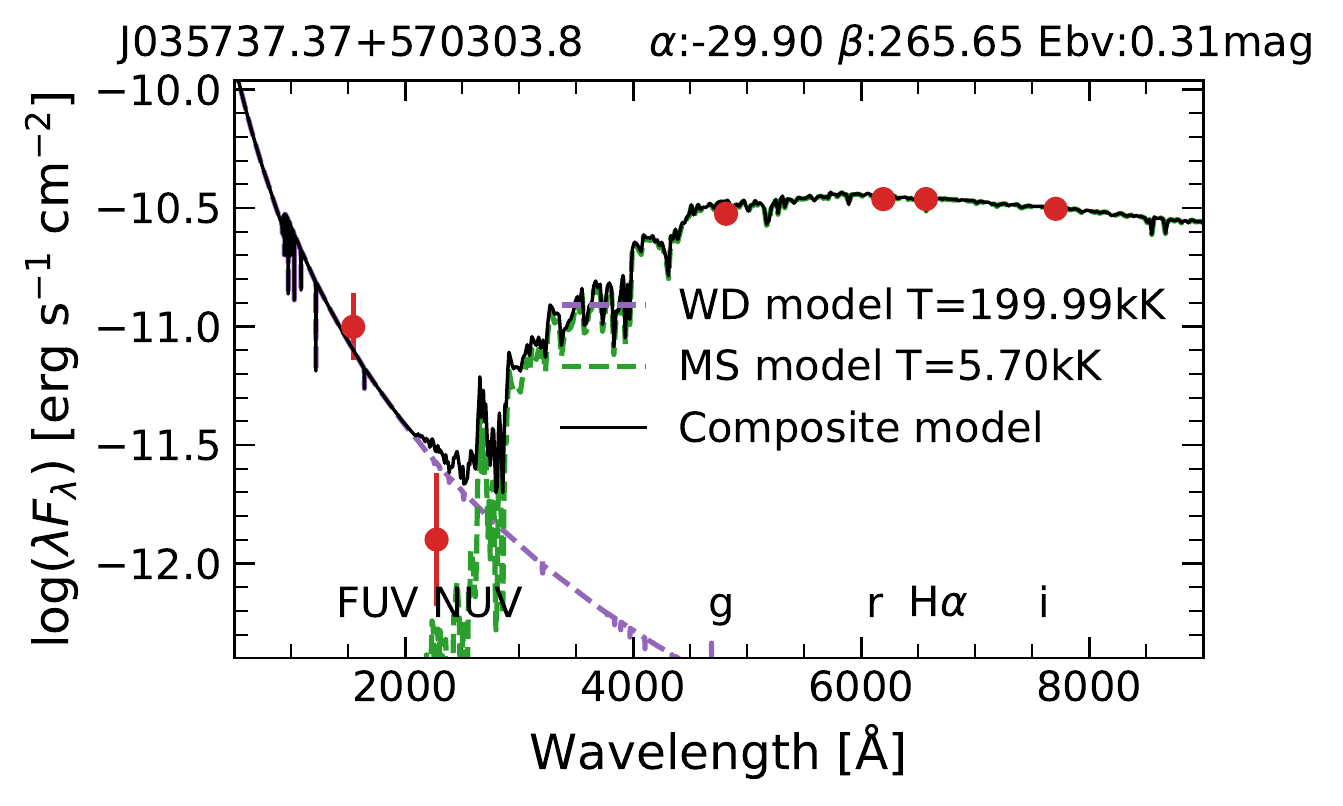}
    \caption{SED fit of the \textit{GALEX} \textit{FUV} and \textit{NUV} + IGAPS g, r, i, and H$\alpha$ photometry (red dots) from which
    a binary WD star is infered. Synthetic spectra of the best fit of hot and cool star is shown as purple  and green dashed lines, respectively. The composite model spectrum is also shown (black solid line).}
    \label{fig:binary_wd1}
\end{figure*}

\begin{figure*}
\centering
\includegraphics[width=0.33\textwidth]{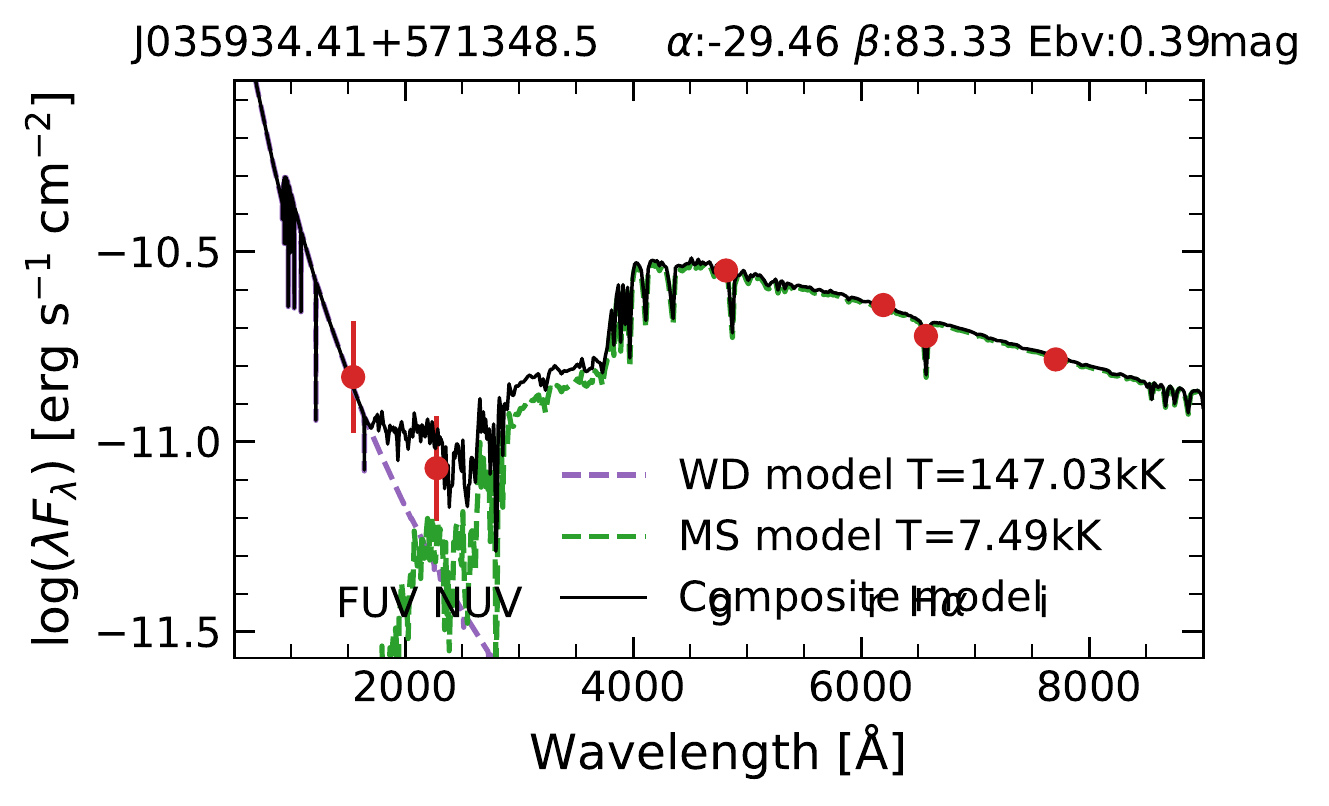}
\includegraphics[width=0.33\textwidth]{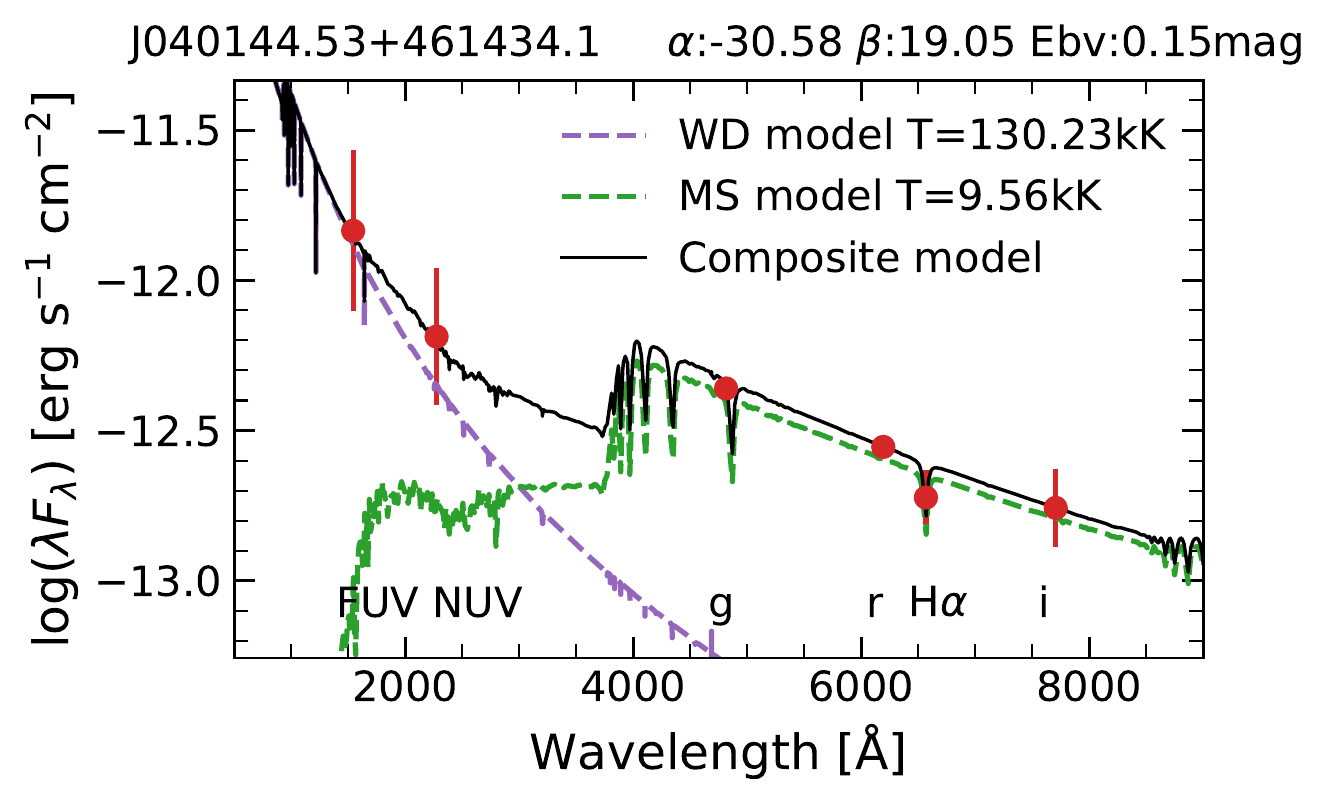}
\includegraphics[width=0.33\textwidth]{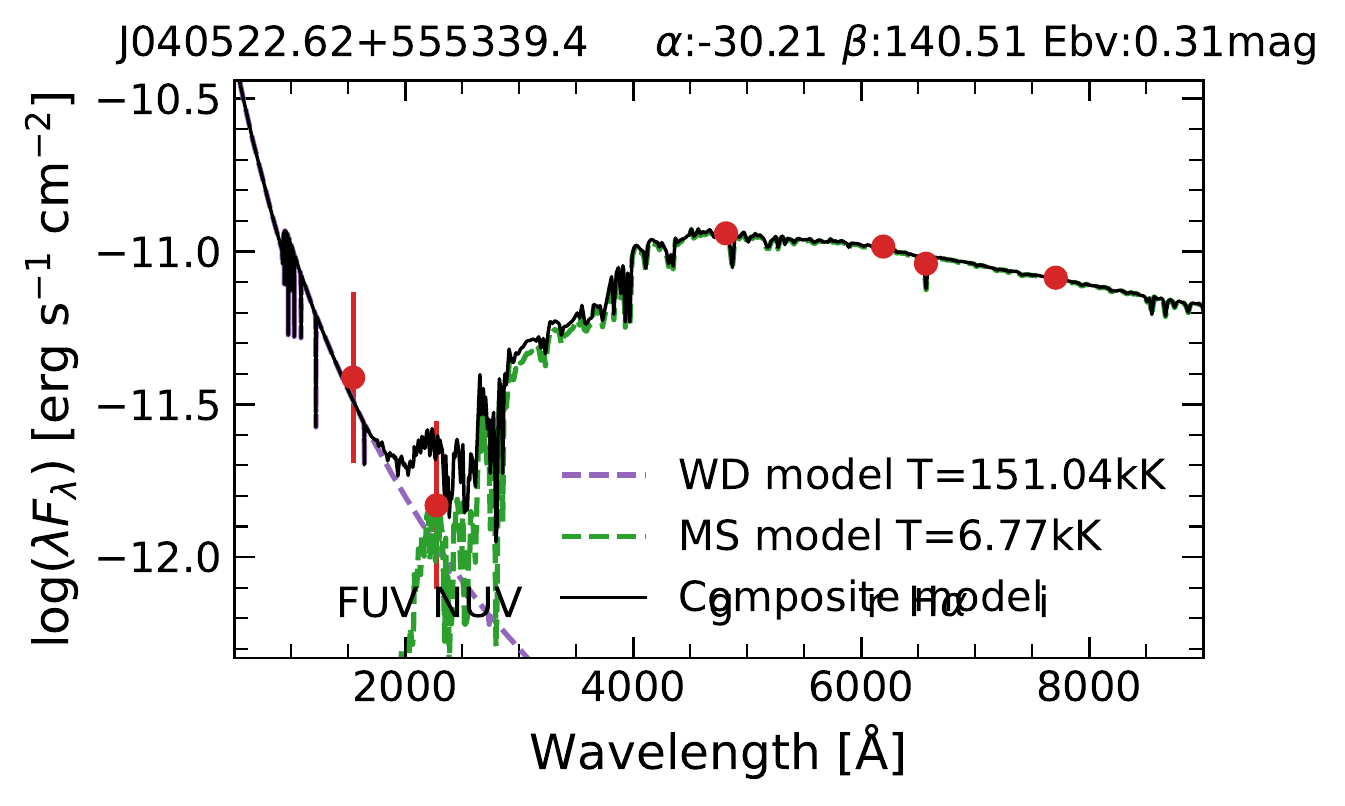}
\includegraphics[width=0.33\textwidth]{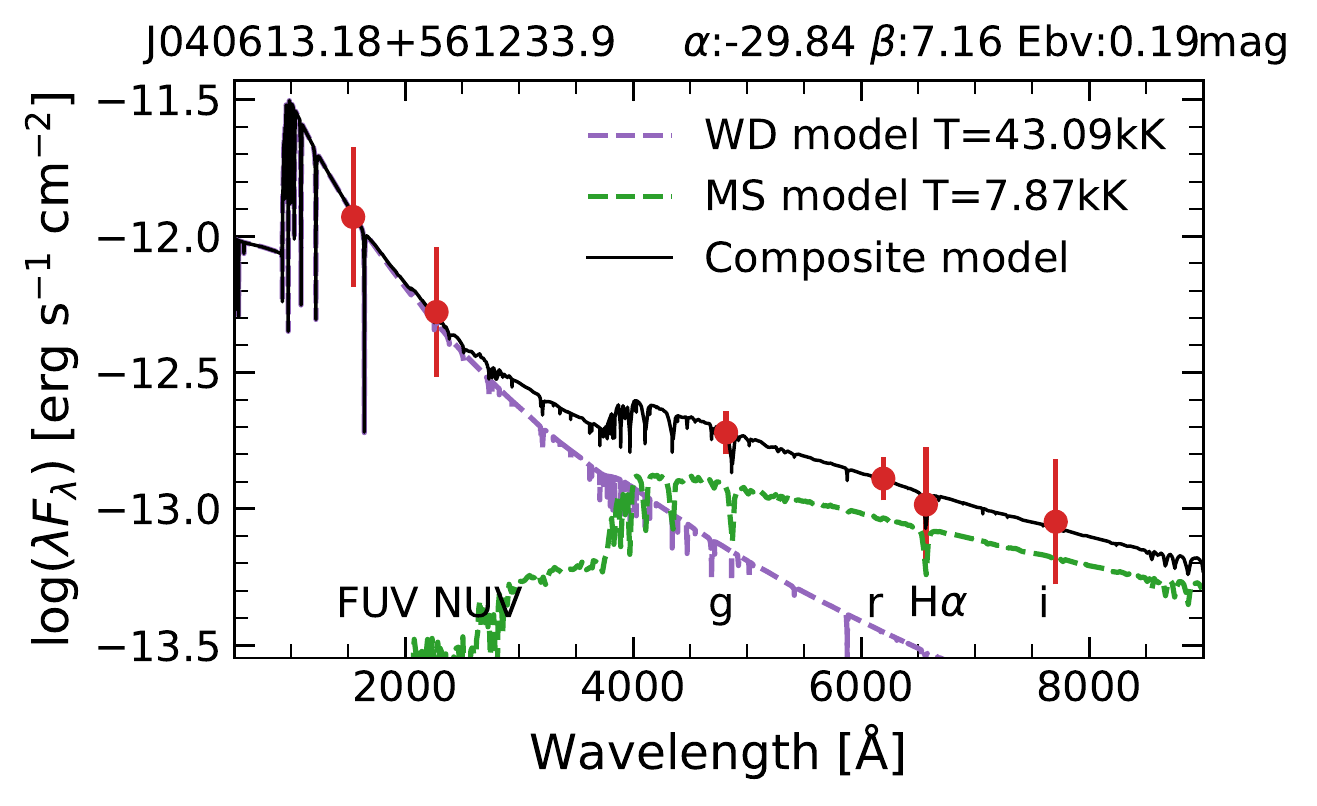}
\includegraphics[width=0.33\textwidth]{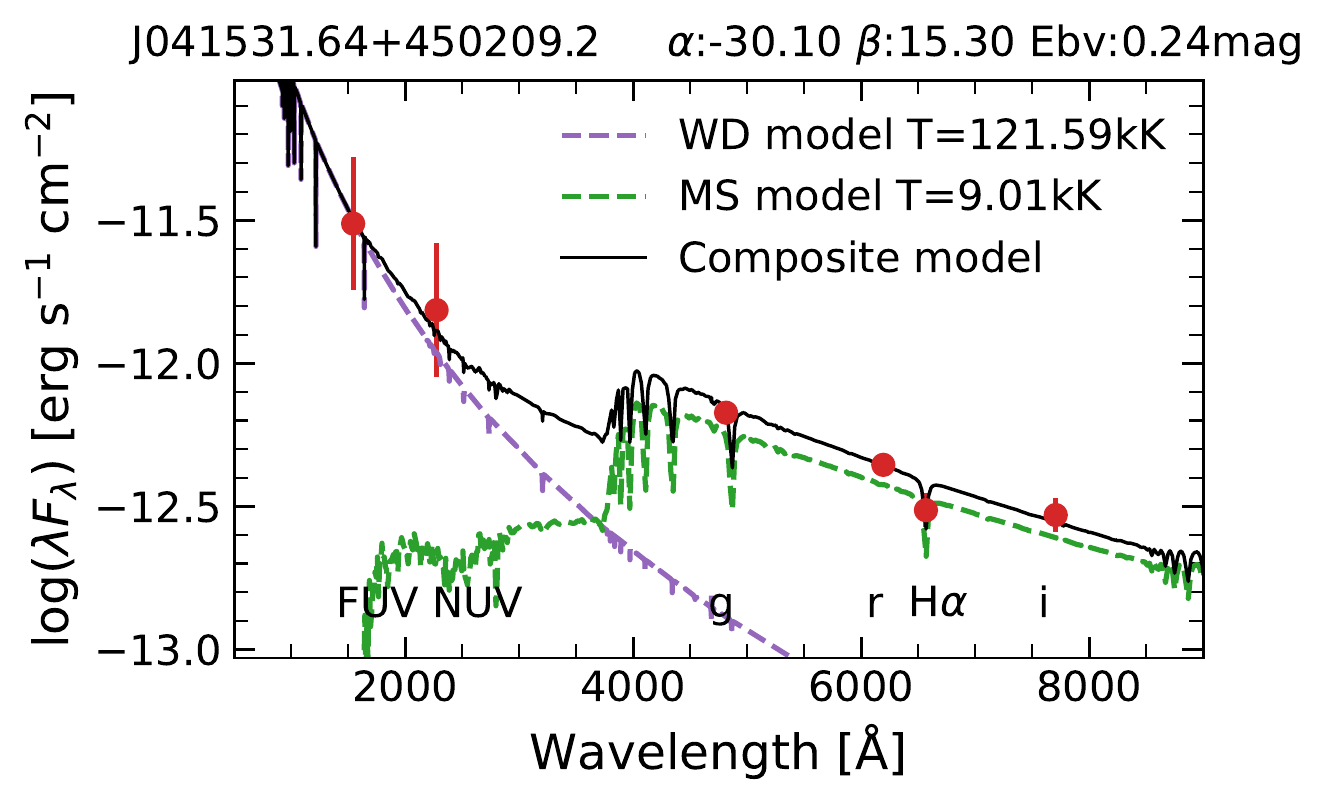}
\includegraphics[width=0.33\textwidth]{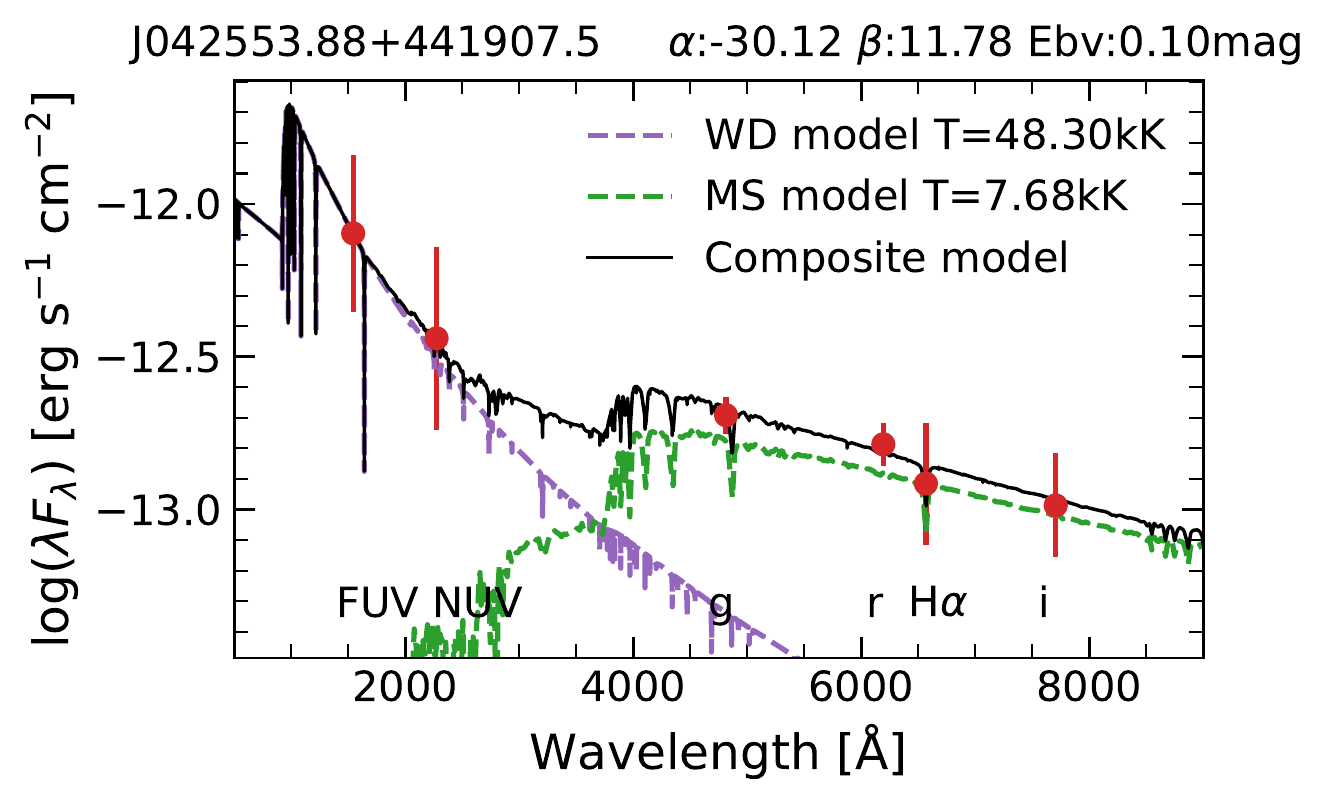}
\includegraphics[width=0.33\textwidth]{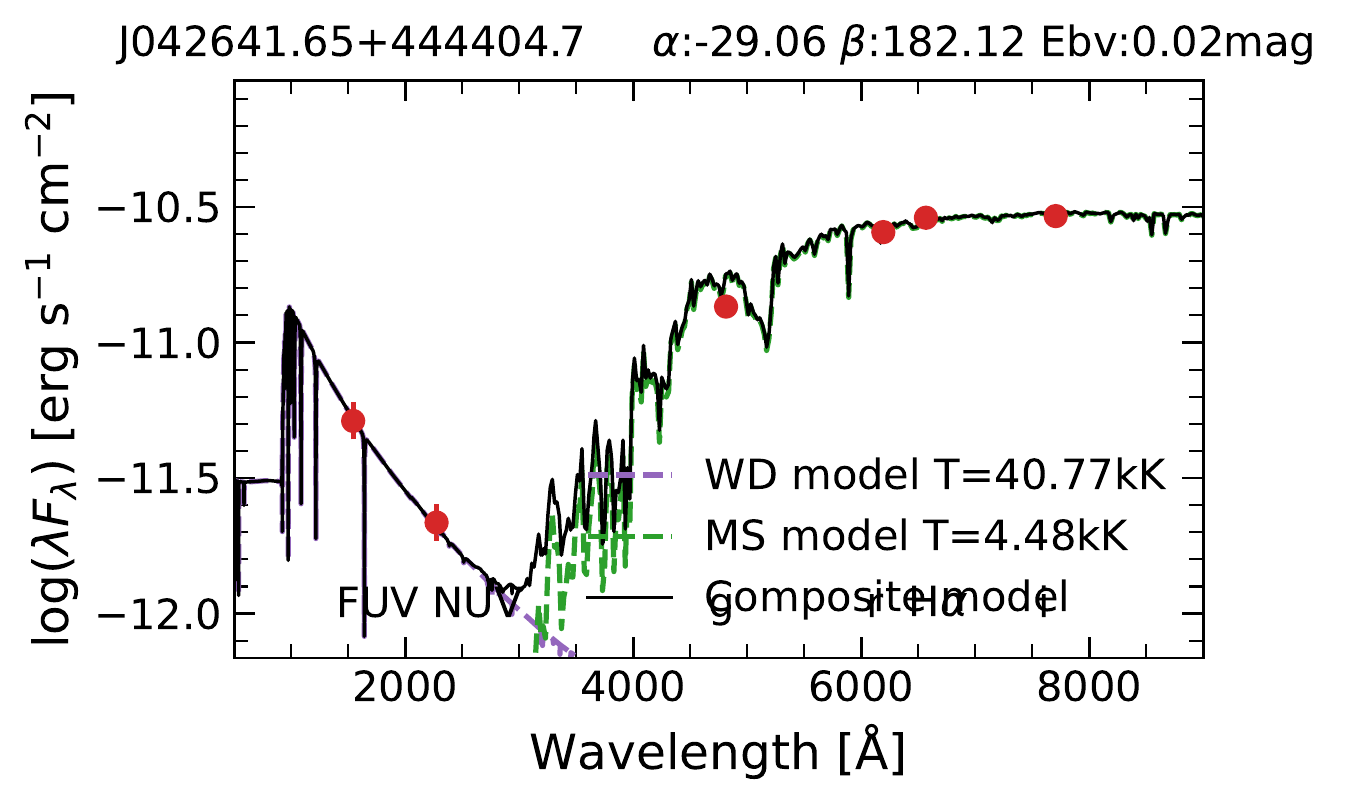}
\includegraphics[width=0.33\textwidth]{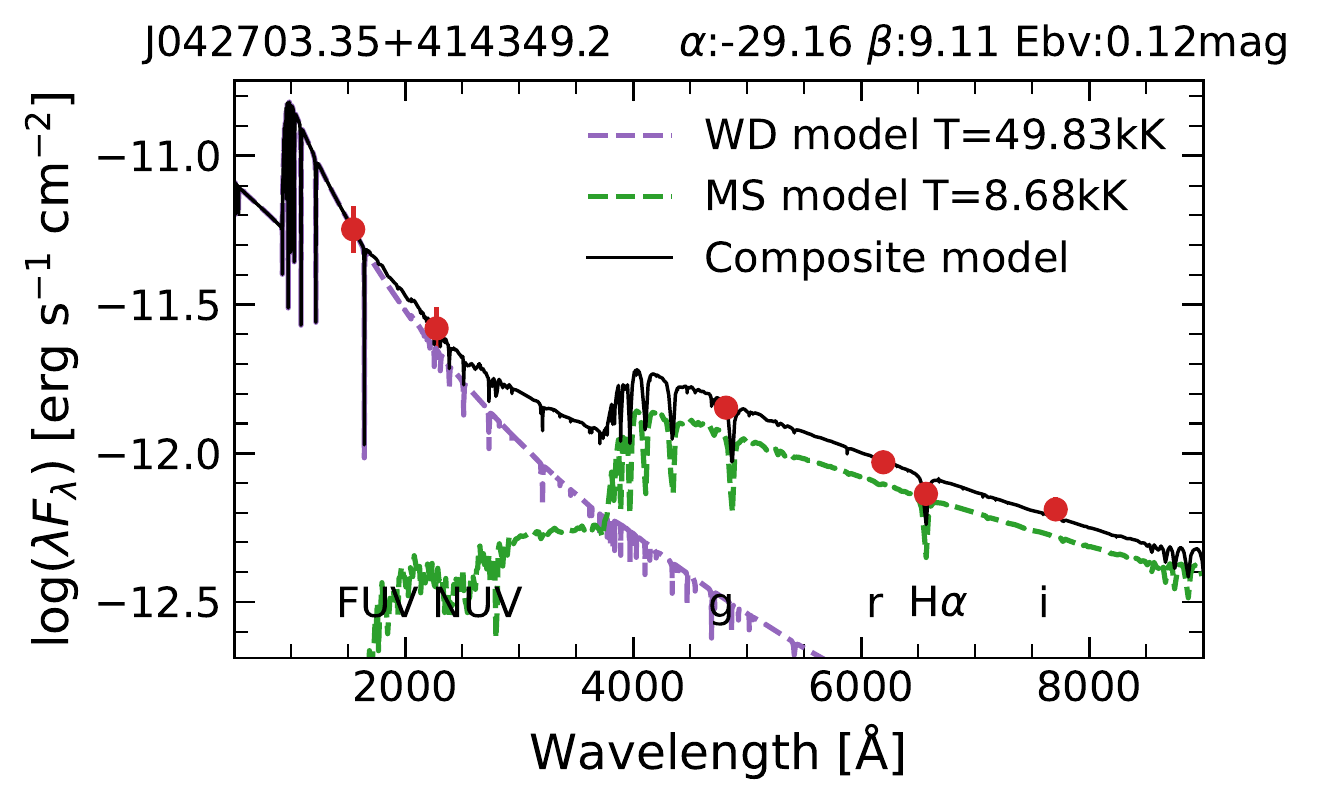}
\includegraphics[width=0.33\textwidth]{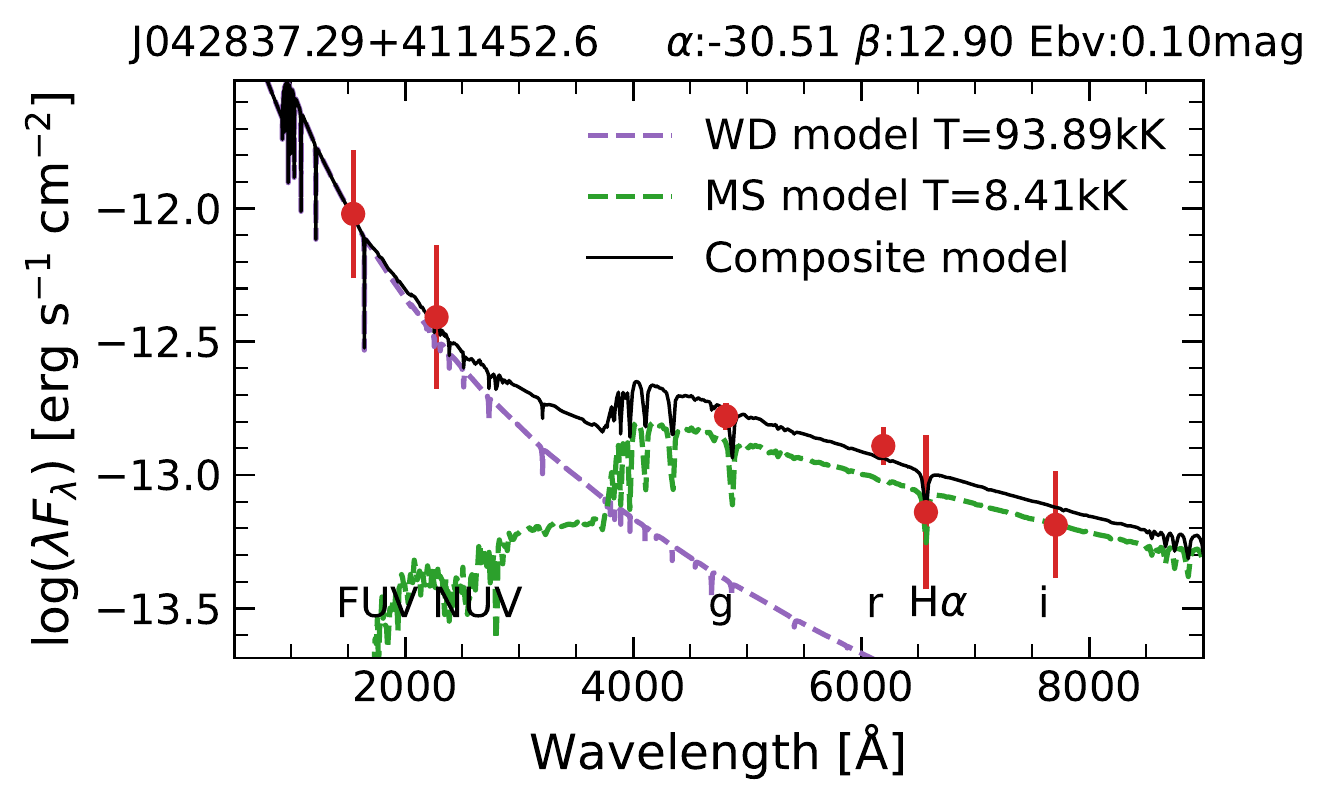}
\includegraphics[width=0.33\textwidth]{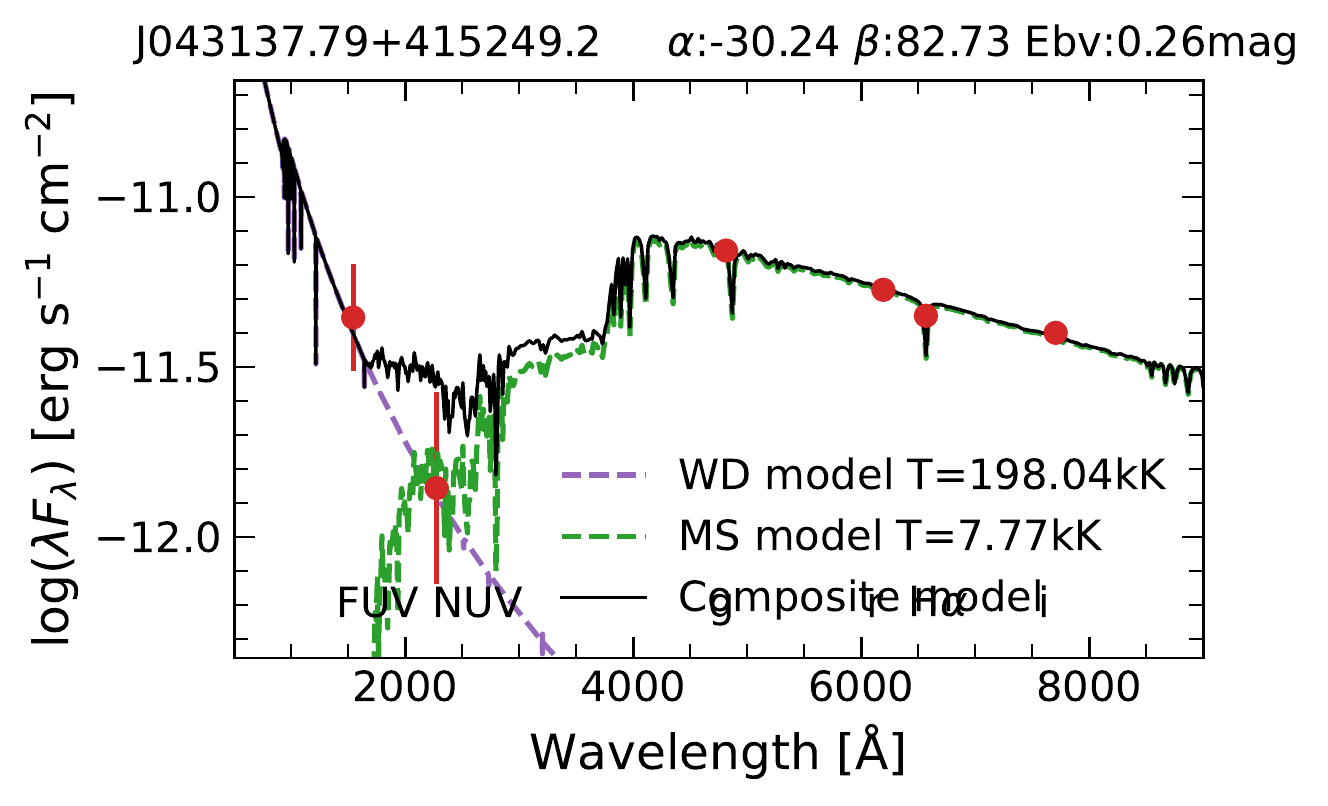}
\includegraphics[width=0.33\textwidth]{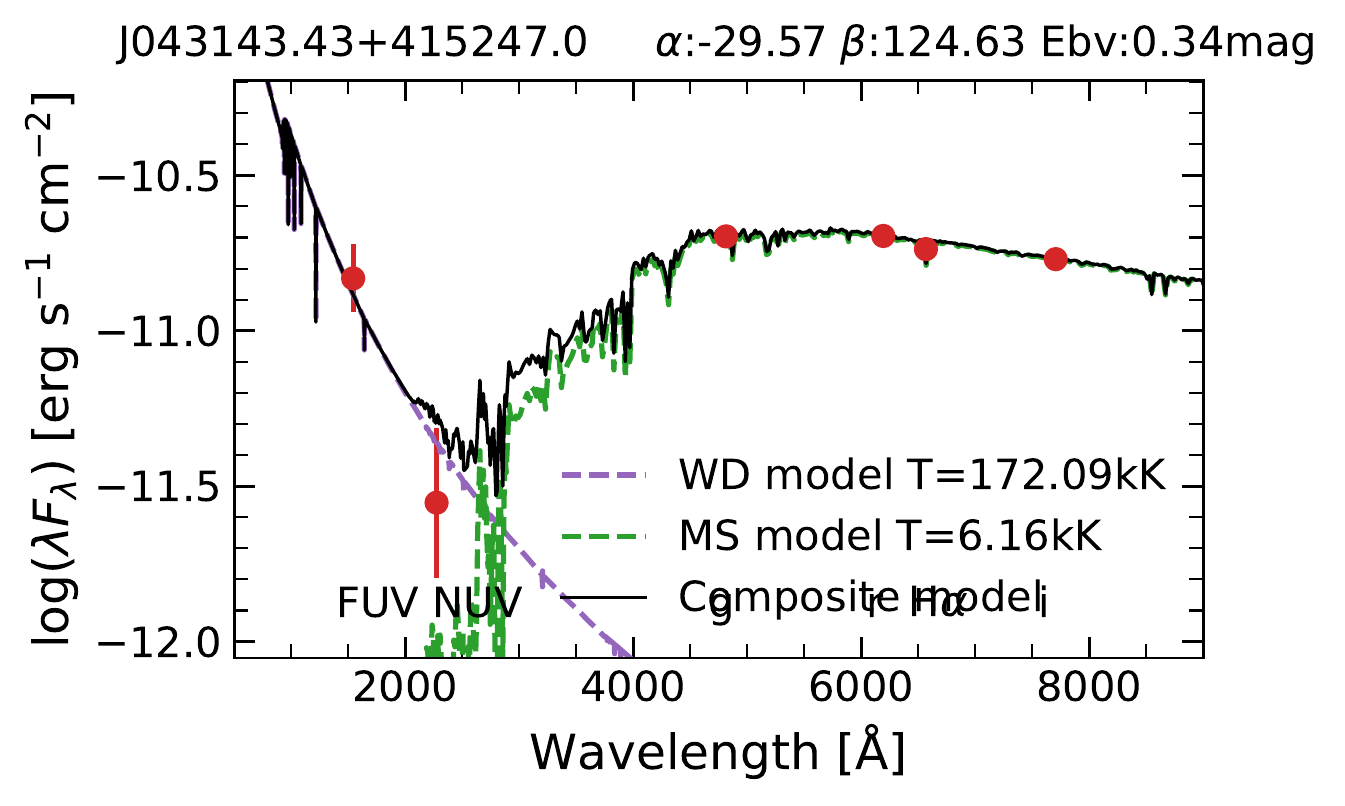}
\includegraphics[width=0.33\textwidth]{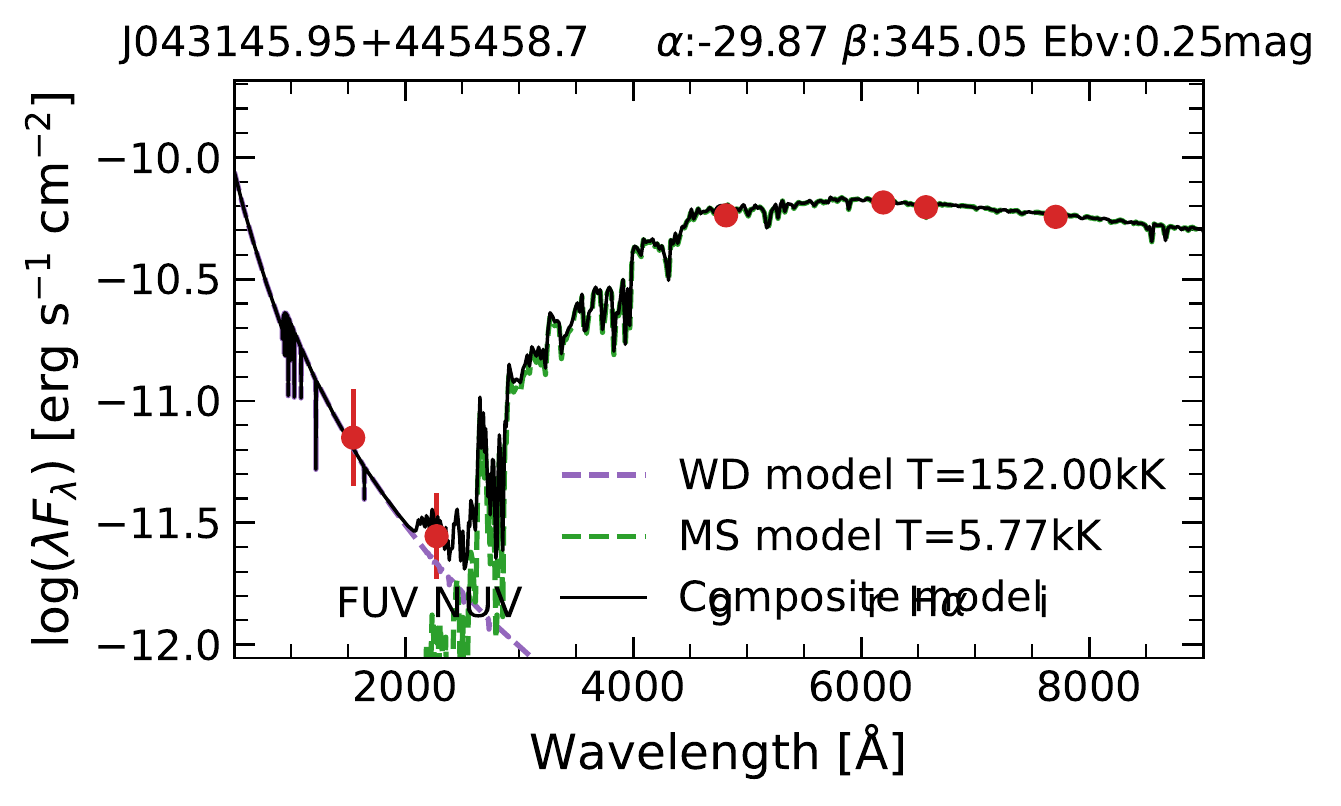}
\includegraphics[width=0.33\textwidth]{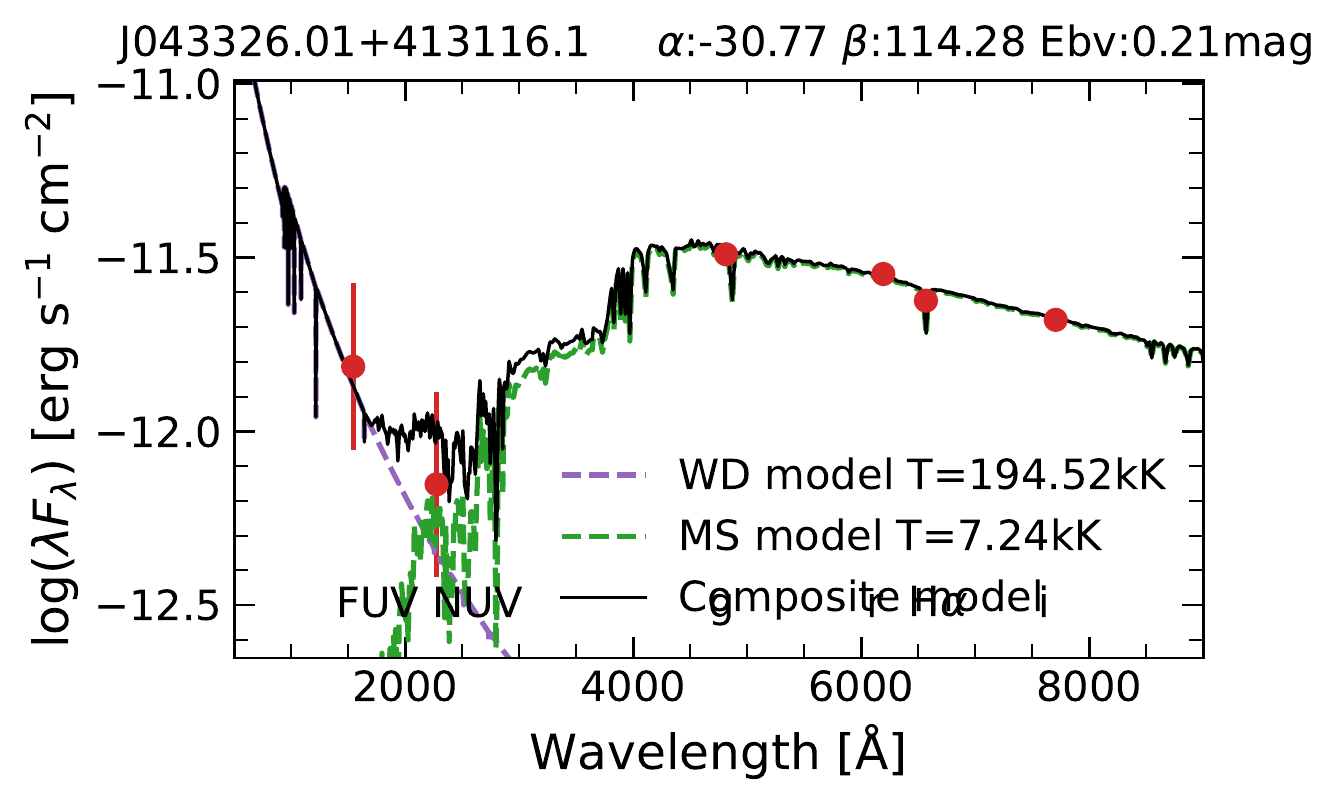}
\includegraphics[width=0.33\textwidth]{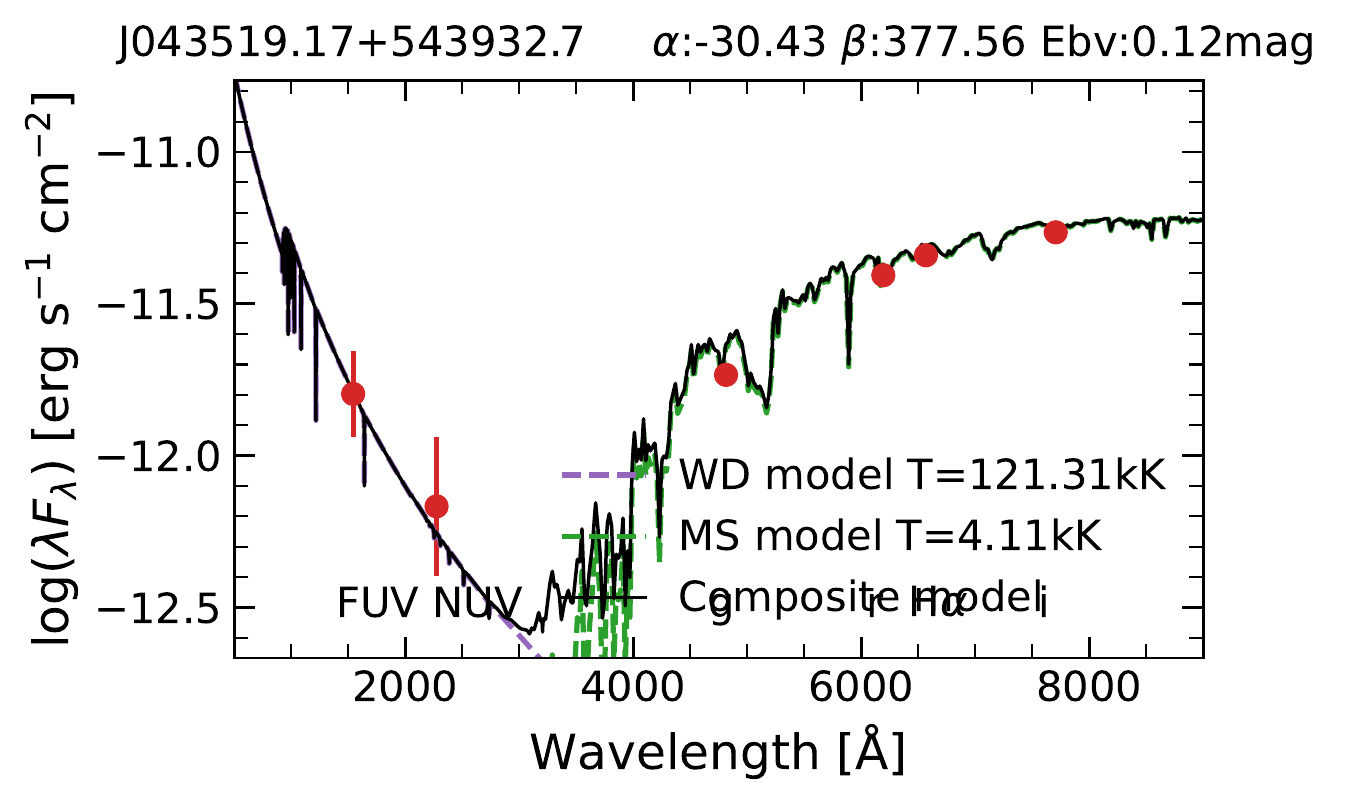}
\includegraphics[width=0.33\textwidth]{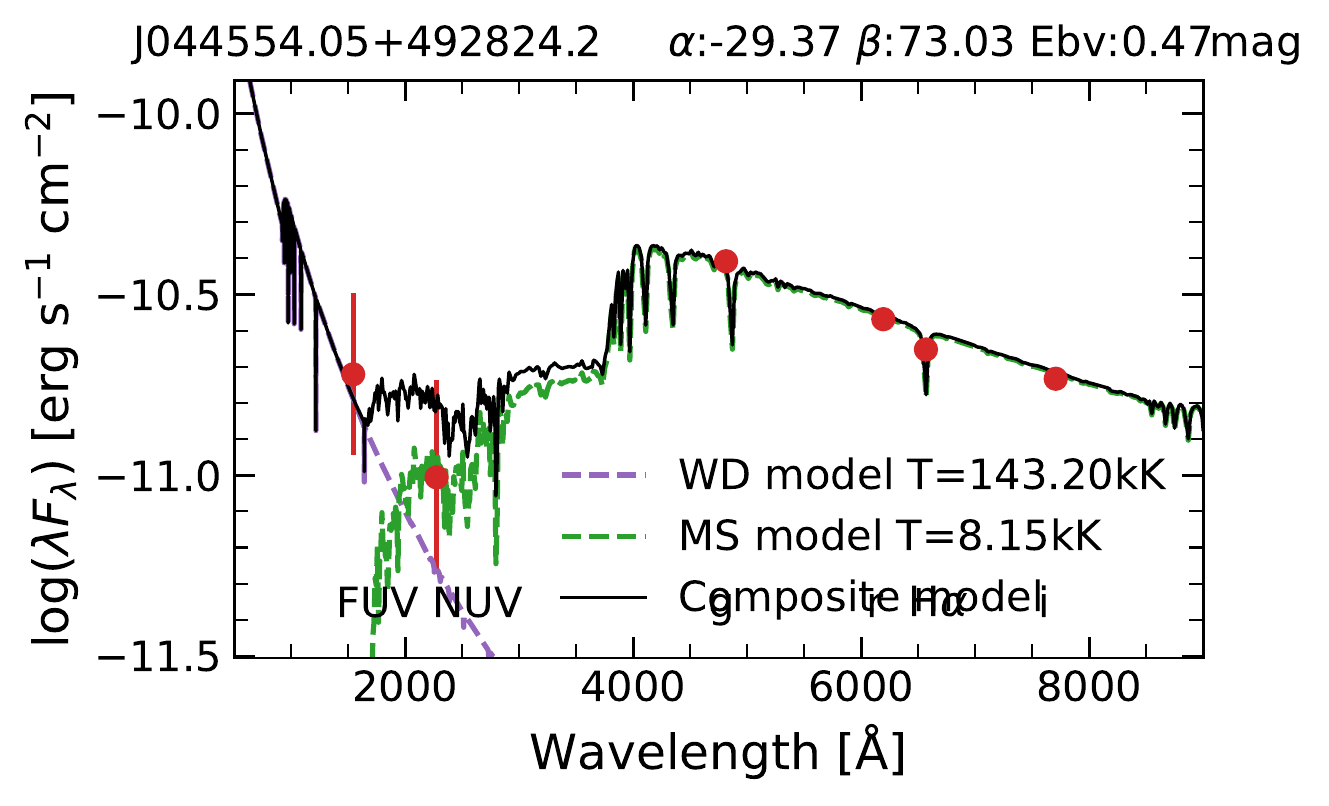}
    \caption{Similar to Figure~\ref{fig:binary_wd1}.}
    \label{fig:binary_wd2}
\end{figure*}

\begin{figure*}
\centering
\includegraphics[width=0.33\textwidth]{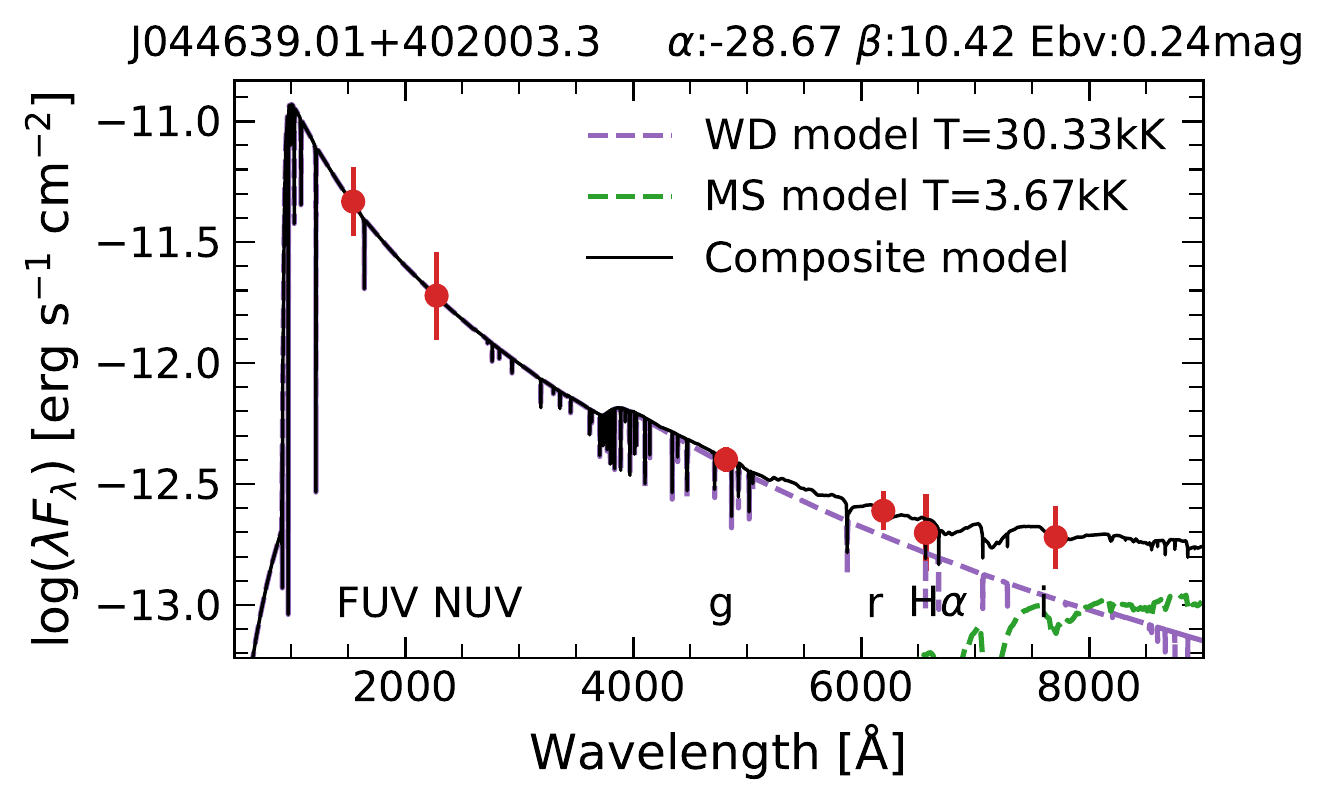}
\includegraphics[width=0.33\textwidth]{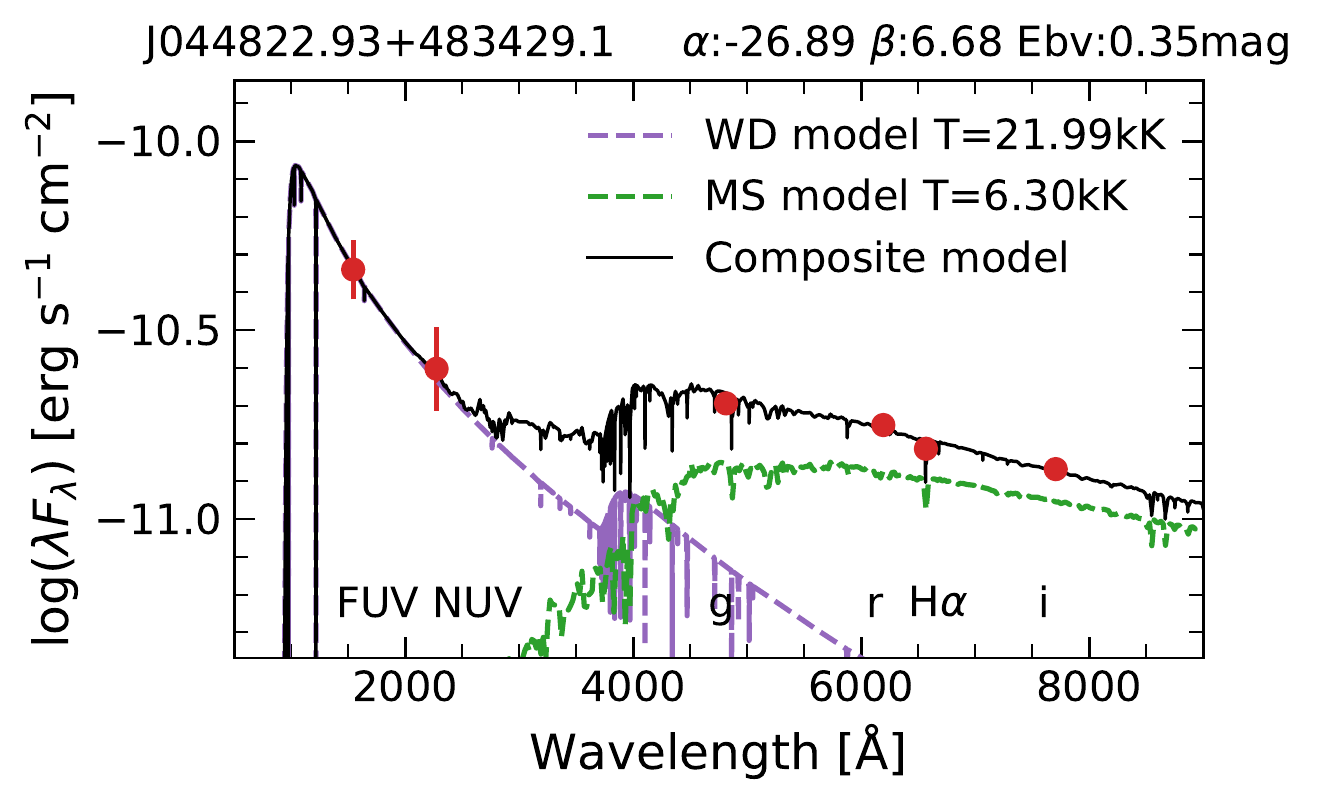}
\includegraphics[width=0.33\textwidth]{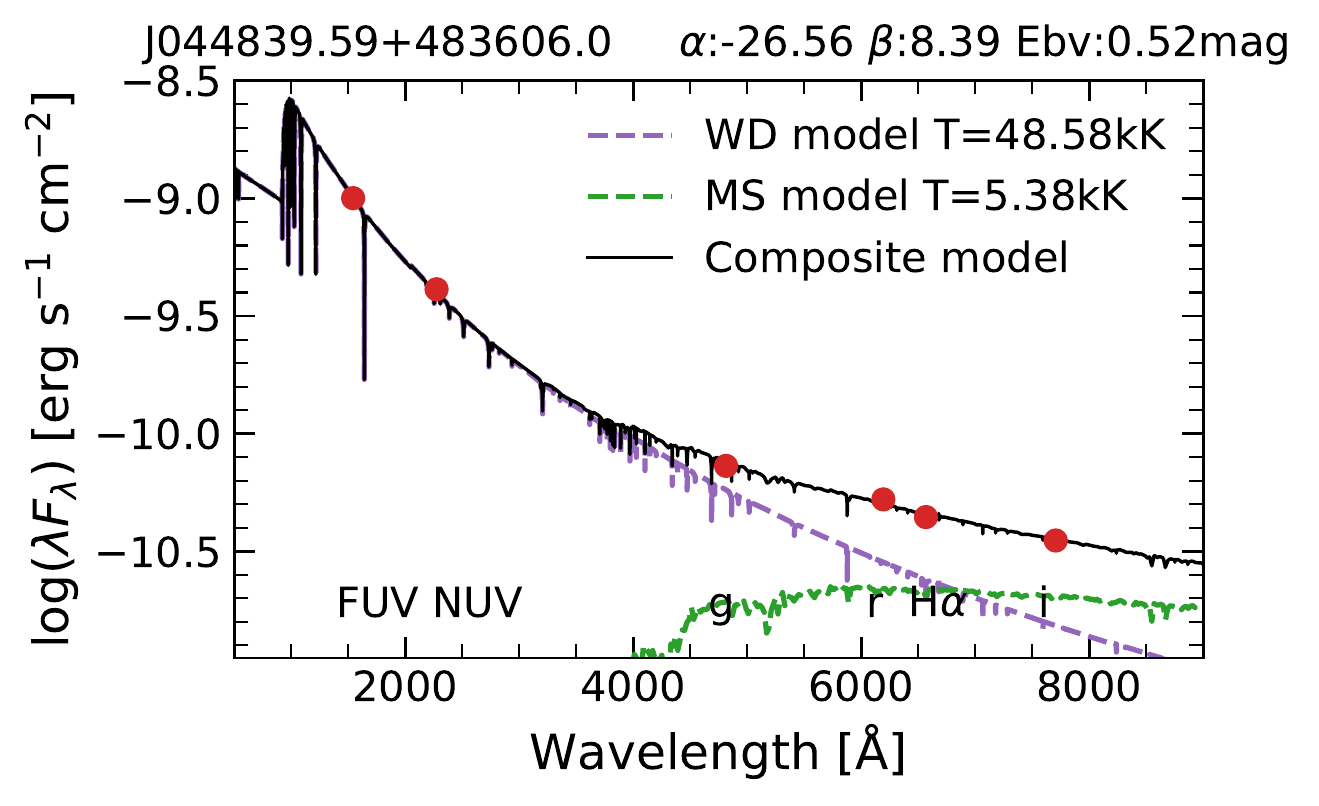}
\includegraphics[width=0.33\textwidth]{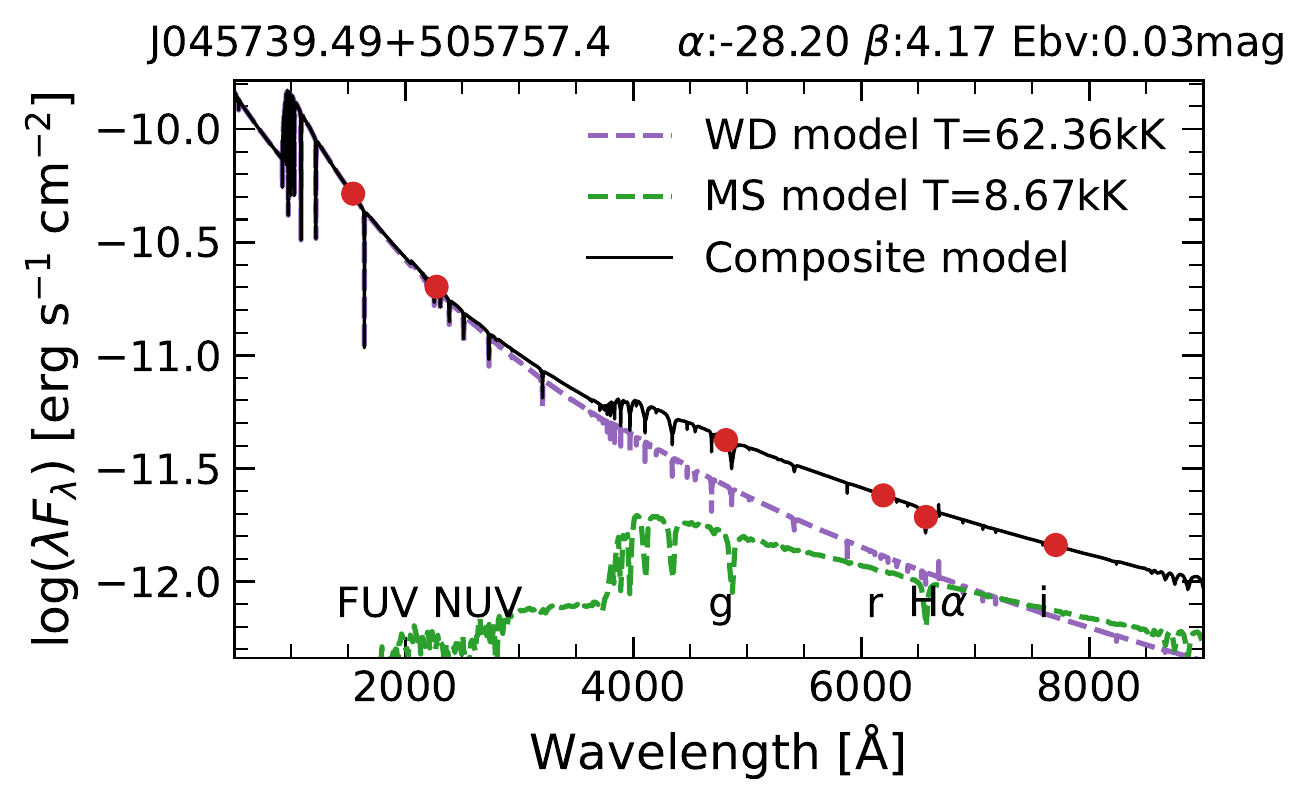}
\includegraphics[width=0.33\textwidth]{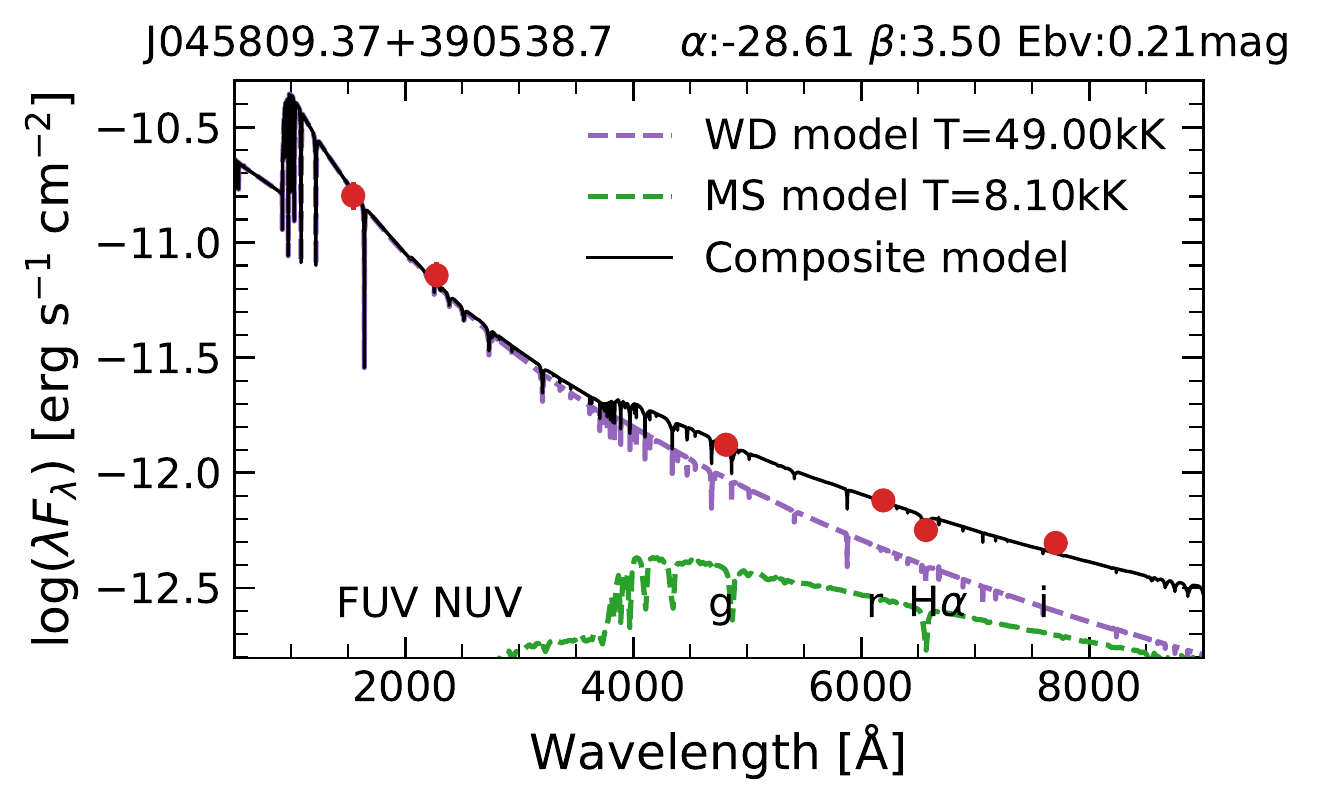}
\includegraphics[width=0.33\textwidth]{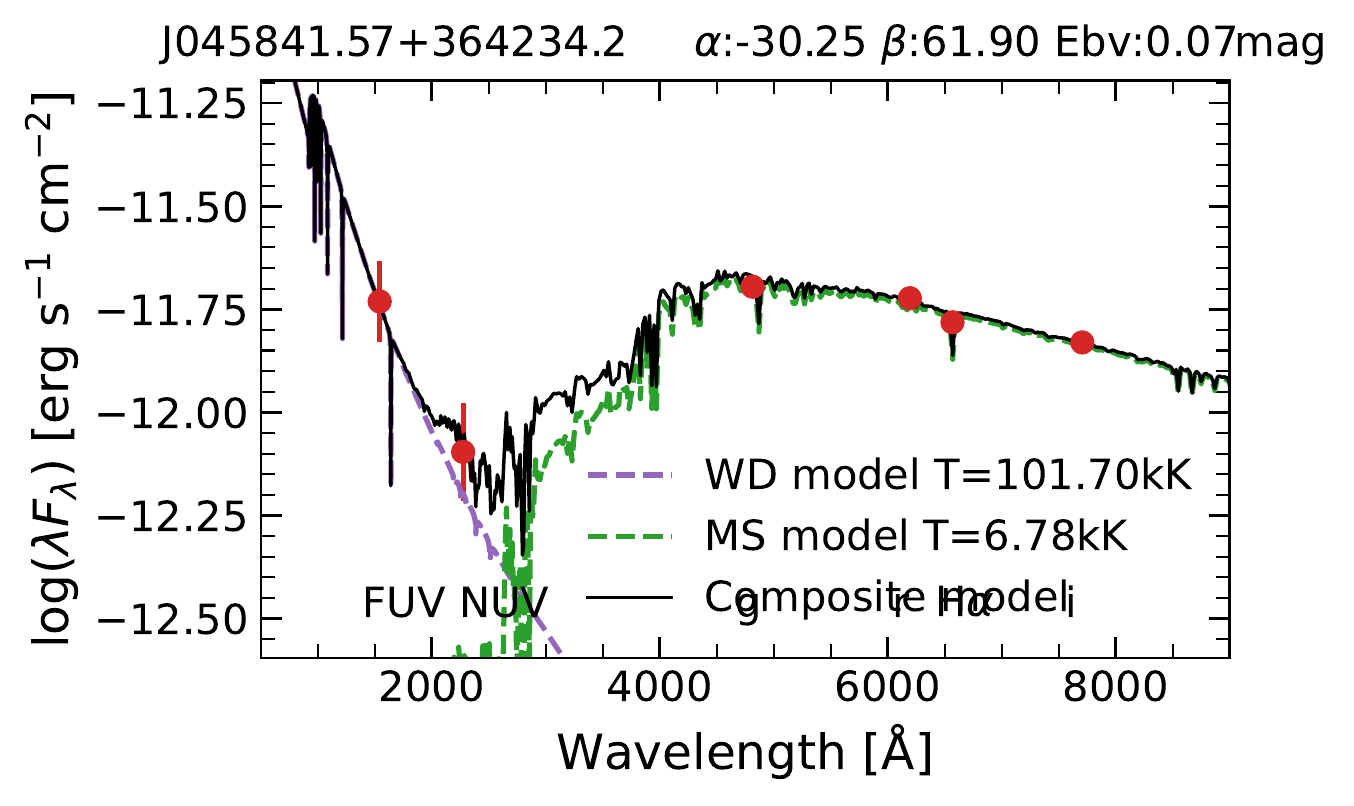}
\includegraphics[width=0.33\textwidth]{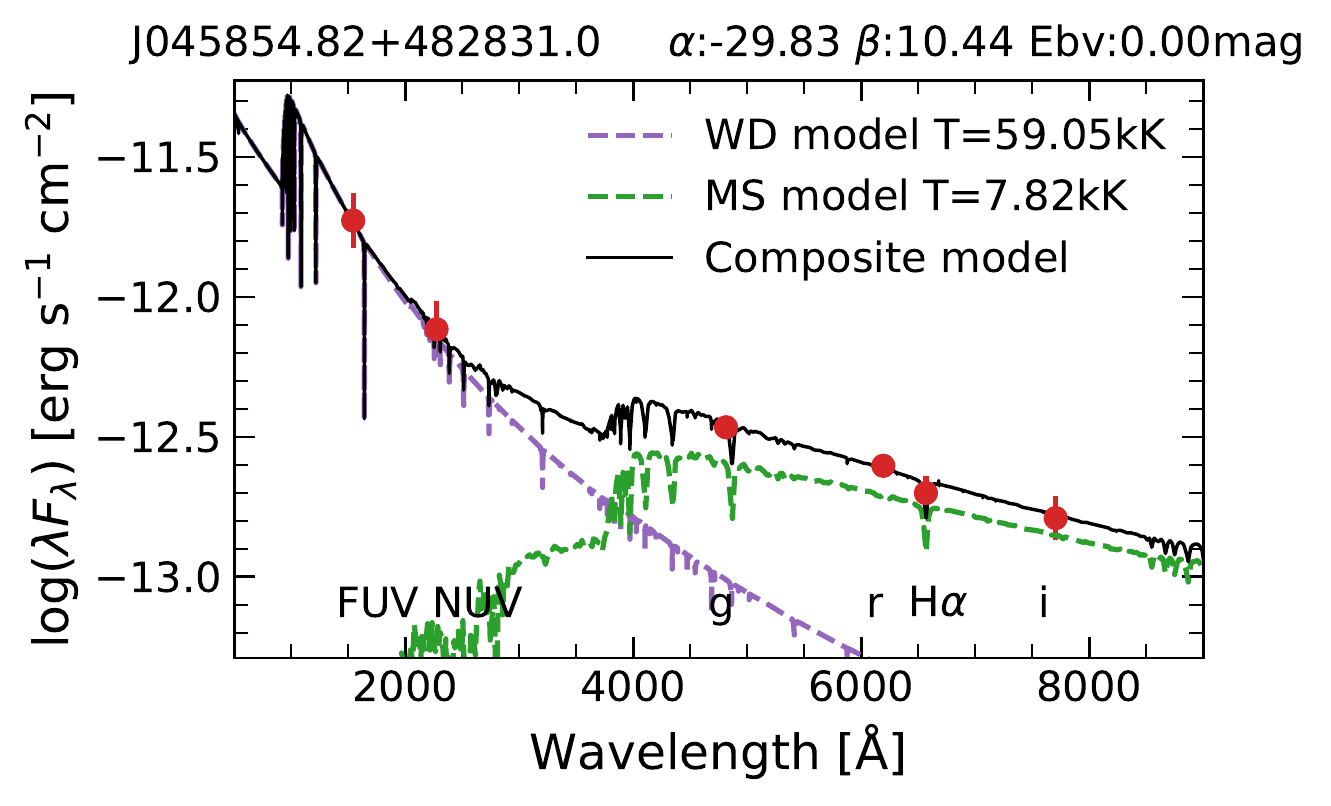}
\includegraphics[width=0.33\textwidth]{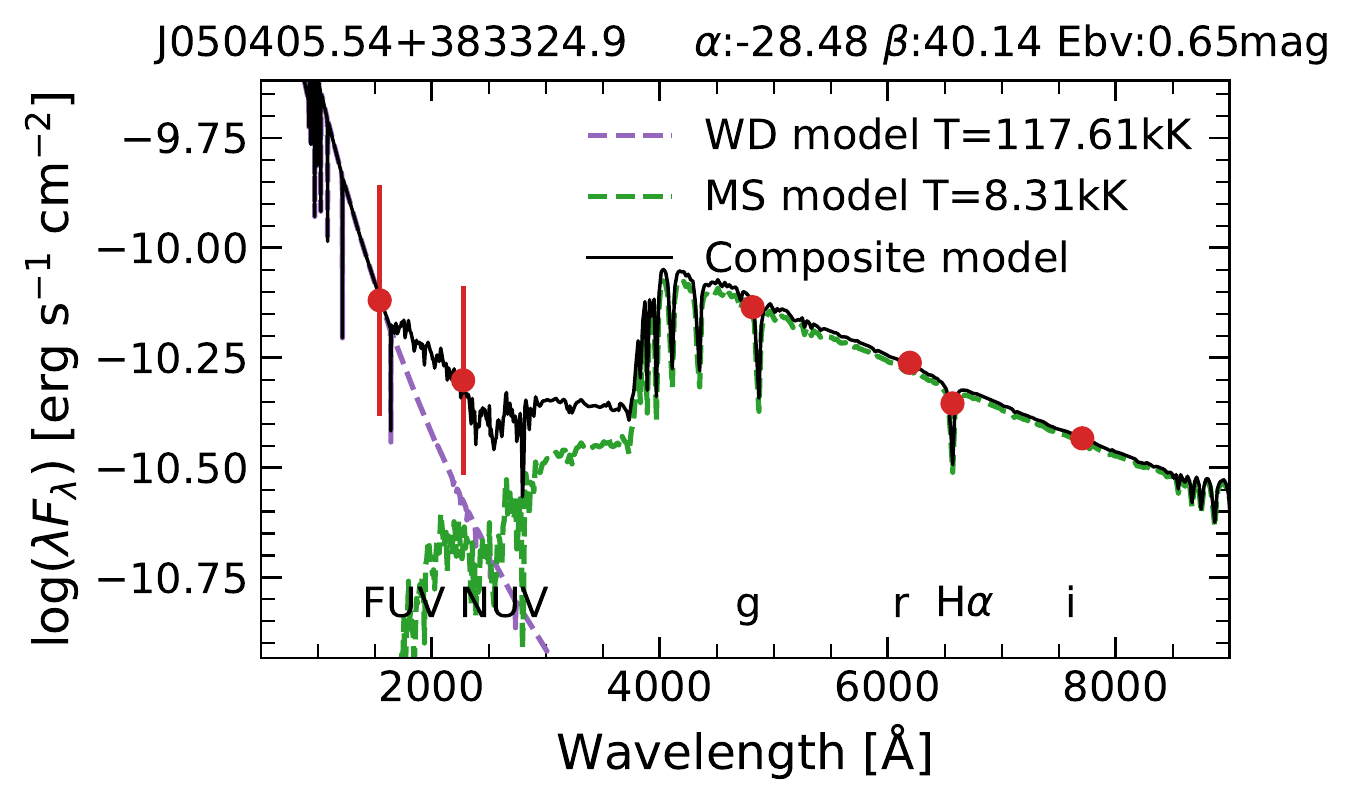}
\includegraphics[width=0.33\textwidth]{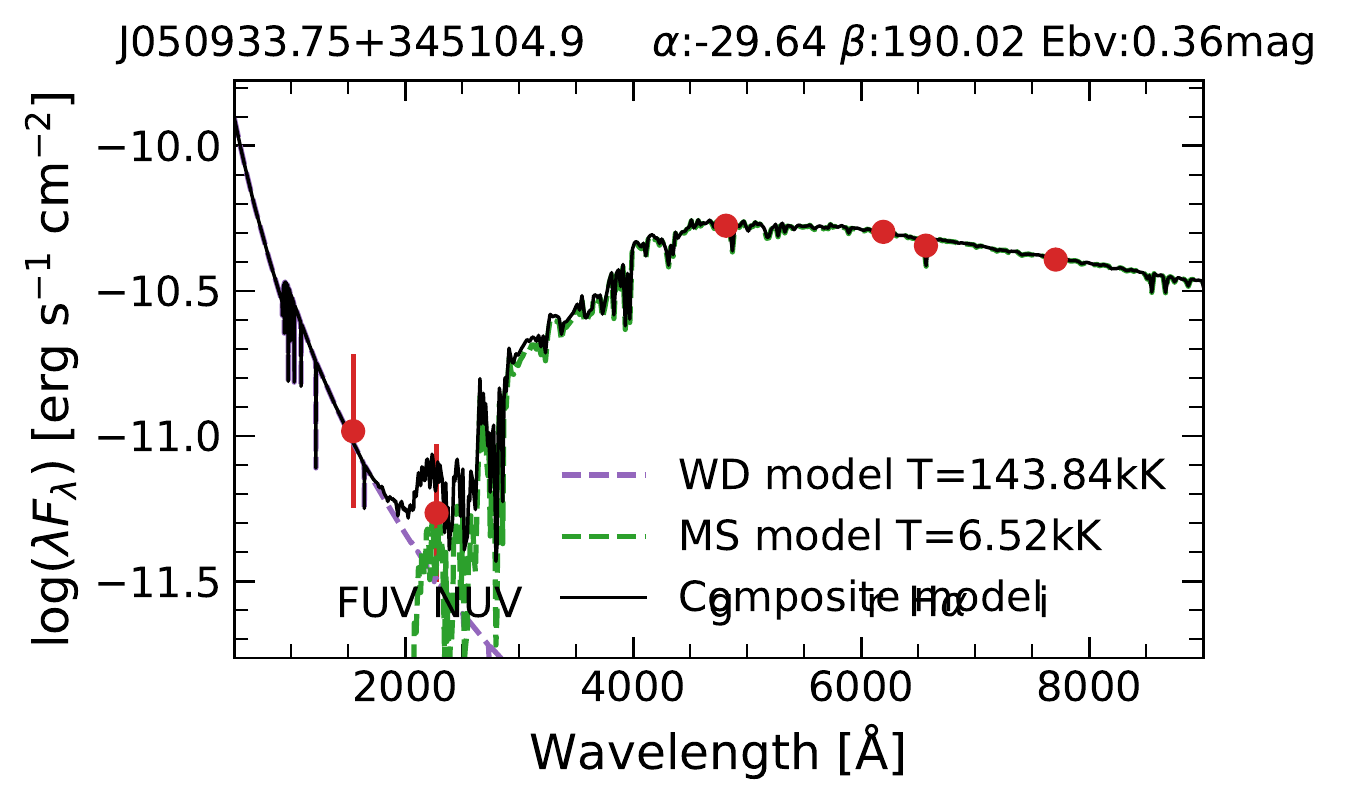}
\includegraphics[width=0.33\textwidth]{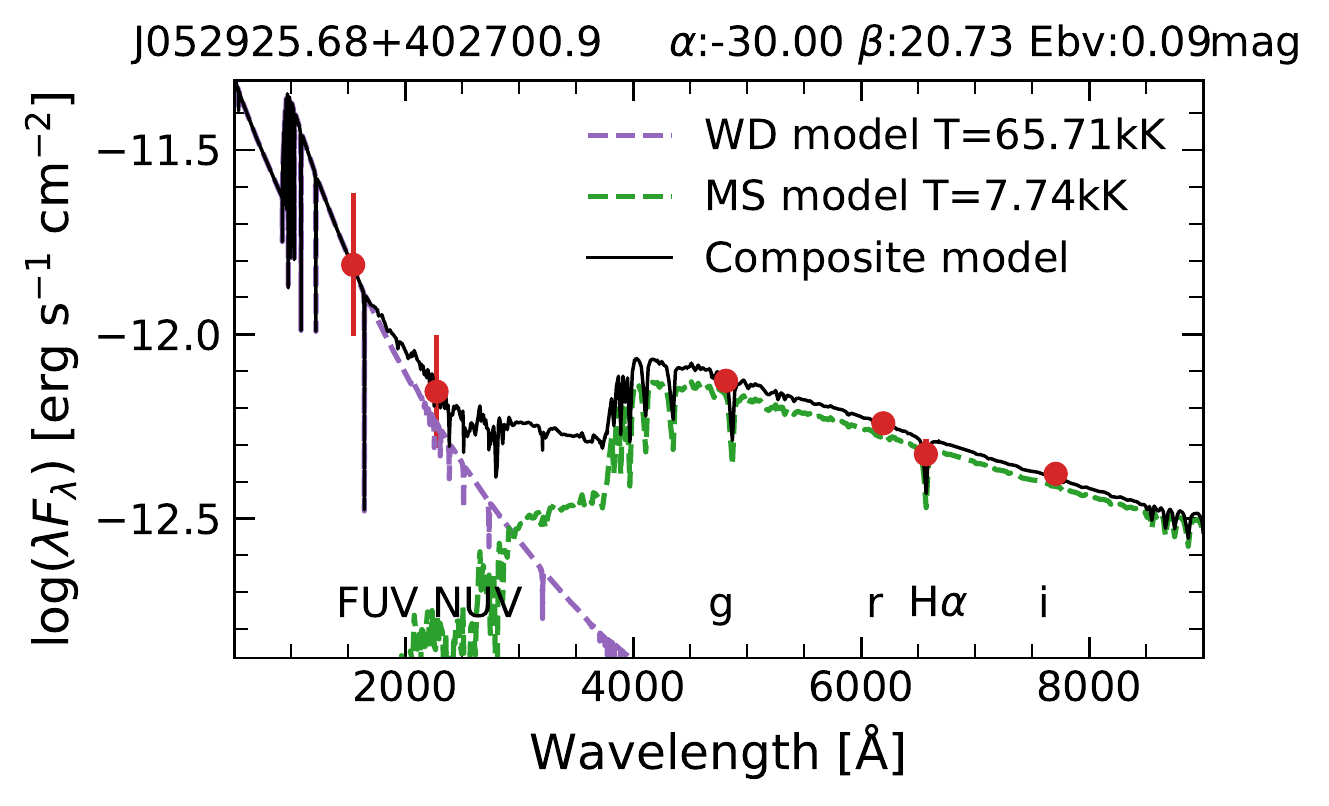}
\includegraphics[width=0.33\textwidth]{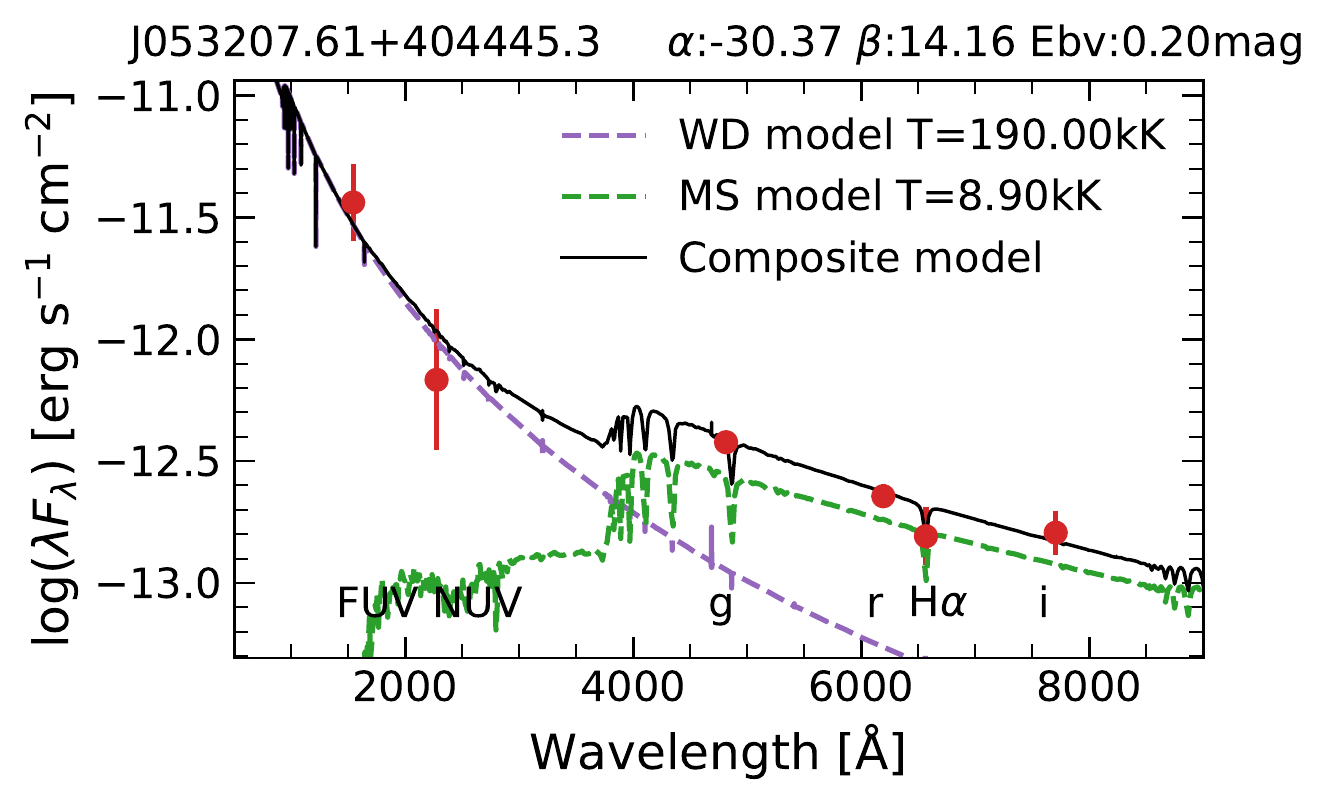}
    \caption{Similar to Figure~\ref{fig:binary_wd1}.}
    \label{fig:binary_wd3}
\end{figure*}

\bsp
\label{lastpage}
\end{document}